\documentclass[journal, final]{IEEEtran}

\usepackage{amssymb}
\usepackage[cmex10]{amsmath}
\usepackage{subeqnarray}
\usepackage{graphicx}
\usepackage{comment}
\usepackage{mathtools}

\newtheorem{thm}{Theorem}
\newtheorem{example}{Example}
\newtheorem{cor}{Corollary}
\newtheorem{lem}{Lemma}
\newtheorem{rem}{Remark}
\begin{document}
\title{On the equivalence between Stein and De Bruijn identities}

\author{Sangwoo~Park,~Erchin~Serpedin, and Khalid Qaraqe %}%,~\IEEEmembership{Life~Fellow,~IEEE}% <-this % stops a space
\thanks{Department of Electrical and Computer Engineering, Texas A\&M University, College Station, TX 77843-3128  USA,  e-mail: serpedin@ece.tamu.edu. This work was supported by QNRF-NPRP grant 09-341-2-128.}}
%\IEEEpubid{Copyright (c) 2012 IEEE}
\maketitle

\begin{abstract}
This paper focuses on illustrating 1) the equivalence between Stein's identity and De Bruijn's identity, and 2) two extensions of De Bruijn's identity. First, it is shown that Stein's identity is equivalent to De Bruijn's identity under additive noise channels with specific conditions. Second, for arbitrary but fixed input and noise distributions under additive noise channels, the first derivative of the differential entropy is expressed by a function of the posterior mean, and the second derivative of the differential entropy is expressed in terms of a function of Fisher information. Several applications over a number of fields such as signal processing and information theory, are presented to support the usefulness of the  developed results  in this paper.
\end{abstract}


\IEEEpeerreviewmaketitle

% Introduction

\section{Introduction}\label{sec_int}
\IEEEPARstart{S}{tein's} identity (or lemma) was first established in 1956 \cite{Stein:first}, and since then it has been widely used by many researchers (e.g., \cite{heat:brown}, \cite{stein:katt}, \cite{stein_idt:hudson}). Due to its applications in the James-Stein estimation technique, empirical Bayes methods, and numerous other fields, Stein's identity has attracted a lot of interest (see e.g., \cite{JStein:Manton}, \cite{JStein2:Manton}, \cite{Eldar:SURE}).

Recently, another identity, De Bruijn's identity, has attracted increased interest due to its applications in estimation and turbo (iterative) decoding schemes. De Bruijn's identity shows a link between two fundamental concepts in information theory: entropy and Fisher information \cite{entropy:barron}, \cite{inf:johnson}, \cite{inf:cover}. Verd$\mathrm{\acute{u}}$ and his collaborators  conducted a series of studies \cite{inf_gauss:guo}, \cite{Palomar:LinVecGaussGradient}, \cite{Palomar:LinVecGauss} to analyze the relationship between the input-output mutual information and the
minimum mean-square error (MMSE), a result referred to as the I-MMSE identity for additive Gaussian noise channels, studies which were later extended to non-Gaussian channels in \cite{inf_nongauss:guo}, \cite{Palomar:NonGauss}. Also, the equivalence between De Bruijn's identity and I-MMSE identity was shown in \cite{inf_gauss:guo}.

The main theme of this paper is to study how Stein's identity (Theorem \ref{thm2}) is related to De Bruijn's identity (Theorem \ref{thm1}). To compare Stein's identity with De Bruijn's identity, additive noise channels of the following form are considered in this paper:
\begin{eqnarray}\label{int_eq1_1}
Y & = & X + \sqrt{a}W,
\end{eqnarray}
where input signal $X$ and additive noise $W$ are arbitrary random variables, $X$ and $W$ are independent of each other, and parameter $a$ is assumed nonnegative. First, when additive noise $W$ is Gaussian with zero mean and unit variance, the equivalence between the generalized Stein's identity (Theorem \ref{thm2}) and De Bruijn's identity (Theorem \ref{thm1}) is proved. Since the standard-form Stein's identity in (\ref{stein_eq8_1}) requires both random variables $X$ and $W$ to be Gaussian, instead of the standard-form Stein's identity, the generalized version of Stein's identity in (\ref{stein_eq7_1}) is used. If we further assume that input signal $X$ is also Gaussian, then both random variables $X$ and $W$ are Gaussian, and the output signal $Y$ is Gaussian. In this case, not only Stein's and De Bruijn's identities are equivalent, but also they are equivalent to the heat equation identity, proposed in \cite{heat:brown}.

The second major question that we will address in this paper is how De Bruijn's identity could be extended. De Bruijn's identity shows the relationship between the differential entropy and the Fisher information of the output signal $Y$ under additive Gaussian noise channels. Therefore, under additive non-Gaussian noise channels, we cannot use De Bruijn's identity. However, we will  derive a similar form of De Bruijn's identity for additive non-Gaussian noise channels. Considering additive arbitrary noise channels, the first derivative of the differential entropy of output signal $Y$ will be expressed by the posterior mean, while the second derivative of the differential entropy of output signal $Y$ will be represented by a function of Fisher information. Even though some of these relationships do not include the Fisher information, they still show relationships among basic concepts in information theory and estimation theory, and these relationships hold  for
arbitrary noise channels.

\IEEEpubidadjcol

Based on the results mentioned above, we introduce several applications dealing with both estimation theoretic and information theoretic aspects. In the estimation theory field, the Fisher information inequality, the Bayesian Cram\'{e}r-Rao lower bound (BCRLB), and a new lower bound for the mean square error (MSE)  in Bayesian estimation are derived. The surprising result is that the newly derived lower bound for MSE is tighter than the BCRLB. The proposed new bound overcomes the main drawback of BCRLB, i.e., its looseness in the low Signal-to-Noise Ratio (SNR) regime, since it provides a  tighter bound than BCRLB especially at low SNRs. Even though some of the proposed applications have already been proved before, in this paper we show not only alternative ways to prove them, but also new relationships among them. In the information theory realm, Costa's entropy power inequality- previously proved in \cite{Costa:EPI}, \cite{EPI:Guo}- is derived in two different ways based on our results. Both proposed methods show novel, simple, and alternative ways to prove Costa's entropy power inequality. Finally, applications in other areas are briefly mentioned.

The rest of this paper is organized as follows. Various relationships between Stein's identity and De Bruijn's identity are established in Section \ref{sec_stein}. Some extensions of De Bruijn's identity are provided in Section \ref{sec_bruijn}. In Section \ref{sec_appl}, several applications based on the proposed novel results  are supplied. Finally, conclusions are mentioned in Section \ref{sec_con}. All the detailed mathematical derivations for the proposed  results are given in appendices.

\section{Preliminary Results}\label{sec_pre}
In this section, several definitions and preliminary theorems are provided. First, the concept of Fisher information is defined as follows.

Fisher information of a deterministic parameter $\theta$ is defined as
\begin{eqnarray}\label{F_eq1_1}
J_{\theta}(Y) & = & \int_{-\infty}^{\infty} f_{Y}(y;\theta) \left(\frac{d}{d\theta} \log f_Y(y;\theta) \right)^2 dy\nonumber\\
& = & \mathbb{E}_Y \left[S_{Y_{\theta}}(Y)^2\right],
\end{eqnarray}
where $S_{Y_{\theta}}(Y)$ denotes a score function and is defined as $(d/d\theta) \log f_Y(y;\theta)$. Under a regularity condition,
\begin{eqnarray}
\mathbb{E}_{Y} \left[S_{Y_{\theta}}(Y)\right] & = & \int_{-\infty}^{\infty} \frac{d}{d \theta} f_Y(y;\theta) dy\nonumber\\
& = & 0,\nonumber
\end{eqnarray}
the Fisher information in (\ref{F_eq1_1}) is equivalently expressed as
\begin{eqnarray}
J_{\theta}(Y) & = & - \int_{-\infty}^{\infty} f_{Y}(y;\theta) \frac{d^2}{d\theta^2} \log f_Y(y;\theta)  dy\nonumber\\
& = & - \mathbb{E}_Y \left[\frac{d}{d \theta}S_{Y_{\theta}}(Y)\right].
\end{eqnarray}

This is a general definition of Fisher information in signal processing, and Fisher information provides a lower bound, called the Cram\'{e}r-Rao lower bound, for mean square error of any unbiased estimator. Like other concepts, such as entropy and mutual information, in information theory, Fisher information also shows information about uncertainty. However, it is difficult to directly adopt the definition of Fisher information in information theory despite the fact that it has been commonly used in statistics. Instead, a more specific definition of Fisher information is proposed as follows.

If $\theta$ is assumed to be a location parameter, then
\begin{eqnarray}
\frac{d}{d\theta} f_Y(y;\theta) & = & - \frac{d}{dy} f_Y(y-\theta; \theta).
\end{eqnarray}
Therefore, the definition of Fisher information in (\ref{F_eq1_1}) is changed as follows:
\begin{eqnarray}\label{F_eq3_1}
J_{\theta}(Y) & = & \int_{-\infty}^{\infty} f_{Y}(y;\theta) \left(\frac{d}{d\theta} \log f_Y(y;\theta) \right)^2 dy\nonumber\\
& = & \int_{-\infty}^{\infty} f_{Y}(y-\theta;\theta) \left(-\frac{d}{dy} \log f_Y(y-\theta;\theta) \right)^2 dy\nonumber\\
& = & \int_{-\infty}^{\infty} f_{\tilde{Y}}(\tilde{y};\theta) \left(-\frac{d}{d\tilde{y}} \log f_{\tilde{Y}}(\tilde{y};\theta) \right)^2 d\tilde{y}\nonumber\\
& = & \mathbb{E}_{\tilde{Y}} \left[S(\tilde{Y})^2\right],
\end{eqnarray}
where $S(\tilde{Y})$ denotes a score function, and it is defined as $(d/d\tilde{y}) \log f_{\tilde{Y}} (\tilde{y};\theta)$.
In equation (\ref{F_eq3_1}), since we only consider a location parameter, we refer to Fisher information in (\ref{F_eq3_1}) as Fisher information with respect to a location (or translation) parameter, and it is denoted as $J(\tilde{Y})$ (even though the definition of Fisher information with respect to a location parameter in (\ref{F_eq3_1}) is derived from the definition of Fisher information in (\ref{F_eq1_1}), the definition in (\ref{F_eq3_1}) is more commonly used in information theory, and we do not distinguish  random variable $\tilde{Y} = Y - \theta$ from random variable $Y$).

Given the channel model in (\ref{int_eq1_1}), by substituting the parameter $a$ for the unknown parameter $\theta$, the expressions of Fisher information in (\ref{F_eq1_1}) and (\ref{F_eq3_1}) are respectively given by
\begin{eqnarray}\label{stein_eq2_1}
J(Y) & = & \int_{-\infty}^{\infty} f_Y(y;a) \left( \frac{d}{dy} \log f_Y(y;a) \right)^2 dy\nonumber\\
& = & \mathbb{E}_Y \left[S_Y(Y)^2\right],
\end{eqnarray}
and
\begin{eqnarray} \label{stein_eq3_1}
J_a(Y) & = & \int_{-\infty}^{\infty} f_Y(y;a) \left( \frac{d}{da} \log f_Y(y;a) \right)^2 dy\nonumber\\
& = & \mathbb{E}_Y \left[S_{Y_a}(Y)^2\right].
\end{eqnarray}

Second, two fundamental concepts, differential entropy and entropy power,  are defined as follows.
Differential entropy of random variable $Y$, $h(Y)$, is defined as
\begin{eqnarray}\label{stein_eq1_1}
h(Y) & = & -\int_{-\infty}^{\infty} f_Y(y;a) \log f_Y(y;a) dy,
\end{eqnarray}
where $f_Y(y;a)$ denotes the probability density function (pdf) of random variable $Y$, $\log$ denotes the natural logarithm, and $a$ is a deterministic parameter in the pdf. Similarly, the conditional entropy of random variable $Y$ given random variable $X$, $h(Y|X)$ is defined as
\begin{eqnarray}
h(Y|X) = - \int_{\scriptscriptstyle -\infty}^{\scriptscriptstyle \infty} \int_{\scriptscriptstyle -\infty}^{\scriptscriptstyle \infty} \hspace{-1.5mm} f_{X,\hspace{-.5mm}Y} (x,y;a) \log \hspace{-1mm} f_{Y|X} (y|x;a) dx dy,\hspace{-.5mm}
\end{eqnarray}
where $f_{X,Y}(x,y;a)$ denotes the joint pdf of  random variables $X$ and $Y$, $f_{Y|X}(y|x;a)$ is the conditional pdf of random variable $Y$ given random variable $X$.

Entropy power of random variable $Y$, $N(Y)$, and (conditional) entropy power of random variable $Y$ given random variable $X$, $N(Y|X)$ are respectively defined as
\begin{eqnarray}
N(Y) & = & \frac{1}{2 \pi e} \exp(2 h(Y)),\nonumber\\
N(Y|X) & = & \frac{1}{2 \pi e} \exp (2 h(Y|X)).
\end{eqnarray}

Based on the definitions mentioned above, three preliminary theorems- De Bruijn's, Stein's, and heat equation identities- are introduced next.
%Theorem 1
\begin{thm}[De Bruijn's Identity \cite{inf:cover}, \cite{EPI:Stam}] \label{thm1}
Given the additive noise channel $Y=X+\sqrt{a}W$, let $X$ be an arbitrary random variable with a finite second-order moment, and $W$ be independent normally distributed with zero mean and unit variance. Then,
\begin{eqnarray}\label{deBruijn}
\frac{d}{da} h(Y) & = & \frac{1}{2}J(Y).
\end{eqnarray}
\begin{proof}
See \cite{inf:cover}.
\end{proof}
\end{thm}

%Theorem 2
\begin{thm} [Generalized Stein's Identity \cite{stein:katt}] \label{thm2}
Let $Y$ be an absolutely continuous random variable. If the probability density function $f_Y(y)$ satisfies the following equations,
\begin{eqnarray}
\lim_{y\rightarrow \pm \infty}k(y)f_Y(y) & = & 0,\nonumber
\end{eqnarray}
and
\begin{eqnarray}
\frac{\frac{d}{dy}f_Y(y)}{f_Y(y)} & = & -\frac{\frac{d}{dy}k(y)}{k(y)}+\frac{\left(\nu-t(y)\right)}{k(y)}\nonumber
\end{eqnarray}
for some function $k(y)$, then
\begin{eqnarray}\label{stein_eq7_1}
\mathbb{E}_Y\left[r(Y)\left(t(Y)-\nu\right)\right] & = & \textstyle \mathbb{E}_Y \left[\frac{d}{dY}r(Y) k(Y) \right],
\end{eqnarray}
for any function $r(Y)$ which satisfies $\mathbb{E}_Y\left[\left|r(Y)t(Y)\right|\right]<\infty$, $\mathbb{E}_Y\left[r(Y)^2\right]<\infty$, and
$\mathbb{E}_Y\left[\left|k(Y)\frac{d}{dY}r(Y)\right|\right]<\infty$. $\mathbb{E}_Y[\cdot]$ denotes the expectation with respect to the pdf of random variable $Y$. In particular, when random variable $Y$ is normally distributed with mean $\mu_y$ and variance $\sigma_y^2$, equation (\ref{stein_eq7_1}) simplifies to
\begin{eqnarray}\label{stein_eq8_1}
\mathbb{E}_Y\left[r(Y)\left(Y-\mu_y\right)\right] & = & \textstyle\sigma_y^2 \mathbb{E}_Y \left[\frac{d}{dY}r(Y)\right].
\end{eqnarray}
Equation (\ref{stein_eq8_1})  is the well-known classic Stein's identity.
\begin{proof}
See \cite{stein:katt}.
\end{proof}
\end{thm}

%Theorem 3
\begin{thm}[Heat Equation Identity {\cite{heat:brown}}] \label{thm3}
Let $Y$ be normally distributed with mean $\mu$ and variance $1+a$. Assume $g(y)$ is a twice continuously differentiable function, and both $g(y)$ and $|\frac{d}{dy} g(y)
|$ are\footnote{$O(\cdot)$ denotes the limiting behavior of the function, i.e., $g(y)=O(q(y))$ if and only if there exist positive real numbers $K$ and $y^{\ast}$ such that $g(y)\leq K |q(y)|$ for any $y$ which is greater than $y^{\ast}$.} $O(e^{c|y|})$ for some $0\leq c <\infty$. Then,
\begin{eqnarray}\label{stein_eq9_1}
\frac{d}{da} \mathbb{E}_Y \left[ g(Y) \right] & = & \frac{1}{2} \mathbb{E}_Y \left[\frac{d^2}{dY^2} g(Y) \right].
\end{eqnarray}
\begin{proof}
See \cite{heat:brown}.
\end{proof}
\end{thm}

% Relationships between Stein's Identity and De Bruijn's identity
\section{Relationships between Stein's Identity and De Bruijn's Identity}\label{sec_stein}

In Section \ref{sec_pre}, Theorems \ref{thm1}, \ref{thm2}, and \ref{thm3} share an analogy: an identity between expectations of functions, which include derivatives. Especially, the heat equation identity admits the same form as De Bruijn's identity by choosing function $g(y)$ as $-\log f_Y(y;a)$. If De Bruijn's identity is equivalent to the heat equation identity, it is also equivalent to Stein's identity, since the equivalence between the heat equation identity and Stein's identity was proved in \cite{heat:brown}. However, there are two critical issues that stand in the way of the equivalence between Stein's and De Bruijn's identities: first, the function $g(y)$ in Theorem \ref{thm3} must be independent of the parameter $a$, which is not true when $g(y)=-\log f_Y(y;a)$. Second,  in the heat equation identity, random variable $Y$ must be Gaussian, which may not be true in De Bruijn's identity.

Due to the difficulties mentioned above, we will directly compare De Bruijn's identity (Theorem \ref{thm1}) with the generalized Stein's identity (Theorem \ref{thm2}).
%Theorem 5
\begin{thm}\label{thm5}
Given the channel model (\ref{int_eq1_1}), let $X$ be an arbitrary random variable with a finite second-order moment, and  let $W$ be normally distributed with zero mean and unit variance. Independence between random variables $X$ and $W$ is also assumed.
Then, De Bruijn's identity (\ref{deBruijn}) is equivalent to the generalized Stein's identity in (\ref{stein_eq7_1}) under specific conditions, i.e.,
\begin{eqnarray}
\frac{d}{da}h(Y) & = & \frac{1}{2} J(Y)\nonumber\\
\Longleftrightarrow \mathbb{E}_Y\left[r(Y;a)\left(t(Y;a)-\nu\right)\right] & = & \textstyle \mathbb{E}_Y \left[\frac{d}{dY}r(Y;a) k(Y;a) \right],\nonumber
\end{eqnarray}
with
\begin{eqnarray}
r(y;a) = -\frac{d}{dy} \log f_Y(y;a), \quad k(y) = 1,\nonumber\\
t(y;a) = -\frac{\frac{d}{dy}f_Y(y;a)}{f_Y(y;a)}, \quad \text{and} \quad \nu = 0,
\end{eqnarray}
where $\Longleftrightarrow$ denotes the equivalence between before and after the notation.
\begin{proof}
See Appendix \ref{AppA}.
\end{proof}
\end{thm}

Now, when random variable $Y$ is Gaussian, i.e., both random variables $X$ and $W$ are Gaussian, we can derive relationships among three identities, De Bruijn, Stein, and heat equation, as a special case of Theorem \ref{thm5}.

%Theorem 4
\begin{thm}\label{thm4}
Given the channel model (\ref{int_eq1_1}), let random variable $X$ be normally distributed with mean $\mu$ and unit variance. Assume $W$ is independent normally distributed with zero mean and unit variance. If we define the functions in (\ref{stein_eq7_1}) as follows:
\begin{eqnarray}
&&r(y;a) = -\frac{d}{dy} \log f_Y(y;a), \quad k(y;a) = \frac{1}{a}, \nonumber\\
&&t(y;a) = y,\quad \text{and} \quad \nu = \mu,\nonumber
\end{eqnarray}
 then Stein's identity is equivalent to De Bruijn's identity. Moreover, if we define $g(y;a)$ as
\begin{eqnarray}
g(y;a) & = & -\log f_Y(y;a)\nonumber
\end{eqnarray}
in (\ref{stein_eq9_1}), then De Bruijn's identity is also equivalent to the heat equation identity.
\begin{proof}
In Theorem \ref{thm5}, given the channel model (\ref{int_eq1_1}) with an arbitrary but fixed random variable $X$ and a Gaussian random variable $W$, the equivalence between De Bruijn's identity and the generalized Stein's identity was proved (cf.  Appendix \ref{AppA}). Here, by choosing random variable $X$ as Gaussian, this is a special case of Theorem \ref{thm5}. Therefore, the equivalence between the two identities is trivial, and the details of the proof is omitted in this paper. The only thing to prove is the second part of this theorem, namely, the equivalence between De Bruijn's identity and the heat equation identity. Since the equivalence between Stein's identity and the heat equation identity is proved in \cite{heat:brown}, this also proves the second part of the theorem, and the proof is completed.
\end{proof}
\end{thm}

The functions $k(y;a)$, $r(y;a)$, $t(y;a)$, and $g(y;a)$ are the same as $k(y)$, $r(y)$, and $t(y)$ in Theorem \ref{thm2} and $g(y)$ in Theorem \ref{thm3}, respectively. To show the dependence on  parameter $a$, the functions $k(y;a)$, $r(y;a)$, $t(y;a)$, and $g(y;a)$ are used instead of $k(y)$, $r(y)$, $t(y)$, and $g(y)$, respectively.

% Extension of De Bruijn's Identity
\section{Extension of De Bruijn's Identity}\label{sec_bruijn}

De Bruijn's identity is derived from the attribute of Gaussian density functions, which satisfy the heat equation. However, in general, probability density functions do not satisfy the heat equation. Therefore, to extend De Bruijn's identity to  additive non-Gaussian noise channels, a general relationship between differentials of a probability density function with respect to $y$ and $a$ of the form:
\begin{eqnarray}\label{appd_eq17_1}
\frac{d}{da}f_{Y|X}(y|x;a) = -\frac{1}{2a} \frac{d}{dy}\left((y-x) f_{Y|X}(y|x;a)\right),
\end{eqnarray}
is  required, a result that it is obtained in Appendix \ref{AppH} by exploiting the assumptions (\ref{bruijn_eq1_1}). The relationship (\ref{appd_eq17_1})   represents the key ingredient  in establishing the link
between the derivative of differential entropy and posterior mean, as described by the following theorem.

% Theorem 6
% Extension of De Bruijn (first derivative)
\begin{thm}\label{thm6}
Consider the channel model (\ref{int_eq1_1}), where $X$ and $W$  are arbitrary random variables independent of each other. Given the following assumptions:
\begin{subeqnarray}\label{bruijn_eq1_1}
&&\hspace{-8mm}\frac{d}{dy} \mathbb{E}_X\left[f_{Y|X}(y|X;a)\right] = \mathbb{E}_X \left[ \frac{d}{dy} f_{Y|X}(y|X;a)\right],\nonumber\\
&&\hspace{-8mm}\frac{d}{da}\mathbb{E}_X \left[f_{Y|X}(y|X;a) \right] = \mathbb{E}_X \left[ \frac{d}{da} f_{Y|X}(y|X;a) \right],\\
\label{bruijn_eq1_2}
&&\hspace{-8mm}\frac{d}{da} \int_{-\infty}^{\infty}  f_Y(y;a) \log f_{Y}(y;a) dy\nonumber\\
&&=  \int_{-\infty}^{\infty} \frac{d}{da} \Big(f_Y(y;a) \log f_{Y}(y;a)\Big) dy, \\
\label{bruijn_eq1_3}
&&\hspace{-8mm}\lim\limits_{\scriptscriptstyle y \rightarrow \pm \infty} \mathbb{E}_X \left[X f_{Y|X}(y|X;a)\right]= \mathbb{E}_X \left[ \lim\limits_{\scriptscriptstyle y \rightarrow \pm \infty}X f_{Y|X}(y|X;a)\right], \nonumber\\
&&\hspace{-8mm} \lim\limits_{\scriptscriptstyle y \rightarrow \pm \infty} \mathbb{E}_X \left[ f_{Y|X}(y|X;a)\right] = \mathbb{E}_X \left[ \lim\limits_{\scriptscriptstyle y \rightarrow \pm \infty} f_{Y|X}(y|X;a)\right], \nonumber\\
&&\hspace{-8mm}\lim_{\scriptscriptstyle y \rightarrow \pm \infty} y^2f_{Y}(y;a)=0,\\
%&& \hspace{-13mm} \left|\mathbb{E}_{X|Y} \left[X|Y=y\right]\right| < \infty,
&& \hspace{-8mm} \left|\frac{\mathbb{E}_{X} \left[X f_{Y|X}(y|X;a)\right]}{\sqrt{f_Y(y;a)}}\right| < \infty,
\end{subeqnarray}
where $\mathbb{E}_{X|Y}[\cdot|\cdot]$ denotes the posterior mean, the first derivative of the differential entropy is expressed as
\begin{eqnarray}\label{bruijn_eq2_1}
\frac{d}{da} h(Y) & = &  \frac{1}{2a} \left\{1-\mathbb{E}_Y \left[\frac{d}{dY} \mathbb{E}_{X|Y} \left[X|Y\right] \right]\right\}.
\end{eqnarray}
\begin{proof}
See Appendix \ref{AppB}.
\begin{rem}
This is equivalent to the results in \cite{inf_nongauss:guo}.
\end{rem}
\end{proof}
\end{thm}
It can be observed that the conditions (\ref{bruijn_eq1_1})  are required in  the dominated convergence theorem and Fubini's theorem to ensure the interchangeability between a limit and an integral, and are not that restrictive. Also, the condition $\lim_{y \rightarrow \pm \infty} y^2f_{Y}(y;a)=0$
is not restrictive at all, and it is satisfied by all noise distributions of interest in practice.
% Corollary 1
% Gaussian
\begin{cor}[De Bruijn's identity]\label{cor1}
Given the channel model in (\ref{int_eq1_1}) with  an arbitrary but fixed random variable $X$ with a finite second moment and a Gaussian random variable $W$ with zero mean and unit variance,
\begin{eqnarray}
\frac{d}{da} h(Y) & = &  \frac{1}{2}J(Y).\nonumber
\end{eqnarray}
\begin{rem}
This is the well-known De Bruijn's identity \cite{EPI:Stam}. Therefore, De Bruijn's identity is a special case of Theorem \ref{thm6} when  random variable $W$ is normally distributed. When random variable $W$ is Gaussian, assumptions in (\ref{bruijn_eq1_1})  are simplified to the existence of a finite second-order moment.
\end{rem}
\end{cor}

% Corollary 2
% exponential
\begin{cor}\label{cor2}
Given the channel model in (\ref{int_eq1_1}) with an arbitrary but fixed non-negative random variable $X$ whose moment generating function exists and its pdf is bounded, and an exponential random variable $W$ with unit value of the parameter (i.e., $f_W(w)=exp(-w)\mathrm{U}(w)$, where $\mathrm{U}(\cdot)$ denotes the unit step function),
\begin{eqnarray}
\frac{d}{da} h(Y) \hspace{-.4mm} = \hspace{-.4mm}  \frac{1}{2a\sqrt{a}}\left\{\sqrt{a}- \mathbb{E}_X \hspace{-1.5mm}\left[X\right]+ \mathbb{E}_X\hspace{-1.5mm}\left[\mathbb{E}_{X|Y}\hspace{-1mm}\left[X|Y\right]|Y\hspace{-.7mm}=\hspace{-.7mm}X\right]\right\}.\nonumber
\end{eqnarray}
\end{cor}

When the random variable $W$ is exponentially distributed,  assumptions in (\ref{bruijn_eq1_1})  are reduced  to the existence of the moment generating function of $X$, as explained in Appendix \ref{AppI}. Therefore, the assumptions in (\ref{bruijn_eq1_1}) for an exponential random variable are as simple as the assumptions (\ref{bruijn_eq1_1}) for a Gaussian random variable.

% Corollary 3
% gamma
\begin{cor}\label{cor3}
Given the channel model in (\ref{int_eq1_1}) with an arbitrary but fixed non-negative random variable $X$ whose moment generating function exists and  a gamma random variable $W$ with a shape parameter $\alpha$   $(\alpha \geq 2)$ and an inverse scale parameter $\beta$  $(\beta=1)$,
\begin{eqnarray}
\frac{d}{da} h(Y)& = &\frac{1}{2a\sqrt{a}}\Big\{\sqrt{a}-\mathbb{E}_X\left[X\right]\nonumber\\
 &&\hspace{3mm}+ \mathbb{E}_{Y_{\alpha-1}} \left[\mathbb{E}_{X|Y}\left[X|Y\right] |Y=Y_{\alpha-1}\right] \Big\},\nonumber
\end{eqnarray}
where $Y_{k}=X+\sqrt{a}W_{k}$, and $W_k$ denotes a gamma random variable with  shape parameter $k$. Notation $Y_{\alpha}$  stands for  $Y$. As explained in Appendix \ref{AppI}, the assumptions (\ref{bruijn_eq1_1}) are  quite simplified in the presence of the moment generating function of random variable $X$.
\end{cor}

For additive non-Gaussian noise channels, the differential entropy  cannot be expressed  in terms of the Fisher information. Instead, the differential entropy is expressed by the posterior mean as shown in Theorem \ref{thm6}. Fortunately, several noise distributions of interest in communication problems satisfy the required assumptions (\ref{bruijn_eq1_1}) in Theorem \ref{thm6} (e.g., Gaussian, gamma, exponential, chi-square with restrictions on parameters, Rayleigh, etc.). Therefore, Theorem \ref{thm6} is quite powerful. If the posterior mean $\mathbb{E}_{X|Y}[X|Y]$ is expressed by a polynomial function of $Y$, e.g.,  $X$ and $W$ are independent Gaussian random variables in equation (\ref{int_eq1_1}) or  random variables belonging to the natural exponential family of distributions  \cite{morris:naturalexp}, then equation (\ref{bruijn_eq2_1}) can be expressed in simpler forms.

\begin{example}\label{exam1}
Consider an additive white Gaussian noise (AWGN) channel. Given the channel model (\ref{int_eq1_1}), let $X$ and $W$ be normally distributed with zero mean and unit variance. Assume $X$ and $W$ are independent of each other. Then, the posterior mean is expressed as
\begin{eqnarray}
\mathbb{E}_{X|Y}  \left[X|Y=y\right] & = & \frac{1}{1+a}y,\nonumber
\end{eqnarray}
which is linear to $y$. Therefore, equation (\ref{bruijn_eq2_1}) is expressed as
\begin{eqnarray}
\frac{d}{da} h(Y) & = &  \frac{1}{2a}\left\{1- \mathbb{E}_Y \left[\frac{d}{dY}\mathbb{E}_{X|Y} \left[X|Y\right]\right]\right\}\nonumber\\
& = & \frac{1}{2(1+a)}.\nonumber
\end{eqnarray}
\end{example}

Now, we consider the second derivative of the differential entropy. One interesting property of the second derivative of the differential entropy is that it can always be expressed as a function of the Fisher information (\ref{stein_eq3_1}).

%Theorem 7
% extension of De Bruijn (second derivative)
\begin{thm}\label{thm7}
Given the channel model (\ref{int_eq1_1}), let $X$ and $W$ be arbitrary random variables, independent of each other. Given the following assumptions:
\begin{subeqnarray}\label{bruijn_eq12_1}
&&\hspace{-8mm} \frac{d^2}{dy^2} \mathbb{E}_X\left[f_{Y|X}(y|X;a)\right] \hspace{-1mm}= \hspace{-1mm} \mathbb{E}_X \left[ \frac{d^2}{dy^2} f_{Y|X}(y|X;a)\right],\nonumber\\
&&\hspace{-8mm} \frac{d^2}{da^2}\mathbb{E}_X \left[f_{Y|X}(y|X;a) \right] \hspace{-1mm} = \hspace{-1mm} \mathbb{E}_X \left[ \frac{d^2}{da^2} f_{Y|X}(y|X;a) \right],\\
\label{bruijn_eq12_2}
&&\hspace{-8mm} \frac{d^2}{da^2} \int_{-\infty}^{\infty}  f_Y(y;a) \log f_{Y}(y;a) dy\nonumber\\
&&=  \int_{-\infty}^{\infty} \frac{d^2}{da^2} \Bigg(f_Y(y;a) \log f_{Y}(y;a)\Bigg) dy,\\
&&\hspace{-8mm} \lim\limits_{\scriptscriptstyle y \rightarrow \pm \infty} \mathbb{E}_X \left[X^2 \frac{\frac{d}{dy} f_{Y|X}(y|X;a)}{\sqrt{f_Y(y;a)}}\right]\nonumber\\
&&= \mathbb{E}_X \left[ \lim\limits_{\scriptscriptstyle y \rightarrow \pm \infty}X^2 \frac{\frac{d}{dy} f_{Y|X}(y|X;a)}{\sqrt{f_Y(y;a)}}\right],\nonumber\\
&&\hspace{-8mm}\lim\limits_{\scriptscriptstyle y \rightarrow \pm \infty} \mathbb{E}_X \left[X f_{Y|X}(y|X;a)\right]\nonumber\\
&&= \mathbb{E}_X \left[ \lim\limits_{\scriptscriptstyle y \rightarrow \pm \infty}X f_{Y|X}(y|X;a)\right],\\
\label{bruijn_eq12_3}
&&\hspace{-8mm} \lim\limits_{\scriptscriptstyle y \rightarrow \pm \infty} \mathbb{E}_X \left[f_{Y|X}(y|X;a)\right] = \mathbb{E}_X \left[ \lim\limits_{\scriptscriptstyle y \rightarrow \pm \infty}f_{Y|X}(y|X;a)\right], \nonumber\\
&&\hspace{-8mm}\lim\limits_{\scriptscriptstyle y \rightarrow \pm \infty} y^8 f_Y(y;a) =0,\\
&& \hspace{-8mm} \left|\frac{\mathbb{E}_{X} \left[X^2 f_{Y|X} (y|X;a)\right]}{(f_Y(y;a))^{3/4}}  \right| < \infty,
\end{subeqnarray}
where $\mathbb{E}_{X|Y}[\cdot|\cdot]$ denotes the posterior mean, the following identity holds:
\begin{eqnarray}
\frac{d^2}{da^2}h(Y) & = & -J_a(Y)- \frac{1}{2a}\frac{d}{da}h(Y)\nonumber\\
&&-\frac{1}{4a^2}\mathbb{E}_Y\left[\frac{d}{dY}S_Y(Y)\mathbb{E}_{X|Y}\left[(Y-X)^2|Y\right]\right],\nonumber
\end{eqnarray}
or equivalently,
\begin{eqnarray}\label{bruijn_eq13_2}
\hspace{-3mm}\frac{d^2}{da^2}h(Y)\hspace{-3mm} & = & \hspace{-2mm}-J_a(Y)- \frac{1}{4a^2}\mathbb{E}_Y\left[\frac{d}{dY}\mathbb{E}_{X|Y}\left[(Y-X)|Y\right]\right]\nonumber\\
\hspace{-3mm} \hspace{-3mm}&&\hspace{-2mm}- \frac{1}{4a^2}\mathbb{E}_Y\left[\frac{d}{dY}S(Y) \mathbb{E}_{X|Y} \left[(Y-X)^2|Y\right]\right].
\end{eqnarray}
\begin{proof}
See Appendix \ref{AppC}.
\end{proof}
\end{thm}

Similar to the corollaries of Theorem \ref{thm6}, by specifying a noise distribution and manipulating equation (\ref{bruijn_eq13_2}) in Theorem \ref{thm7}, we derive the following corollaries.

% Corollary 5 (Gaussian)
\begin{cor}\label{cor5}
Given the channel (\ref{int_eq1_1}), let $X$ be an arbitrary but fixed random variable with a finite second-order moment, and let $W$ be independent normally distributed with zero mean and unit variance. Then,
\begin{eqnarray}
\frac{d^2}{da^2} h(Y) & = &  -J_a(Y) - \frac{1}{4a} J(Y)\nonumber\\
&&- \frac{1}{4a^2}\mathbb{E}_Y \left[ \frac{d}{dY} S_Y(Y) \mathbb{E}_{X|Y} \left[ (Y-X)^2|Y \right] \right]\nonumber\\
& = & - \frac{1}{2}\mathbb{E}_Y\left[\left(\frac{d}{dY}S_Y(Y)\right)^2\right].\nonumber
\end{eqnarray}
\begin{rem}
This result is a scalar version of the result reported in \cite{Palomar:LinVecGauss}. At the same time, this result is a special case, when $X$ is a Gaussian random variable, of the general result in Theorem \ref{thm7}.
\end{rem}
\end{cor}

% Corollary 6 (Exponential)
\begin{cor}\label{cor6}
Under the channel (\ref{int_eq1_1}), let $X$ be an arbitrary but fixed non-negative random variable with a finite moment generating function, and its pdf is bounded. Let $W$ be independent exponentially distributed with unit value as the parameter ($\lambda$) of the distribution. Namely, $f_W(w)=exp(-w)\mathrm{U}(w)$, where $\mathrm{U}(\cdot)$ denotes the unit step function. Then,
\begin{eqnarray}
\frac{d^2}{da^2} h(Y) \hspace{-3mm}& = & \hspace{-3mm} -J_a(Y) + \frac{3}{4a^2\sqrt{a}} \mathbb{E}_X \left[\mathbb{E}_{X|Y}\left[Y-X|Y\right]|Y\hspace{-1mm}=\hspace{-1mm}X\right]\nonumber\\
 \hspace{-3mm}&&\hspace{-3mm} +\frac{1}{4a^2}  -\frac{1}{4a^3} \mathbb{E}_X \left[\mathbb{E}_{X|Y} \left[(Y-X)^2|Y\right] |Y\hspace{-1mm}=\hspace{-1mm}X\right].\nonumber
\end{eqnarray}
\end{cor}

% Corollary 7 (Gamma)
\begin{cor}\label{cor7}
Under the channel (\ref{int_eq1_1}), let $X$ be an arbitrary but fixed non-negative random variable with a finite moment generating function, and $W$ be an independent gamma random variable with parameters $\alpha$ ($\alpha \geq 3$) and $\beta$ ($\beta=1$), i.e., $f_W(w)=\beta^{\alpha}w^{\alpha-1} \exp(-\beta w) \mathrm{U}(w) / \Gamma(\alpha)$, where $\mathrm{U}(\cdot)$ denotes the unit step function and $\Gamma(\cdot)$ stands for the gamma function. Then,
\begin{eqnarray}
\hspace{-7mm}&&\frac{d^2}{da^2} h(Y)  =  -\frac{1}{4a^3} \mathbb{E}_{\scriptscriptstyle Y_{\alpha-2}}\left[ \mathbb{E}_{\scriptscriptstyle X|Y} \left[(Y-X)^2 |Y\right]|Y=Y_{\alpha-2}\right]\nonumber\\
\hspace{-7mm}&&\hspace{7mm}-\frac{1}{4a^2\sqrt{a}} \mathbb{E}_{\scriptscriptstyle Y_{\alpha-1}} \left[\mathbb{E}_{\scriptscriptstyle X|Y} \left[X|Y\right] |Y=Y_{\alpha-1}\right] \nonumber\\
\hspace{-7mm}&&\hspace{7mm} + \frac{(\alpha-1)}{4a^2\sqrt{a}} \mathbb{E}_{\scriptscriptstyle Y_{\alpha-1}} \left[\frac{\mathbb{E}_{\scriptscriptstyle X|Y}\left[(Y-X)^2|Y\right]}{\mathbb{E}_{\scriptscriptstyle X|Y_{\alpha-1}} \left[Y_{\alpha-1}-X|Y_{\alpha-1}\right]}\Bigg|Y=Y_{\alpha-1}\right]\nonumber\\
\hspace{-7mm}&&\hspace{7mm}-J_a(Y)- \frac{1}{4a^2\sqrt{a}} \left(\sqrt{a}-\mathbb{E}_{\scriptscriptstyle X} \left[X\right] \right) ,\nonumber
\end{eqnarray}
where $Y_{\alpha}=X+\sqrt{a}W_{\alpha}$, and $W_{\alpha}$ denotes a gamma random variable with  a shape parameter $\alpha$. %Notation $Y_{\alpha}$ stands for $Y$. %In addition, $f_{Y_k}(y;a)$ must be bounded for any $y$ and for $k=\alpha-1, \alpha, \alpha+1, \alpha+2$.
\end{cor}

Like Corollaries \ref{cor1}, \ref{cor2}, and \ref{cor3}, the assumptions (\ref{bruijn_eq12_1}) reduce to simplified forms in Corollaries \ref{cor5}, \ref{cor6}, and \ref{cor7}. Even though we have not enumerated all possible probability density functions for Theorem \ref{thm6} and Theorem \ref{thm7}, many of the probability density functions that present an exponential term satisfy the assumptions (\ref{bruijn_eq1_1}) and (\ref{bruijn_eq12_1}), since such a condition proves to be sufficient for the required interchange between a limit and a integral.

% Applications
\section{Applications}\label{sec_appl}
As mentioned in \cite{inf_gauss:guo} and \cite{Rioul:EPI}, De Bruijn's identity has been widely used in a variety of areas such as information theory, estimation theory, and so on. Similarly, De Bruijn-type identities mentioned in this paper can be adopted in many applications. Here, we introduce several applications from the  estimation theory realm as well as from the information theory field.

\subsection{Applications in Estimation Theory}\label{subsec_est}
% First Application
In estimation theory, there exist two fundamental lower bounds: Cram\'{e}r-Rao lower bound (CRLB) and Bayesian Cram\'{e}r-Rao lower bound (BCRLB). CRLB is a lower bound for the estimation error of any unbiased estimator, and it is derived from a frequentist perspective. This lower bound is tight when the output distribution of the channel is Gaussian. CRLB and its tightness can be justified using Cauchy-Schwarz inequality \cite{Est:Kay}. On the other hand, BCRLB is a lower bound for the estimation error of any estimator, and it is calculated from a  Bayesian perspective. BCRLB does not require unbiasedness of estimators unlike CRLB; however, BCRLB requires prior knowledge (i.e., distribution) of random parameters. BCRLB is also tight when all random variables are Gaussian \cite{Est:Tree}.

Surprisingly, assuming a Gaussian additive noise channel, both of these lower bounds can be derived using De Bruijn-type identities, and there exist counterparts both in information theory and estimation theory. Since CRLB and its counterpart, the worst additive noise lemma, are derived in \cite{Rioul:EPI}, we will only show the derivation of BCRLB and its counterpart in this paper.

\begin{lem}[Bayesian Cram\'{e}r-Rao Lower Bound]\label{lem1}
Given the channel (\ref{int_eq1_1}), let $\hat{X}$ be an arbitrary estimator of $X$ in a Bayesian estimation framework. Then, the mean square error (MSE) of $\hat{X}$ is lower bounded as follows:
\begin{eqnarray}
MSE(\hat{X}) & \geq & \frac{1}{\mathbb{E}_X \left[J(Y|X)\right]+J(X)},\nonumber
\end{eqnarray}
where $X$ is an arbitrary but fixed random variable with a finite second-order
 moment, $W$ is a Gaussian random variable with zero mean and  unit variance, and
\begin{eqnarray}\label{appl_eq2_1}
\hspace{-5mm}J(Y|X) & = & \int_{-\infty}^{\infty}\left(\frac{d}{dx} \log f_{Y|X}(y|x)\right)^2 f_{Y|X}(y|x) dy.
\end{eqnarray}
\begin{proof}
See Appendix \ref{AppD}.
\end{proof}
\end{lem}

Interestingly, there exists a counterpart, based on differential entropies, of BCRLB in information theory, and this counterpart is a tighter lower bound than BCRLB.

\begin{lem}\label{lem4}
Under  the same conditions as in Lemma \ref{lem1},
\begin{eqnarray}\label{appl_eq17_1}
MSE(\hat{X}) & \geq & N(X|Y),
\end{eqnarray}
where $N(X|Y)=(1/2\pi e) \exp(2 h(X|Y))$, $Y=X+\sqrt{a}W$, $a\geq0$, and $X$ and $W$ are independent of each other.
\begin{proof}
See Appendix \ref{AppE}.
\begin{rem}
Lemma \ref{lem4} seems to be similar to the estimation counterpart of Fano's inequality  \cite[p. 255, Theorem 8.6.6]{inf:cover}. However, the current result is completely different than \cite[p. 255, Theorem 8.6.6]{inf:cover}. In \cite{inf:cover}, to satisfy the inequality (\ref{appl_eq17_1}), the hidden assumption  is
\begin{eqnarray}\label{rev_eq1_0}
Var(X|Y) & = &Var(X_G|Y_G),
\end{eqnarray}
where $Var(X|Y)$ and $Var(X_G|Y_G)$ denote posterior variances for random variables $X$ and $Y$, and Gaussian random variables $X_G$ and $Y_G$, respectively. With the assumption (\ref{rev_eq1_0}), the following relations hold:
\begin{eqnarray}
\mathbb{E}_{X, Y} \left[ \left(X- \mathbb{E}_{X|Y}[X|Y] \right)^2\right] & = & Var(X|Y) \nonumber\\
& = & Var(X_G|Y_G) \nonumber\\
\label{rev_eq1_3}& = & \frac{1}{2 \pi e} \exp(2h(X_G|Y_G)) \nonumber\\
\label{rev_eq1_4}& \geq & \frac{1}{2 \pi e} \exp(2h(X|Y)) \nonumber\\
& = & N(X|Y)\nonumber.
\end{eqnarray}
This is nothing but the entropy maximizing theorem, i.e., the Gaussian random variable being the one that maximizes the entropy among all real-valued distributions with fixed mean and variance.

However,  under the assumptions $Var(X)=Var(X_G)$ and $Var(Y)=Var(Y_G)$, which are common assumptions in signal processing problems, (\ref{rev_eq1_0}) may not be always true due to the following fact. Given the additive Gaussian noise channel, $Y = X + \sqrt{a}W_G,$ where $X$ is an arbitrary non-Gaussian random variable whose variance is identical to that of  Gaussian random variable $X_G$, and $W_G$ is a Gaussian random variable with zero mean and unit variance,
\begin{eqnarray}\label{rev_eq1_2_1}
Var(X|Y) & < & Var(X_G|Y_G),
\end{eqnarray}
where $Y_G$ is a Gaussian random variable whose variance is identical to that of $Y$. Equation (\ref{rev_eq1_2_1}) violates the assumption (\ref{rev_eq1_0}). Therefore, the result in \cite[p. 255, Theorem 8.6.6]{inf:cover} cannot be adopted under the assumptions, $Var(X)=Var(X_G)$ and $Var(Y)=Var(Y_G)$, which are common in signal processing problems.

On the other hand, the inequality in Lemma \ref{lem4} is obtained not by imposing identical posterior variances but by assuming identical second-order  moments. Thus, (\ref{appl_eq17_1}) represents a lower bound on the mean square error similar to  BCRLB. Therefore,  Lemma \ref{lem4} illustrates a novel lower bound on the mean square error  from an  information theoretic perspective.
\end{rem}
\end{proof}
\end{lem}
Surprisingly, this lower bound is tighter than BCRLB as the following lemma indicates.
\begin{lem}\label{lem5}
Under the same conditions as in Lemma \ref{lem4},
\begin{eqnarray}\label{appl_eq17_2}
N(X|Y) & \geq & \frac{1}{\mathbb{E}_X \left[J(Y|X)\right]+J(X)},
\end{eqnarray}
where $Y=X+\sqrt{a}W$, $a$ is nonnegative, $X$ is an arbitrary but fixed random variable with a finite second-order  moment, $W$ is a Gaussian random variable with zero mean and unit variance, and $J(Y|X)$ is defined as equation (\ref{appl_eq2_1}). The equality holds if the  random variable $X$ is Gaussian.
\begin{proof}
See Appendix \ref{AppF}.
\end{proof}
\end{lem}
Figure \ref{Fig1} illustrates  how tighter the new lower bound (\ref{appl_eq17_1}) is compared to BCRLB when $X$ is a student-t random variable, and $W$ is a Gaussian random variable. The degrees of freedom of $X$ is 3, and the variance of $W$ is 1. As shown in Figure \ref{Fig1}, the new lower bound is much tighter than BCRLB especially in low SNRs where the BCRLB is generally loose. Also, Figure \ref{Fig1} shows how tight the new lower bound is with respect to  the minimum mean square error.
\begin{figure}[hb]
\begin {center}
    \includegraphics[width=0.5\textwidth]{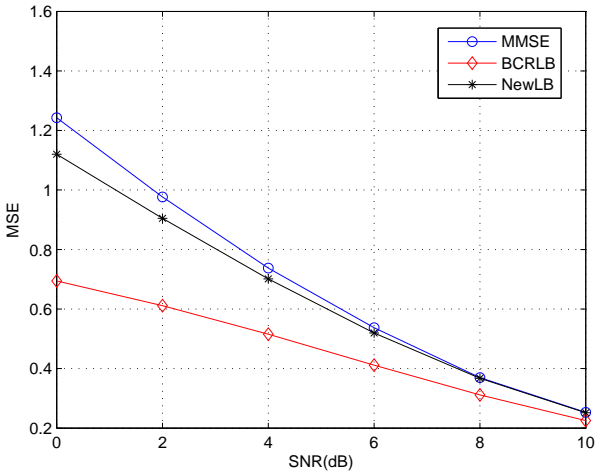}\\
  \caption{Comparison of MMSE, BCRLB, and new lower bound (New LB) in (\ref{appl_eq17_1}) with respect to SNR.}
\label{Fig1}
\end{center}
\end{figure}

\subsection{Applications in Information Theory}\label{subsec_inf}

In information theory, the entropy power inequality (EPI) is one of the most important inequalities since it is helps to prove the channel capacity under several different circumstances, e.g., the capacity of scalar Gaussian broadcast channel \cite{CapBroad:Bergmans}, the capacity of Gaussian MIMO broadcast channel \cite{CapBroad:Shamai}, \cite{Extremal:Liu}, the secrecy capacity of Gaussian wire-tap channel \cite{SecrecyCap:Liu}, \cite{Extrem:SPark} and so on. The channel capacity can be proved not by EPI alone but by EPI in conjunction with Fano's inequality. Depending on the channel model, an additional technique, channel enhancement technique \cite{CapBroad:Shamai}, is required. Therefore, various versions of the EPI such as a classical EPI \cite{EPI:Stam}, \cite{Math:Shannon}, \cite{EPI:Blachman},  Costa's EPI \cite{Costa:EPI}, and an extremal inequality \cite{Extremal:Liu} were proposed  by several different authors.
In this section, we  will prove Costa's entropy power inequality, a stronger version of a classical EPI using Theorem \ref{thm7}.

\begin{lem}[Costa's EPI]\label{lem7}
For a Gaussian random variable $W$ with zero mean and unit variance,
\begin{eqnarray}\label{appl_eq27_1}
N(X+\sqrt{a}W) & \geq & (1-a) N(X) +a N(X+W),
\end{eqnarray}
where $0\leq a \leq 1$, $X$ and $W$ are independent of each other, and the entropy power $N(X)$ is defined as $N(X)=(1/2\pi e)\exp(2h(X))$. Alternatively, the inequality (\ref{appl_eq27_1}) is expressed as
\begin{eqnarray}\label{appl_eq28_1}
\frac{d^2}{da^2}N(X+\sqrt{a}W) & \leq & 0,
\end{eqnarray}
i.e.,  $N(X+\sqrt{a}W)$ is a concave function of $a$ \cite{Costa:EPI}.
\begin{proof}
See Appendix \ref{AppG}.
\end{proof}
\end{lem}

\subsection{Applications in Other Areas}\label{subsec_oth}

There are  many other applications of the proposed results. First, since Theorem \ref{thm6} is equivalent to Theorem 1 in \cite{inf_nongauss:guo}, Theorem \ref{thm6} can be used for  applications such as generalized EXIT charts and power allocation in systems with parallel non-Gaussian noise channels as mentioned in \cite{inf_nongauss:guo}. Second, by Theorem \ref{thm5}, we showed the equivalence among Stein, De Bruijn, and heat equation identities. Therefore, a broad range of problems (in probability, decision theory, Bayesian statistics and graph theory) as described in \cite{heat:brown} could be considered as additional potential  applications of Theorems \ref{thm5} and  \ref{thm6}.

\section{Conclusions} \label{sec_con}
This paper mainly disclosed three information-estimation relationships. First, the equivalence between Stein identity and De Bruijn identity was proved. Second, it was proved that the first derivative of the differential entropy with respect to the parameter $a$ can be expressed in terms of the posterior mean. Second, this paper showed that the second derivative of the differential entropy with respect to the parameter $a$ can be expressed in terms of the Fisher information. Finally, several applications based on the three main results listed above were provided. The suggested applications illustrate that the proposed  results are useful not only in information theory  but also in the estimation theory field and other fields.

% Appendix
\appendices

% proof of equivalence
% proof of Theorem 5
\section{A Proof of Theorem \ref{thm5}}\label{AppA}
Since Theorem \ref{thm4} is considered as a special case of Theorem \ref{thm5}, we only show the proof of Theorem \ref{thm5} in this paper.
\begin{proof}\label{pf_thm5} [Theorem \ref{thm5}]

Prior to proving Theorem \ref{thm5}, we first introduce the following relationships in Lemma \ref{lem6}, which are required for the proof.

\begin{lem}\label{lem6}
For random variables $W$, $X$ and $Y$ defined in equation (\ref{int_eq1_1}) when Gaussian random variable $W$ has zero mean and unit variance and random variable $X$ has finite second-order moment, the following identities are satisfied:
%\begin{figure*}[!t]
\begin{eqnarray}\label{appd_eq1_1}
\hspace{-3mm}&\text{i)}& \hspace{-2mm}\frac{d}{da} \log f_{\scriptscriptstyle Y}(y;a) \Bigg|_{\scriptscriptstyle y=u+\sqrt{a}w}\nonumber\\
\hspace{-3mm} && \hspace{-2mm}= \frac{1}{2a^2} \left(\frac{\mathbb{E}_{\scriptscriptstyle X }\left[(y-X)^2 f_{\scriptscriptstyle Y|X}(y|X;a)\right]}{ f_{\scriptscriptstyle Y}(y;a)} -a \right)\Bigg|_{\scriptscriptstyle y=u+\sqrt{a}w},\nonumber\\
\hspace{-3mm}&\text{ii)}  &\hspace{-2mm} \frac{d}{da}\log f_{\scriptscriptstyle Y}(u+\sqrt{a}w;a)\nonumber\\
\hspace{-3mm} &&\hspace{-2mm}= \frac{1}{2a^2} \hspace{-0.5mm}\left(\hspace{-0.5mm} \frac{\mathbb{E}_{\scriptscriptstyle X}\hspace{-1mm} \left[(u\hspace{-0.2mm}-\hspace{-0.2mm}X)(y\hspace{-0.2mm}-\hspace{-0.2mm}X) f_{\scriptscriptstyle Y|X}\hspace{-0.2mm}(y|X;a)\right]}{ f_{\scriptscriptstyle Y}(y;a)} \hspace{-0.2mm} - \hspace{-0.2mm} a\hspace{-0.2mm}\right)\hspace{-1mm}\Bigg|_{\scriptscriptstyle y=u+\sqrt{a}w},\nonumber\\
\hspace{-3mm}&\text{iii)}  &\hspace{-2mm} \frac{d}{dy} \log f_{\scriptscriptstyle Y}(y;a) \Bigg|_{\scriptscriptstyle y=u+\sqrt{a}w}\nonumber\\
\hspace{-3mm} &&\hspace{-2mm}= - \frac{\mathbb{E}_{\scriptscriptstyle X} \left[(y-X) f_{\scriptscriptstyle Y|X}(y|X;a)\right]}{a f_{\scriptscriptstyle Y}(y;a)} \Bigg|_{\scriptscriptstyle y=u+\sqrt{a}w},\nonumber\\
\hspace{-3mm}&\text{iv)}  &\hspace{-2mm} \frac{w}{2\sqrt{a}} \frac{d}{dy} \log f_{\scriptscriptstyle Y}(y;a) \Bigg|_{\scriptscriptstyle y=u+\sqrt{a}w}\nonumber\\
\hspace{-3mm} &&\hspace{-2mm} = \frac{d}{da}\log f_{\scriptscriptstyle Y}(u+\sqrt{a}w;a) -\left[ \frac{d}{da} \log f_{\scriptscriptstyle Y}(y;a) \right]_{\scriptscriptstyle y=u+\sqrt{a}w},\nonumber
\end{eqnarray}
%\end{figure*}[!t]
where $f(y)|_{y=a}$ denotes $\lim_{y\rightarrow a} f(y)$. In some cases, to avoid confusion, $[f(y)]_{y=a}$ is used instead of $f(y)|_{y=a}$.
\begin{proof}
Since $f_{Y|X}(y|x;a)$ is normally distributed with mean $x$ and variance $a$, the following relationships hold:
\begin{eqnarray}
\label{appd_eq1_1_1}
\hspace{-8mm}&&\hspace{-5mm}f_{\scriptscriptstyle Y|X}(y|x;a) = \frac{1}{\sqrt{2\pi a}} \exp\left(-\frac{(y-x)^2}{2a}\right),\\
\label{appd_eq1_1_3}
\hspace{-8mm}&&\hspace{-5mm}\frac{d}{dy}f_{\scriptscriptstyle Y|X}(y|x;a) = -\frac{1}{ a}(y-x)f_{\scriptscriptstyle Y|X}(y|x;a),\\
\label{appd_eq1_1_2}
\hspace{-8mm}&&\hspace{-5mm}\frac{d}{da}f_{\scriptscriptstyle Y|X}(y|x;a) = \left(-\frac{1}{2a} + \frac{1}{2a^2} (y-x)^2 \right)f_{\scriptscriptstyle Y|X}(y|x;a),\\
\label{appd_eq1_1_4}
\hspace{-8mm}&&\hspace{-5mm}\frac{d}{da}f_{\scriptscriptstyle Y|X}(u+\sqrt{a}w|x;a)=  f_{\scriptscriptstyle Y|X}(u+\sqrt{a}w|x;a)\nonumber\\
\hspace{-8mm} &&\hspace{15mm}\times \left(-\frac{1}{ 2a} + \frac{1}{2a^2} (u+\sqrt{a}w-x)(u-x)\right).
\end{eqnarray}
Equation (\ref{appd_eq1_1_4}) is true since
\begin{eqnarray}
\hspace{-8mm}&& \hspace{-3mm}\frac{d}{da}f_{Y|X}(u+\sqrt{a}w|x;a) \nonumber\\
\hspace{-8mm}& = & \hspace{-3mm}\frac{d}{da} \left[\frac{1}{\sqrt{2\pi a}} \exp \left(-\frac{1}{2a}(u+\sqrt{a}w-x)^2 \right)\right]\nonumber\\
\hspace{-8mm}& = & \hspace{-3mm}-\frac{1}{2a} \left(\frac{1}{\sqrt{2\pi a}} \exp \left( -\frac{1}{2a} (u+\sqrt{a}w-x)^2 \right) \right)\nonumber\\
\hspace{-8mm}&& \hspace{-3mm} + \left(\frac{1}{\sqrt{2\pi a}} \exp \left( -\frac{1}{2a} (u+\sqrt{a}w-x)^2 \right) \right) \nonumber\\
\hspace{-8mm}&& \times \left( -\frac{2(u+\sqrt{a}w-x)(\frac{w}{2\sqrt{a}})a-(u+\sqrt{a}w-x)^2}{2a^2}\right)\nonumber\\
\hspace{-8mm}& = & \hspace{-3mm}-\frac{1}{2a} f_{Y|X}(u+\sqrt{a}w|x;a)\nonumber\\
\hspace{-8mm}&& \hspace{-3mm}+ f_{Y|X}(u+\sqrt{a}w|x;a) \left( -\frac{(u+\sqrt{a}w-x)(u-x)}{2a^2}\right).\nonumber
\end{eqnarray}

Based on equation (\ref{appd_eq1_1_2}), i) is proved by following these calculations:
\begin{eqnarray}
\hspace{-8mm}&&\hspace{-2mm}\frac{d}{da} \log f_Y (y;a) \Bigg|_{y=u+\sqrt{a}w} \nonumber\\
\hspace{-8mm}& = & \hspace{-2mm}\frac{\mathbb{E}_X \left[ \frac{d}{da} f_{Y|X}(y|X;a) \right]}{f_Y(y;a)} \Bigg|_{y=u+\sqrt{a}w}\nonumber\\
\label{appd_eq1_1_7}
\hspace{-8mm}& = & \hspace{-2mm}\frac{1}{2a^2} \left( \frac{\mathbb{E}_X \left[ (y-X)^2 f_{Y|X}(y|X;a) \right]}{f_Y(y;a)}-a \right) \Bigg|_{y=u+\sqrt{a}w}.
\end{eqnarray}

Second, equation ii) is proved by the following calculations:
\begin{eqnarray}
\hspace{-8mm}&&\hspace{-3mm} \frac{d}{da} \log f_{\scriptscriptstyle Y}(u+\sqrt{a}w;a)\nonumber\\
\hspace{-8mm}& = & \hspace{-3mm}\frac{\mathbb{E}_{\scriptscriptstyle X} \left[\frac{d}{da} f_{\scriptscriptstyle Y|X} (u+\sqrt{a}w|X;a) \right]}{f_{\scriptscriptstyle Y}(u+\sqrt{a}w;a)}\nonumber\\
\label{appd_eq1_1_5}
\hspace{-8mm}& = & \hspace{-3mm}\frac{\mathbb{E}_{\scriptscriptstyle X} \left[-\frac{1}{2a} f_{\scriptscriptstyle Y|X} (u+\sqrt{a}w|X;a)\right]}{f_{\scriptscriptstyle Y}(u+\sqrt{a}w;a)}\nonumber\\
\hspace{-8mm}&   &\hspace{-3mm} +\hspace{-.5mm}\frac{\mathbb{E}_{\scriptscriptstyle X} \hspace{-1mm}\left[\frac{1}{2a^2} (u \hspace{-.7mm} + \hspace{-.7mm}\sqrt{a}w \hspace{-.7mm} - \hspace{-.7mm} X)(u \hspace{-.7mm}- \hspace{-.7mm} X) f_{\scriptscriptstyle Y|X} (u \hspace{-.7mm} + \hspace{-.7mm} \sqrt{a}w|X;a) \right]}{f_{\scriptscriptstyle Y}(u+\sqrt{a}w;a)}\nonumber\\
\hspace{-8mm}&&\hspace{-3mm}\\
\hspace{-8mm}& = &\hspace{-3mm} \frac{-a f_{\scriptscriptstyle Y}(u+\sqrt{a}w;a) }{2a^2 f_{\scriptscriptstyle Y}(u+\sqrt{a}w;a)}\nonumber\\
\hspace{-8mm}&  & \hspace{-3mm}+ \frac{\mathbb{E}_{\scriptscriptstyle X} \left[(u\hspace{-.5mm}+\hspace{-.5mm}\sqrt{a}w\hspace{-.5mm}-\hspace{-.5mm}X)(u\hspace{-.5mm}-\hspace{-.5mm}X) f_{\scriptscriptstyle Y|X} (u\hspace{-.5mm}+\hspace{-.5mm}\sqrt{a}w|X;a) \right]}{2a^2 f_{\scriptscriptstyle Y}(u+\sqrt{a}w;a)}\nonumber
\end{eqnarray}
\begin{eqnarray}
\hspace{-6mm}& = & \hspace{-3mm}\frac{1}{2a^2} \hspace{-.7mm}\left(\hspace{-.7mm} \frac{\mathbb{E}_{\scriptscriptstyle X}\hspace{-1mm} \left[(u \hspace{-.7mm} + \hspace{-.7mm} \sqrt{a}w \hspace{-.7mm}- \hspace{-1mm}X)(u \hspace{-.7mm}-\hspace{-1mm}X) f_{\scriptscriptstyle Y|X} (u\hspace{-.7mm}+\hspace{-.7mm}\sqrt{a}w|X;a) \right]}{f_{\scriptscriptstyle Y}(u+\sqrt{a}w;a)} \hspace{-.7mm}-\hspace{-.7mm}a \hspace{-.7mm}\right)\nonumber\\
\label{appd_eq1_1_8}
\hspace{-6mm}& = & \hspace{-3mm}\frac{1}{2a^2}\hspace{-.7mm} \left( \hspace{-.7mm}\frac{\mathbb{E}_{\scriptscriptstyle X} \hspace{-.7mm} \left[(y\hspace{-.7mm}-\hspace{-.7mm}X)(u\hspace{-.7mm}-\hspace{-.7mm}X) f_{\scriptscriptstyle Y|X} (y|X;a) \right]}{f_{\scriptscriptstyle Y}(y;a)} -a \hspace{-.7mm}\right)\hspace{-.7mm}\Bigg|_{\scriptscriptstyle y=u+\sqrt{a}w}.
\end{eqnarray}
The equality in (\ref{appd_eq1_1_5}) is due to equation (\ref{appd_eq1_1_4}).

Third, equation iii) is proved based on equation (\ref{appd_eq1_1_3}) as follows:
\begin{eqnarray}
&&\frac{d}{dy} \log f_Y(y;a) \Bigg|_{y=u+\sqrt{a}w} \nonumber\\
& = & \frac{ \mathbb{E}_X\left[ \frac{d}{dy} f_{Y|X}(y|X;a)\right]}{ f_Y(y;a)}\Bigg|_{y=u+\sqrt{a}w}\nonumber\\
\label{appd_eq1_1_6}& = & \frac{- \mathbb{E}_X\left[ (y-X) f_{Y|X}(y|X;a)\right]}{a f_Y(y;a)}\Bigg|_{y=u+\sqrt{a}w}.
\end{eqnarray}
The equality in (\ref{appd_eq1_1_6}) is due to equation (\ref{appd_eq1_1_3}).

Equation iv) is trivial since equation (\ref{appd_eq1_1_6}) multiplied by $w/2\sqrt{a}$ is equal to equation (\ref{appd_eq1_1_8}) minus equation (\ref{appd_eq1_1_7}), and the proof is completed.
\end{proof}
\end{lem}

Like the proof of Theorem \ref{thm3} in \cite{heat:brown}, the equivalence is proved by showing that each identity is derived from the other one, using  Lemma \ref{lem6}.

First, in the generalized Stein's identity, all necessary functions are defined as follows:
\begin{eqnarray}\label{appd_eq1_2}
r(y;a) = -\frac{d}{dy} \log f_Y(y;a),\quad k(y) = 1, \nonumber\\
t(y;a) = -\frac{\frac{d}{dy}f_Y(y;a)}{f_Y(y;a)}, \quad \textrm{and} \quad \nu = 0.
\end{eqnarray}

Then, De Bruijn's identity is derived from the generalized Stein's identity as follows.
\begin{eqnarray}
\label{appd_eq3_1}
\hspace{-6mm}&& \hspace{-3mm}\frac{1}{2} \mathbb{E}_Y \left[ \frac{d}{dY} r(Y;a) \right]\nonumber\\
\hspace{-6mm}& = & \hspace{-3mm}\frac{1}{2} \mathbb{E}_Y \left[ r(Y;a) t(Y;a) \right] \quad \text{(generalized Stein's identity)}\\
\hspace{-6mm}& = & \hspace{-3mm}-\frac{1}{2} \int_{-\infty}^{\infty} \frac{d}{dy} \mathbb{E}_X \left[f_{Y|X}(y|X;a)\right] r(y;a) dy\nonumber\\
\hspace{-6mm}& = & \hspace{-3mm} - \mathbb{E}_X \left[ \int_{-\infty}^{\infty} \frac{(y-X)}{2a} f_{Y|X}(y|X;a) \frac{d}{dy} \log f_Y(y;a) dy\right]\nonumber\\
\label{appd_eq3_2}
\hspace{-6mm}& = & \hspace{-3mm}-\int_{\scriptscriptstyle -\infty}^{\scriptscriptstyle \infty} \hspace{-1.3mm} f_{\scriptscriptstyle X}(u) \hspace{-.3mm} \underbrace{\int_{\scriptscriptstyle -\infty}^{\scriptscriptstyle \infty} \hspace{-1mm} \frac{(y \hspace{-.7mm} - \hspace{-.7mm} u)}{2a} f_{\scriptscriptstyle Y|X}(y|u;a) \frac{d}{dy} \log f_{\scriptscriptstyle Y}(y;a) dy}_{(A)} du.\nonumber\\
\hspace{-6mm}&&
\end{eqnarray}
The interchangeability among integrals and derivatives are due to the dominated convergence theorem and Fubini's theorem.

Changing the variable as $y=u+\sqrt{a}w$, equation $(A)$ is expressed as
\begin{eqnarray}
\hspace{-5mm}&   & \hspace{-3mm}\int_{\scriptscriptstyle -\infty}^{\scriptscriptstyle \infty} \frac{(y-u)}{2a} f_{\scriptscriptstyle Y|X}(y|u;a) \frac{d}{dy} \log f_{\scriptscriptstyle Y}(y;a) dy \nonumber\\
\hspace{-5mm}& =  &\hspace{-3mm}  \int_{\scriptscriptstyle -\infty}^{\scriptscriptstyle \infty}\hspace{-1.5mm} \frac{\sqrt{a}w}{2a} f_{\scriptscriptstyle Y|X}\hspace{-.7mm}(u \hspace{-.7mm}+\hspace{-.7mm}\sqrt{a}w|u;a) \hspace{-1mm}\left[ \frac{d}{dy}\hspace{-.7mm} \log \hspace{-.7mm}f_{\scriptscriptstyle Y}(y;a) \right]_{\scriptscriptstyle y=u+\sqrt{a}w}\hspace{-3mm}\sqrt{a} dw\nonumber\\
\label{appd_eq4_1}
\hspace{-5mm}& = &\hspace{-3mm} \int_{\scriptscriptstyle -\infty}^{\scriptscriptstyle \infty} f_{\scriptscriptstyle Y|X}(u+\sqrt{a}w|u;a)\Bigg(\frac{d}{da} \log f_{\scriptscriptstyle Y} (u+\sqrt{a}w;a)\nonumber\\
\hspace{-5mm}&  &\hspace{30mm} - \left[ \frac{d}{da} \log f_{\scriptscriptstyle Y}(y;a) \right]_{\scriptscriptstyle y=u+\sqrt{a}w} \Bigg) \sqrt{a} dw\nonumber\\
\hspace{-5mm}&&\hspace{-3mm}\\
\hspace{-5mm}& = &\hspace{-3mm}\int_{\scriptscriptstyle -\infty}^{\scriptscriptstyle \infty} \frac{1}{\sqrt{2\pi}} \exp \left(-\frac{1}{2} w^2 \right) \frac{d}{da} \log f_{\scriptscriptstyle Y} (u+\sqrt{a}w;a) dw\nonumber\\
\hspace{-5mm}&& \hspace{-3mm}- \int_{\scriptscriptstyle -\infty}^{\scriptscriptstyle \infty} \frac{1}{\sqrt{2\pi}} \exp \left(-\frac{1}{2} w^2 \right) \left[ \frac{d}{da} \log f_{\scriptscriptstyle Y}(y;a) \right]_{\scriptscriptstyle y=u+\sqrt{a}w} \hspace{-2mm}dw\nonumber\\
\hspace{-5mm}& = & \hspace{-3mm}\frac{d}{da} \int_{\scriptscriptstyle -\infty}^{\scriptscriptstyle \infty} \frac{1}{\sqrt{2\pi}}  \exp{\left(-\frac{w^2}{2}\right)} \log f_{\scriptscriptstyle Y} (u+\sqrt{a}w;a) dw\nonumber\\
\label{appd_eq4_2}
\hspace{-5mm}&   & \hspace{-3mm} - \int_{\scriptscriptstyle -\infty}^{\scriptscriptstyle \infty}\hspace{-1.5mm} \frac{1}{\sqrt{2\pi}} \exp{\left(-\frac{w^2}{2}\right)} \left[ \frac{d}{da} \log f_{\scriptscriptstyle Y}(y;a) \right]_{\scriptscriptstyle y=u+\sqrt{a}w} \hspace{-2mm} dw.\nonumber\\
\hspace{-5mm}&&\hspace{-3mm}
\end{eqnarray}
The equality in equation (\ref{appd_eq4_1}) is due to Lemma \ref{lem6}, iv).

Re-defining the variable $w=(y-u)/\sqrt{a}$, equation (\ref{appd_eq3_2}) is expressed as
\begin{eqnarray}
\hspace{-6mm} &	&\hspace{-3mm} - \hspace{-1mm} \int_{\scriptscriptstyle -\infty}^{\scriptscriptstyle \infty} \hspace{-2mm} f_{\scriptscriptstyle X}(u) \hspace{-1mm} \left(\int_{\scriptscriptstyle -\infty}^{\scriptscriptstyle \infty}\hspace{-1.5mm} \frac{(y\hspace{-.5mm} - \hspace{-.5mm} u)}{2a} f_{\scriptscriptstyle Y|X}\hspace{-.5mm} (y|u;a) \frac{d}{dy} \hspace{-.5mm} \log\hspace{-.5mm}  f_{\scriptscriptstyle Y}\hspace{-.5mm} (y;a) dy \hspace{-.8mm}  \right) \hspace{-.8mm} du\nonumber\\
\label{appd_eq4_2_1}
\hspace{-6mm} &  = &\hspace{-3mm}  \int_{\scriptscriptstyle -\infty}^{\scriptscriptstyle \infty} f_{\scriptscriptstyle X}(u) \Bigg( \int_{\scriptscriptstyle -\infty}^{\scriptscriptstyle \infty} f_{\scriptscriptstyle Y|X}(y|u;a) \frac{d}{da} \log f_{\scriptscriptstyle Y}(y;a) dy \nonumber\\
\hspace{-6mm} &&\hspace{20mm} - \frac{d}{da} \int_{\scriptscriptstyle -\infty}^{\scriptscriptstyle \infty} f_{\scriptscriptstyle Y|X}(y|u;a) \log f_{\scriptscriptstyle Y}(y;a) dy \Bigg) du\nonumber\\
\hspace{-6mm} & &\hspace{-3mm} \\
\label{appd_eq4_2_2}
\hspace{-6mm} & = &\hspace{-3mm}   \int_{\scriptscriptstyle -\infty}^{\scriptscriptstyle \infty} f_{\scriptscriptstyle Y}(y;a) \frac{d}{da} \log f_{\scriptscriptstyle Y}(y;a) dy \nonumber\\
\hspace{-6mm}  &&\hspace{25mm}  - \frac{d}{da} \int_{-\infty}^{\infty} f_{\scriptscriptstyle Y}(y;a) \log f_{\scriptscriptstyle Y}(y;a) dy\\
\hspace{-6mm} & = &\hspace{-3mm}   \int_{\scriptscriptstyle -\infty}^{\scriptscriptstyle \infty} \frac{d}{da} f_{\scriptscriptstyle Y}(y;a) dy - \frac{d}{da} \int_{\scriptscriptstyle -\infty}^{\scriptscriptstyle \infty} f_{\scriptscriptstyle Y}(y;a) \log f_{\scriptscriptstyle Y}(y;a) dy\nonumber\\
\hspace{-6mm} & = &\hspace{-3mm}   \frac{d}{da} \int_{\scriptscriptstyle -\infty}^{\scriptscriptstyle \infty}  f_{\scriptscriptstyle Y}(y;a) dy - \frac{d}{da} \int_{\scriptscriptstyle -\infty}^{\scriptscriptstyle \infty} f_{\scriptscriptstyle Y}(y;a) \log f_{\scriptscriptstyle Y}(y;a) dy\nonumber\\
\hspace{-6mm} & = &\hspace{-3mm}  - \frac{d}{da} \int_{\scriptscriptstyle -\infty}^{\scriptscriptstyle \infty} f_{\scriptscriptstyle Y}(y;a) \log f_{\scriptscriptstyle Y}(y;a) dy\nonumber\\
\hspace{-6mm} & = &\hspace{-3mm}  \frac{d}{da}h(Y).\nonumber
\end{eqnarray}
The equality in (\ref{appd_eq4_2_1}) is due to the change of variable, and the equality in (\ref{appd_eq4_2_2}) is because of the independence of $f_X(u)$ with respect to $a$.

Since the left-hand side of equation (\ref{appd_eq3_1}) is equal to $J(Y)/2$, we obtain De Bruijn's identity:
\begin{eqnarray}
\frac{1}{2}J(Y) & = & \frac{d}{da}h(Y),\nonumber
\end{eqnarray}
from the generalized Stein's identity.

Second, the generalized Stein's identity is derived from De Bruijn's identity. We define the function
\begin{eqnarray}\label{appd_eq4_3}
g(y;a) & = & \int_{0}^{y} r(u;a) du + q(a),
\end{eqnarray}
where $q(a) = - \log f_Y(y;a) |_{y=0}$. Here, $q(a)$ is always real-valued due to the following:
\begin{eqnarray}\label{qfunction_def_1_1}
f_Y(y;a)\Big|_{y=0} & = & \lim\limits_{y \rightarrow 0} \mathbb{E}_X [f_{Y|X}(y|X;a)]\nonumber\\
& = & \mathbb{E}_X \left[\lim\limits_{y \rightarrow 0}  \frac{1}{\sqrt{2 \pi a}} \exp \left(-\frac{1}{2a}(y-X)^2\right)\right] \nonumber\\
& = & \mathbb{E}_X \left[ \frac{1}{\sqrt{2 \pi a}} \exp \left(-\frac{1}{2a}X^2\right)\right]\nonumber\\
& \leq & \frac{1}{\sqrt{2 \pi a}}.
\end{eqnarray}
The last inequality is due to $\exp (-\frac{1}{2a}X^2) \leq 1$. In addition, equation (\ref{qfunction_def_1_1}) is always greater than zero unless $f_X(x)$ is identical to zero or $a$ is infinite. However, neither case holds. Therefore, $q(a)$ is always mapping to a real-valued number.

Then, the expectation of $g(y;a)$ is expressed as
\begin{eqnarray}\label{appd_eq8_1}
\hspace{-2mm} &   & \hspace{-2mm} \mathbb{E}_Y \left[ g(Y;a) \right]\nonumber\\
\hspace{-2mm} & = & \hspace{-2mm}  \int_{-\infty}^{\infty} f_Y(y;a) \left(\int_{0}^{y} r(u;a) du + q(a)\right) dy\nonumber\\
\hspace{-2mm} & = & \hspace{-2mm}  \int_{0}^{\infty} \int_{0}^{y} f_Y(y;a) r(u;a) du dy\nonumber\\
\hspace{-2mm} &	& \hspace{20mm} + \int_{-\infty}^{0} \int_{0}^{y} f_Y(y;a) r(u;a) du dy + q(a)\nonumber\\
\hspace{-2mm} & = &  \hspace{-2mm} \int_{0}^{\infty} \int_{0}^{y} f_Y(y;a) r(u;a) du dy \nonumber\\
\hspace{-2mm} &	&\hspace{20mm}  - \int_{-\infty}^{0} \int_{y}^{0} f_Y(y;a) r(u;a) du dy + q(a)\nonumber\\
\hspace{-2mm} & = & \hspace{-2mm} \int_{0}^{\infty} \left(\int_{u}^{\infty} f_Y(y;a) dy\right) r(u;a) du \nonumber\\
\hspace{-2mm} &	&\hspace{15mm}  - \int_{-\infty}^{0} \left(\int_{-\infty}^{u} f_Y(y;a) dy \right) r(u;a) du + q(a)\nonumber\\
\hspace{-2mm} & = & \hspace{-2mm} \mathbb{E}_X \left[\int_{0}^{\infty} \left(\int_{u}^{\infty} f_{Y|X}(y|X;a) dy \right) r(u;a) du \right]\nonumber\\
\hspace{-2mm} &	&\hspace{-2mm}  - \mathbb{E}_X \left[ \int_{-\infty}^{0} \left(\int_{-\infty}^{u} f_{Y|X}(y|X;a) dy \right) r(u;a) du \right] +q(a)\nonumber\\
\hspace{-2mm} & = & \hspace{-2mm} \mathbb{E}_X \left[\int_{0}^{\infty} \left( 1 - \Phi\left(\frac{u-X}{\sqrt{a}}\right)\right) r(u;a) du \right]\nonumber\\
\hspace{-2mm} &	& \hspace{-2mm}   -\mathbb{E}_X \left[ \int_{-\infty}^{0} \Phi \left( \frac{u-X}{\sqrt{a}}\right) r(u;a) du\right] +q(a),
\end{eqnarray}
where $\Phi(\cdot)$ denotes the standard normal cumulative density function.

We differentiate both sides of equation (\ref{appd_eq8_1}) with respect to parameter $a$ as follows.
\begin{eqnarray}\label{appd_eq9_1}
\hspace{-8mm} &	& \hspace{-2mm} \frac{d}{da} \mathbb{E}_Y \left[ g(Y;a) \right]\nonumber\\
\hspace{-8mm} & = & \hspace{-2mm} \frac{d}{da} \mathbb{E}_X \left[ \int_{0}^{\infty} \left(1 - \Phi \left(\frac{u-X}{\sqrt{a}}\right)\right) r(u;a) du \right]\nonumber\\
\hspace{-8mm} &	& \hspace{1mm} - \mathbb{E}_X \left[ \int_{-\infty}^{0} \Phi \left( \frac{u-X}{\sqrt{a}}\right) r(u;a) du \right] + \frac{d}{da} q(a)\nonumber\\
\hspace{-8mm} & = & \hspace{-2mm}  -\mathbb{E}_X \left[ \int_{0}^{\infty} \left(\frac{d}{da} \Phi \left(\frac{u-X}{\sqrt{a}}\right)\right) r(u;a) du \right]\nonumber\\
\hspace{-8mm} &	& \hspace{1mm}  + \mathbb{E}_X \left[ \int_{0}^{\infty} \left(1 - \Phi \left(\frac{u-X}{\sqrt{a}}\right)\right) \frac{d}{da} r(u;a) du \right]\nonumber\\
\hspace{-8mm} &   & \hspace{1mm}  - \mathbb{E}_X \left[ \int_{-\infty}^{0}  \left(\frac{d}{da} \Phi \left( \frac{u-X}{\sqrt{a}}\right) \right)r(u;a) du \right] \nonumber\\
\hspace{-8mm} &	& \hspace{1mm}  - \mathbb{E}_X \left[ \int_{-\infty}^{0}  \Phi \left( \frac{u-X}{\sqrt{a}}\right) \frac{d}{da} r(u;a) du \right]+ \frac{d}{da} q(a)\nonumber\\
\hspace{-8mm} & = & \hspace{-2mm} - \mathbb{E}_X \left[ \int_{-\infty}^{\infty} \frac{d}{da} \Phi \left(\frac{u-X}{\sqrt{a}}\right) r(u;a) du \right] \nonumber\\
\hspace{-8mm} &	& \hspace{1mm}  +\underbrace{\mathbb{E}_X \left[\int_{0}^{\infty} \left( 1 - \Phi \left( \frac{u-X}{\sqrt{a}}\right) \right) \frac{d}{da} r(u;a) du \right]}_{(B)}\nonumber\\
\hspace{-8mm} &   & \hspace{1mm}  -\underbrace{\mathbb{E}_X \left[ \int_{-\infty}^{0} \Phi \left(\frac{u-X}{\sqrt{a}}\right) \frac{d}{da} r(u;a) du \right]}_{(C)} + \frac{d}{da} q(a).
\end{eqnarray}
Equations (B) and (C) are further processed as
\begin{eqnarray}
\hspace{-8mm}&	& \hspace{-2mm} \mathbb{E}_X \left[\int_{0}^{\infty} \left( 1 - \Phi \left( \frac{u-X}{\sqrt{a}}\right) \right) \frac{d}{da} r(u;a) du \right] \nonumber\\
\hspace{-10mm}&	& \hspace{6mm}- \mathbb{E}_X \left[ \int_{-\infty}^{0} \Phi \left(\frac{u-X}{\sqrt{a}}\right) \frac{d}{da} r(u;a) du \right]\nonumber\\
\hspace{-10mm}& = & \hspace{-2mm}  \mathbb{E}_X \left[\int_{0}^{\infty} \int_{u}^{\infty} f_{Y|X} (y|X;a) dy \frac{d}{da} r(u;a) du \right]\nonumber\\
\hspace{-10mm}&	& \hspace{6mm}- \mathbb{E}_X \left[ \int_{-\infty}^{0} \int_{-\infty}^{u} f_{Y|X}(y|X;a) dy \frac{d}{da} r(u;a) du \right]\nonumber\\
\hspace{-10mm}& = & \hspace{-2mm}  \mathbb{E}_X \left[\int_{0}^{\infty} \int_{0}^{y} \frac{d}{da} r(u;a) du f_{Y|X} (y|X;a) dy  \right] \nonumber\\
\hspace{-10mm}&	& \hspace{6mm}- \mathbb{E}_X \left[ \int_{-\infty}^{0} \int_{y}^{0}  \frac{d}{da} r(u;a) du f_{Y|X}(y|X;a) dy \right]\nonumber\\
\hspace{-10mm}& = & \hspace{-2mm}  \mathbb{E}_X \left[\int_{0}^{\infty} \int_{0}^{y} \frac{d}{da} r(u;a) du f_{Y|X} (y|X;a) dy\right]\nonumber\\
\hspace{-10mm}&	&  \hspace{6mm}+ \mathbb{E}_X \left[ \int_{-\infty}^{0} \int_{0}^{y}  \frac{d}{da} r(u;a) du f_{Y|X}(y|X;a) dy \right]\nonumber\\
\label{appd_eq4_4}
\hspace{-10mm}& = & \hspace{-2mm}  \mathbb{E}_X \left[\int_{-\infty}^{\infty} \int_{0}^{y} \frac{d}{da} r(u;a) du f_{Y|X} (y|X;a) dy  \right].
\end{eqnarray}
The interchangeability among integrals is due to Fubini's theorem and dominated convergence theorem.

Due to equation (\ref{appd_eq4_3}),
\begin{eqnarray}
\frac{d}{da} g(y;a) & = & \frac{d}{da} \int_{0}^{y} r(u;a) du + \frac{d}{da} q(a),\nonumber
\end{eqnarray}
equation (\ref{appd_eq4_4}) is further simplified as follows:
\begin{eqnarray}
&	& \mathbb{E}_X \left[\int_{-\infty}^{\infty} \int_{0}^{y} \frac{d}{da} r(u;a) du f_{Y|X} (y|X;a) dy  \right]\nonumber\\
& = &  \int_{-\infty}^{\infty} \left(  \frac{d}{da}  \int_{0}^{y} r(u;a) du \right) f_{Y} (y;a) dy \nonumber\\
& =  & \int_{-\infty}^{\infty} f_Y(y;a) \frac{d}{da} g(y;a) dy - \frac{d}{da} q(a) \nonumber\\
\label{appd_eq4_4_1}& = & -\int_{-\infty}^{\infty} f_Y(y;a) \frac{d}{da} \log f_Y(y;a) dy - \frac{d}{da} q(a)\\
& = & -\frac{d}{da} q(a).\nonumber
\end{eqnarray}
The equality in (\ref{appd_eq4_4_1}) holds because $g(y;a) = - \log f_Y(y;a)$.

Therefore, the last three terms in equation (\ref{appd_eq9_1}) vanish, and equation (\ref{appd_eq9_1}) is expressed as
\begin{eqnarray}
\hspace*{-3mm}&   &\hspace*{-3mm} -\mathbb{E}_X \left[ \int_{-\infty}^{\infty} \frac{d}{da} \Phi \left(\frac{u-X}{\sqrt{a}}\right) r(u;a) du \right]\nonumber\\
\hspace*{-3mm}& = &\hspace*{-3mm} \mathbb{E}_X \left[ \int_{-\infty}^{\infty} \frac{(u-X)}{2a\sqrt{a}} \left[ \frac{d}{dy} \Phi  \left(y\right)\right]_{y=\frac{u-X}{\sqrt{a}}} r(u;a) du \right]\nonumber\\
\hspace*{-3mm}& = &\hspace*{-3mm} \mathbb{E}_X \left[ \int_{-\infty}^{\infty} \frac{(u-X)}{2a\sqrt{a}} \phi \left(\frac{u-X}{\sqrt{a}}\right) r(u;a) du \right]\nonumber\\
\hspace*{-3mm}& = &\hspace*{-3mm} \frac{1}{2} \int_{-\infty}^{\infty} \hspace*{-3mm} \mathbb{E}_X \hspace*{-1mm} \left[  \frac{(u-X)}{a} \frac{1}{\sqrt{2\pi a}} \exp\left(-\frac{(u-X)^2}{2a}\right) \right] r(u;a) du \nonumber\\
\hspace*{-3mm}& = &\hspace*{-3mm} - \frac{1}{2} \int_{-\infty}^{\infty} \hspace*{-3mm} \mathbb{E}_X \hspace*{-1mm} \left[ \frac{d}{dy} f_{Y|X}(y|X;a) \right] r(u;a) du \nonumber\\
\hspace*{-3mm}& = &\hspace*{-3mm} -\frac{1}{2} \int_{-\infty}^{\infty} \frac{\frac{d}{du}f_Y(u;a)}{f_Y(u;a)} r(u;a) f_Y(u;a) du\nonumber\\
\hspace*{-3mm}& = &\hspace*{-3mm} \frac{1}{2} \mathbb{E}_Y \left[ t(Y;a) r(Y;a) \right], \nonumber
\end{eqnarray}
where $\phi(\cdot)$ denotes the standard normal probability density function, and $t(y;a) = - (\frac{d}{dy}f_Y(y;a))/f_Y(y;a)$.

Since
\begin{eqnarray}
\frac{d}{da}h(Y) & = & \frac{d}{da} \mathbb{E}_Y \left[g(Y;a)\right]\nonumber\\
& = & \frac{1}{2} \mathbb{E}_Y \left[ t(Y;a)r(Y;a) \right],\nonumber
\end{eqnarray}
and
\begin{eqnarray}
\frac{1}{2}J(Y) & = & \frac{1}{2} \mathbb{E}_Y \left[ \frac{d}{dY} r(Y;a) \right],\nonumber
\end{eqnarray}
from De Bruijn's identity, we derive the generalized Stein's identity:
\begin{eqnarray}
\frac{d}{da}h(Y) & = & \frac{1}{2}J(Y)\nonumber\\
%\Longleftrightarrow \frac{1}{2} \mathbb{E}_Y \left[ -\frac{\frac{d}{dY}f_Y(Y;a)}{f_Y(Y;a)} r(Y;a) \right] & = & \frac{1}{2} \mathbb{E}_Y  \left[\frac{d}{dY}r(Y;a)\right]\nonumber\\
\Longleftrightarrow \mathbb{E}_Y \left[t(Y;a)r(Y;a)\right] & = & \mathbb{E}_Y \left[\frac{d}{dY}r(Y;a) \right],\nonumber
\end{eqnarray}
where $\Longleftrightarrow$ denotes equivalence between before and after the notation.
\end{proof}

% proof of theorem 6 (extension of De Bruijn)
\section{A Proof of Theorem \ref{thm6}}\label{AppB}
Based on equation (\ref{appd_eq17_1}), Theorem \ref{thm6} is proved next using integration
by parts and the dominated convergence theorem.

\begin{proof}\label{pf_thm6} [Theorem \ref{thm6}]
\begin{eqnarray}
\hspace{-8mm} &&\hspace{-3mm} \frac{d}{da}h(Y) \nonumber\\
\hspace{-8mm} & = &\hspace{-3mm}  - \int_{\scriptscriptstyle -\infty}^{\scriptscriptstyle \infty} \left(1+\log f_{\scriptscriptstyle Y}(y;a)\right) \frac{d}{da} f_{\scriptscriptstyle Y}(y;a) dy\nonumber\\
\label{appd_eq18_2}
\hspace{-8mm} & = &\hspace{-3mm}  - \int_{\scriptscriptstyle -\infty}^{\scriptscriptstyle \infty} \hspace{-1mm} \frac{d}{da} f_{\scriptscriptstyle Y}(y;a) dy - \int_{\scriptscriptstyle -\infty}^{\scriptscriptstyle \infty}\hspace{-1mm}  \log f_{\scriptscriptstyle Y}(y;a) \frac{d}{da} f_{\scriptscriptstyle Y}(y;a) dy \\
\hspace{-8mm} & = &\hspace{-3mm}  - \int_{\scriptscriptstyle -\infty}^{\scriptscriptstyle \infty} \log f_{\scriptscriptstyle Y}(y;a) \frac{d}{da} \mathbb{E}_{\scriptscriptstyle X} \left[ f_{\scriptscriptstyle Y|X}(y|X;a)\right] dy\nonumber\\
\label{appd_eq18_4}
\hspace{-8mm} & = &\hspace{-3mm}  - \int_{\scriptscriptstyle -\infty}^{\scriptscriptstyle \infty} \log f_{\scriptscriptstyle Y}(y;a) \mathbb{E}_{\scriptscriptstyle X} \left[ \frac{d}{da} f_{\scriptscriptstyle Y|X}(y|X;a)\right] dy.
\end{eqnarray}
The interchangeability between integral and derivative is due to assumptions (\ref{bruijn_eq1_1}a) and (\ref{bruijn_eq1_1}b).

Using equation (\ref{appd_eq17_1}),  equation (\ref{appd_eq18_4}) is expressed as
\begin{eqnarray}
\hspace{-8mm}&   &\hspace{-3mm} - \int_{\scriptscriptstyle -\infty}^{\scriptscriptstyle \infty} \hspace{-1mm} \log f_{\scriptscriptstyle Y}(y;a) \mathbb{E}_{\scriptscriptstyle X} \left[ \frac{d}{da} f_{\scriptscriptstyle Y|X}(y|X;a)\right] dy\nonumber\\
\hspace{-8mm}& = &\hspace{-3mm}   \frac{1}{2a} \int_{\scriptscriptstyle -\infty}^{\scriptscriptstyle \infty} \hspace{-1mm}\log f_{\scriptscriptstyle Y}(y;a) \mathbb{E}_{\scriptscriptstyle X} \left[ \frac{d}{dy} \left((y-X)f_{\scriptscriptstyle Y|X} (y|X;a)\right) \right]dy\nonumber\\
\label{appd_eq19_2}
\hspace{-8mm}& = &\hspace{-3mm}  \frac{1}{2a} \int_{\scriptscriptstyle -\infty}^{\scriptscriptstyle \infty}\hspace{-1mm} \log f_{\scriptscriptstyle Y}(y;a) \frac{d}{dy} \mathbb{E}_{\scriptscriptstyle X} \left[ (y-X)f_{\scriptscriptstyle Y|X}(y|X;a)\right] dy\\
\label{appd_eq19_2_1}
\hspace{-8mm}& = &\hspace{-3mm}  \frac{1}{2a} \log f_{\scriptscriptstyle Y}(y;a)  \mathbb{E}_{\scriptscriptstyle X} \left[ (y-X)f_{\scriptscriptstyle Y|X}(y|X;a)\right] \Bigg|_{\scriptscriptstyle y=-\infty}^{\scriptscriptstyle \infty}\nonumber\\
\hspace{-8mm}&   &\hspace{-3mm}  - \frac{1}{2a} \int_{\scriptscriptstyle -\infty}^{\scriptscriptstyle \infty} \hspace{-1mm} \frac{d}{dy} \log f_{\scriptscriptstyle Y}(y;a) \mathbb{E}_{\scriptscriptstyle X} \left[ (y-X)f_{\scriptscriptstyle Y|X}(y|X;a)\right] dy\\
\label{appd_eq19_3}
\hspace{-8mm}& = &\hspace{-3mm}   - \frac{1}{2a}\int_{\scriptscriptstyle -\infty}^{\scriptscriptstyle \infty} \hspace{-1mm}\frac{d}{dy} \log f_{\scriptscriptstyle Y}(y;a) \mathbb{E}_{\scriptscriptstyle X} \left[(y-X)f_{\scriptscriptstyle Y|X}(y|X;a)\right] dy\\
\label{appd_eq19_4}
\hspace{-8mm}& = &\hspace{-3mm}   - \frac{1}{2a}\int_{\scriptscriptstyle -\infty}^{\scriptscriptstyle \infty} \hspace{-1mm} \frac{d}{dy} f_{\scriptscriptstyle Y}(y;a) \mathbb{E}_{\scriptscriptstyle X} \left[(y-X)\frac{f_{\scriptscriptstyle Y|X}(y|X;a)}{f_{\scriptscriptstyle Y}(y;a)}\right] dy,
\end{eqnarray}
where $f(y)|_{y = a_1}^{a_2}$ denotes $\lim\limits_{y\rightarrow a_2} f(y) - \lim\limits_{y\rightarrow a_1} f(y)$.

The first term in equation (\ref{appd_eq19_2_1}) vanishes due to the following relationship:
\begin{eqnarray}\label{appd_eq19_4_1}
\hspace{-10mm} &   & \log f_Y(y;a)  \mathbb{E}_X \left[ (y-X)f_{Y|X}(y|X;a)\right] \Big|_{y=-\infty}^{\infty}\nonumber\\
\hspace{-10mm} &  = & y f_Y(y;a) \log f_Y(y;a)\Big|_{y=-\infty}^{\infty} \nonumber\\
\hspace{-10mm} && \hspace{10mm}- \mathbb{E}_X \left[ X f_{Y|X}(y|X;a)\right] \log f_Y(y;a)\Big|_{y=-\infty}^{\infty}.
\end{eqnarray}
The first term in (\ref{appd_eq19_4_1}) is expressed as
\begin{eqnarray}\label{appd_eq19_4_2}
\hspace{-10mm} &&y f_Y(y;a) \log f_Y(y;a)\Big|_{y=-\infty}^{\infty}\nonumber\\
\hspace{-10mm} & = & 2 y \sqrt{f_Y(y;a)} \sqrt{f_Y(y;a)} \log \sqrt{f_Y(y;a)}\Big|_{y=-\infty}^{\infty}.
\end{eqnarray}
Due to assumptions (\ref{bruijn_eq1_1}d), $y \sqrt{f_Y(y;a)}$ converges to zero as $y$ goes to $\pm\infty$. Since $x \log x$ becomes zero as $x$ goes to zero and $f_Y(y;a)$ converges to zero as $y$ goes to $\pm\infty$, $\sqrt{f_Y(y;a)} \log \sqrt{f_Y(y;a)}$ in (\ref{appd_eq19_4_2}) also becomes zero as $y$ approaches $\pm\infty$.

Similarly, the second term in (\ref{appd_eq19_4_1}) is re-written as
\begin{eqnarray}\label{appd_eq19_4_3}
\hspace{-8mm}&&\hspace{-3mm}\mathbb{E}_{\scriptscriptstyle X} \left[ X f_{\scriptscriptstyle Y|X}(y|X;a)\right] \log f_{\scriptscriptstyle Y}(y;a)\Big|_{\scriptscriptstyle y=-\infty}^{\scriptscriptstyle \infty} \nonumber\\
\hspace{-8mm}&=&\hspace{-3mm} \underbrace{\frac{\mathbb{E}_{\scriptscriptstyle X} \left[ X f_{\scriptscriptstyle Y|X}(y|X;a)\right]}{\sqrt{f_{\scriptscriptstyle Y}(y;a)}} }_{(a_1)} 2 \underbrace{ \sqrt{f_{\scriptscriptstyle Y}(y;a)} \log \sqrt{f_{\scriptscriptstyle Y}(y;a)}}_{(a_2)}\Bigg|_{\scriptscriptstyle y=-\infty}^{\scriptscriptstyle \infty}.
%& = & \underbrace{\mathbb{E}_{X|Y} \left[ X|Y=y\right]}_{(a_1)} \underbrace{f_Y(y;a) \log f_Y(y;a)}_{(a_2)}.
\end{eqnarray}
Since factor $(a_2)$ tends to zero as $y$ approaches $\pm\infty$, and factor $(a_1)$ is bounded due to assumption (\ref{bruijn_eq1_1}d), the right-hand side of equation (\ref{appd_eq19_4_3}) approaches zero as $y$ goes to $\pm \infty$. Therefore, the first term in equation (\ref{appd_eq19_2_1}) is zero, and the equality in  (\ref{appd_eq19_3}) is verified.

Again, using integration by parts, equation (\ref{appd_eq19_4}) is expressed as
\begin{eqnarray}
\hspace{-8mm}&  &\hspace{-3mm}  - \frac{1}{2a}\int_{-\infty}^{\infty} \frac{d}{dy} f_Y(y;a) \mathbb{E}_X \left[(y-X)\frac{f_{Y|X}(y|X;a)}{f_Y(y;a)}\right] dy\nonumber\\
\hspace{-8mm}& = & \hspace{-3mm} - \frac{1}{2a} f_Y(y;a) \mathbb{E}_X \left[(y-X)\frac{f_{Y|X}(y|X;a)}{f_Y(y;a)} \right] \Bigg|_{y=-\infty}^{\infty} \nonumber\\
\label{appd_eq19_5}
\hspace{-8mm}&	&\hspace{-3mm} + \frac{1}{2a}\int_{-\infty}^{\infty}  f_Y(y;a) \frac{d}{dy} \mathbb{E}_X \left[(y-X)\frac{f_{Y|X}(y|X;a)}{f_Y(y;a)}\right] dy \\
\label{appd_eq19_6}
\hspace{-8mm}& = &\hspace{-3mm} \frac{1}{2a} \int_{-\infty}^{\infty}  f_Y(y;a) \frac{d}{dy} \mathbb{E}_X \left[(y-X)\frac{f_{Y|X}(y|X;a)}{f_Y(y;a)}\right] dy\\
\hspace{-8mm}& = &\hspace{-3mm} \frac{1}{2a} \int_{-\infty}^{\infty}  f_Y(y;a) \frac{d}{dy} \left(y - \mathbb{E}_X \left[X\frac{f_{Y|X}(y|X;a)}{f_Y(y;a)}\right]\right) dy\nonumber\\
\hspace{-8mm}& = &\hspace{-3mm} \frac{1}{2a} \left\{1- \mathbb{E}_Y \left[ \frac{d}{dY}\mathbb{E}_{X|Y} \left[X|Y\right] \right] \right\}.
\end{eqnarray}

The equality in (\ref{appd_eq19_6}) is verified by the following procedure: the first part of equation (\ref{appd_eq19_5}) is re-written as
\begin{eqnarray}
\hspace{-8mm}&	& - \frac{1}{2a} f_Y(y;a) \mathbb{E}_X \left[(y-X)\frac{f_{Y|X}(y|X;a)}{f_Y(y;a)} \right] \Bigg|_{y=-\infty}^{\infty} \nonumber\\
\label{appd_eq19_7}
\hspace{-8mm}& = &- \frac{1}{2a}\left( y f_Y(y;a) - \mathbb{E}_X \left[Xf_{Y|X}(y|X;a) \right]\right) \Bigg|_{y=-\infty}^{\infty}\\
\hspace{-8mm}& = & 0.\nonumber
\end{eqnarray}
Due to assumptions (\ref{bruijn_eq1_1}c) and (\ref{bruijn_eq1_1}d), both terms $y f_Y(y;a)$ and $\mathbb{E}_X [Xf_{Y|X}(y|X;a) ]$ become zero as $y$ goes to $\pm\infty$, and equation (\ref{appd_eq19_7}) is zero.

Therefore,
\begin{eqnarray}
\frac{d}{da} h(Y) = \frac{1}{2a} \left\{1- \mathbb{E}_Y \left[ \frac{d}{dY}\mathbb{E}_{X|Y} \left[X|Y\right] \right] \right\},\nonumber
\end{eqnarray}
and the proof is completed.
\end{proof}

% Proof of Theorem 7 (extension of De Bruijn: second derivative)
\section{A Proof of Theorem \ref{thm7}}\label{AppC}

\begin{proof}\label{pf_thm7} [Theorem \ref{thm7}]

From equation (\ref{appd_eq18_2}), we know
\begin{eqnarray}
\hspace{-4mm}&	&\hspace{-3mm} \frac{d}{da}h(Y) \nonumber\\
\hspace{-4mm}& = &\hspace{-3mm}  - \int_{\scriptscriptstyle -\infty}^{\scriptscriptstyle \infty}\frac{d}{da} f_Y(y;a) dy  - \int_{\scriptscriptstyle -\infty}^{\scriptscriptstyle \infty} \log f_Y(y;a) \frac{d}{da} f_Y(y;a) dy\nonumber\\
\hspace{-4mm}& = &\hspace{-3mm}   - \int_{\scriptscriptstyle -\infty}^{\scriptscriptstyle \infty} \log f_Y(y;a) \frac{d}{da} f_Y(y;a) dy.\nonumber
\end{eqnarray}
Therefore, the second derivative of differential entropy is expressed as
\begin{eqnarray}
\hspace{-5mm} \frac{d^2}{da^2}h(Y) \hspace{-3mm} & = & \hspace{-3mm} - \int_{\scriptscriptstyle -\infty}^{\scriptscriptstyle \infty} \frac{d}{da}\log f_Y(y;a) \frac{d}{da} f_Y(y;a) dy\nonumber\\
\hspace{-5mm} \hspace{-3mm} && \hspace{8mm} - \int_{\scriptscriptstyle -\infty}^{\scriptscriptstyle \infty} \log f_Y(y;a) \frac{d^2}{da^2} f_Y(y;a) dy,\nonumber\\
\label{appd_eq22_2}
\hspace{-5mm} \hspace{-3mm} & = &\hspace{-3mm}  - J_a(Y) - \int_{\scriptscriptstyle -\infty}^{\scriptscriptstyle \infty} \log f_Y(y;a) \frac{d^2}{da^2} f_Y(y;a) dy.
\end{eqnarray}
The last equality is due to the definition of Fisher information with respect to parameter $a$ in (\ref{stein_eq3_1}).

From equation (\ref{appd_eq17_1}), we derive  an  additional relationship between the second order differentials with respect to $y$ and $a$:
\begin{eqnarray}
&	& \frac{d^2}{da^2} f_{Y|X}(y|x;a)\nonumber\\
& = & \frac{d}{da} \left(-\frac{1}{2a}\frac{d}{dy}\left((y-x)f_{Y|X}(y|x;a)\right)\right)\nonumber\\
& = & \frac{1}{2a^2} \frac{d}{dy}\left((y-x)f_{Y|X}(y|x;a)\right)\nonumber\\
&	& +\frac{1}{4a^2}\frac{d}{dy}\left((y-x)\left(\frac{d}{dy}\left((y-x)f_{Y|X}(y|x;a)\right)\right)\right). \nonumber
\end{eqnarray}
Since
\begin{eqnarray}
\hspace{-3mm} &	& \hspace{-3mm} \frac{d^2}{dy^2} \left((y-x)^2f_{\scriptscriptstyle Y|X}(y|x;a)\right) \nonumber\\
\hspace{-3mm} & =  & \hspace{-3mm} \frac{d^2}{dy^2} \left[(y-x)\left((y-x)f_{\scriptscriptstyle Y|X}(y|x;a)\right)\right]\nonumber\\
\hspace{-3mm} & = & \hspace{-3mm} \frac{d}{dy} \hspace{-1mm}\left(\hspace{-.5mm}(y \hspace{-.7mm} - \hspace{-.7mm} x)f_{\scriptscriptstyle Y\hspace{-.5mm}|\hspace{-.5mm}X} \hspace{-.5mm} (y|x;a)\hspace{-.5mm}\right) \hspace{-.7mm} + \hspace{-.7mm} \frac{d}{dy} \hspace{-.7mm}\left(\hspace{-1mm}(y \hspace{-.7mm} - \hspace{-.7mm} x) \frac{d}{dy} \hspace{-1mm}\left(\hspace{-.5mm}(y \hspace{-.7mm} - \hspace{-.7mm} x) f_{\scriptscriptstyle Y \hspace{-.5mm}|\hspace{-.5mm}X}\hspace{-.5mm}(y|x;a)\hspace{-.5mm} \right) \hspace{-1mm}\right),\nonumber
\end{eqnarray}
we obtain the following relationship:
\begin{eqnarray}\label{appd_eq25_1}
\hspace{-5mm} \frac{d^2}{da^2} f_{Y|X}(y|x;a) \hspace{-1mm} & = &  \hspace{-1mm} \frac{1}{4a^2}\frac{d^2}{dy^2}\left((y-x)^2 f_{Y|X}(y|x;a)\right)\nonumber\\
&& + \frac{1}{4a^2} \frac{d}{dy}\left((y-x)f_{Y|X}(y|x;a)\right).
\end{eqnarray}

Taking the expected value  of both sides of (\ref{appd_eq25_1}),
\begin{eqnarray}\label{appd_eq26_1}
\frac{d^2}{da^2} f_Y(y;a) \hspace{-3mm} & = & \hspace{-3mm}  \frac{1}{4a^2} \Bigg\{ \frac{d^2}{dy^2}\mathbb{E}_{X}\left[(y-X)^2f_{Y|X}(y|X;a)\right] \nonumber\\
&&+ \frac{d}{dy}\mathbb{E}_X \left[(y-X) f_{Y|X} (y|X;a) \right]\Bigg\}.
\end{eqnarray}

After substituting $(d^2 f_Y(y;a) /da^2)$, from equation (\ref{appd_eq26_1}), into  equation (\ref{appd_eq22_2}), the second term of (\ref{appd_eq22_2}) takes the expression:
\begin{eqnarray}
\hspace{-3mm} &	&\hspace{-3mm}  -\int_{\scriptscriptstyle -\infty}^{\scriptscriptstyle \infty} \log f_{\scriptscriptstyle Y} (y;a) \frac{d^2}{da^2}f_{\scriptscriptstyle Y}(y;a) dy \nonumber\\
\hspace{-3mm} & = & \hspace{-3mm} \underbrace{-\frac{1}{4a^2}\int_{\scriptscriptstyle -\infty}^{\scriptscriptstyle \infty}  \hspace{-.7mm} \log f_{\scriptscriptstyle Y}(y;a) \frac{d^2}{dy^2} \mathbb{E}_{\scriptscriptstyle X} \hspace{-.7mm} \left[(y-X)^2 f_{\scriptscriptstyle Y|X}(y|X;a)\right]dy}_{(D)}\nonumber\\
\hspace{-3mm} &	& \hspace{-3mm} \underbrace{- \frac{1}{4a^2} \int_{\scriptscriptstyle -\infty}^{\scriptscriptstyle \infty} \hspace{-.7mm} \log f_{\scriptscriptstyle Y}(y;a) \frac{d}{dy} \mathbb{E}_{\scriptscriptstyle X} \hspace{-.7mm} \left[(y-X)f_{\scriptscriptstyle Y|X}(y|X;a)\right] dy}_{(E)}.\nonumber
\end{eqnarray}
Term $(E)$ is exactly of the same form as (\ref{appd_eq19_2}), and therefore,
\begin{eqnarray}\label{appd_eq28_1}
\hspace{-6mm} && \hspace{-3mm} - \frac{1}{4a^2} \int_{\scriptscriptstyle -\infty}^{\scriptscriptstyle \infty} \log f_{\scriptscriptstyle Y}(y;a) \frac{d}{dy} \mathbb{E}_{\scriptscriptstyle X} \left[(y-X)f_{\scriptscriptstyle Y|X}(y|X;a)\right] dy\nonumber\\
\hspace{-6mm} & = & \hspace{-3mm} -\frac{1}{4a^2} \mathbb{E}_{\scriptscriptstyle Y} \left[ \frac{d}{dY} \mathbb{E}_{\scriptscriptstyle X|Y} \left[Y-X|Y\right]\right]\nonumber\\
\hspace{-6mm} & = & \hspace{-3mm} -\frac{1}{2a}\frac{d}{da}h(Y).
\end{eqnarray}

Term $(D)$ is further simplified by the following procedures:
\begin{eqnarray}
\label{appd_eq29_2}
\hspace{-6mm} && \hspace{-3mm} -\frac{1}{4a^2} \int_{\scriptscriptstyle -\infty}^{\scriptscriptstyle \infty}  \hspace{-1mm} \log f_{\scriptscriptstyle Y}(y;a) \frac{d^2}{dy^2} \mathbb{E}_{\scriptscriptstyle X} \hspace{-1mm}\left[(y\hspace{-.7mm} - \hspace{-.7mm}X)^{\scriptscriptstyle 2} f_{\scriptscriptstyle Y|X}(y|X;a)\right]dy\nonumber\\
\hspace{-6mm} & = &  \hspace{-3mm} -\frac{1}{4a^2}\log f_{\scriptscriptstyle Y}(y;a) \frac{d}{dy} \mathbb{E}_{\scriptscriptstyle X}\hspace{-1mm} \left[(y-X)^{\scriptscriptstyle 2} f_{\scriptscriptstyle Y|X}(y|X;a)\right]\Bigg|_{\scriptscriptstyle y=-\infty}^{\scriptscriptstyle \infty}\nonumber\\
\hspace{-6mm} &	&  \hspace{-3mm} +\frac{1}{4a^2}  \hspace{-1mm}  \int_{\scriptscriptstyle -\infty}^{\scriptscriptstyle \infty} \hspace{-1mm} \frac{d}{dy}\hspace{-.5mm}\log \hspace{-.5mm}f_{\scriptscriptstyle Y}(y;a) \frac{d}{dy} \mathbb{E}_{\scriptscriptstyle X} \hspace{-1mm} \left[(y\hspace{-.7mm}-\hspace{-.7mm}X)^{\scriptscriptstyle 2} f_{\scriptscriptstyle Y\hspace{-.5mm}|\hspace{-.5mm}X}\hspace{-.5mm}(y|X;a)\right]dy.\nonumber\\
\hspace{-6mm}
\end{eqnarray}
The first part of (\ref{appd_eq29_2}) is expressed as
\begin{eqnarray}
\hspace{-7mm} && \hspace{-3mm} -\frac{1}{4a^2}\log f_{\scriptscriptstyle Y}(y;a) \frac{d}{dy} \mathbb{E}_{\scriptscriptstyle X}\hspace{-1mm} \left[(y\hspace{-.7mm} -\hspace{-.7mm} X)^2 f_{\scriptscriptstyle Y|X}(y|X;a)\right]\hspace{-.7mm} \Bigg|_{\scriptscriptstyle y=-\infty}^{\scriptscriptstyle \infty}\nonumber\\
\hspace{-7mm} & = & \hspace{-3mm} -\frac{1}{4a^2}\log f_{\scriptscriptstyle Y}(y;a)\Bigg(\mathbb{E}_{\scriptscriptstyle X} \hspace{-1mm} \left[2(y-X) f_{\scriptscriptstyle Y|X}(y|X;a)\right]\nonumber\\
\hspace{-7mm} &	& \hspace{6mm}+ \mathbb{E}_{\scriptscriptstyle X} \hspace{-1mm} \left[(y^2-2Xy+X^2) \frac{d}{dy} f_{\scriptscriptstyle Y|X}(y|X;a) \right]\Bigg)\Bigg|_{\scriptscriptstyle y=-\infty}^{\scriptscriptstyle \infty}\nonumber\\
\hspace{-7mm} & = & \hspace{-3mm} -\frac{1}{4a^2}\log f_{\scriptscriptstyle Y}(y;a) \Bigg( \hspace{-.7mm} 2y f_{\scriptscriptstyle Y}(y;a) \hspace{-.7mm} -2\mathbb{E}_{\scriptscriptstyle X} \hspace{-1mm} \left[X f_{\scriptscriptstyle Y|X}(y|X;a)\right] \nonumber\\
\hspace{-7mm} &	&\hspace{6mm} + y^2 \frac{d}{dy} f_{\scriptscriptstyle Y}(y;a)  -2y \mathbb{E}_{\scriptscriptstyle X}\hspace{-1mm}  \left[X\frac{d}{dy} f_{\scriptscriptstyle Y|X}(y|X;a) \right] \nonumber\\
\hspace{-7mm} &	&\hspace{6mm} +\mathbb{E}_{\scriptscriptstyle X} \hspace{-1mm} \left[ X^2 \frac{d}{dy} f_{\scriptscriptstyle Y|X}(y|X;a) \right] \Bigg)\Bigg|_{\scriptscriptstyle y=-\infty}^{\scriptscriptstyle \infty}\nonumber\\
\hspace{-7mm} & = & \hspace{-3mm} -\frac{1}{2a^2} \underbrace{\sqrt{f_{\scriptscriptstyle Y}(y;a)} \log \sqrt{f_{\scriptscriptstyle Y}(y;a)}}_{(b_1)} \nonumber\\
\hspace{-7mm} &	& \hspace{3mm}\times \Bigg( 2\underbrace{y \sqrt{f_{\scriptscriptstyle Y}(y;a)}}_{(b_2)} +\mathbb{E}_{\scriptscriptstyle X}\hspace{-1mm}  \Bigg[ \underbrace{X^2 \frac{\frac{d}{dy} f_{\scriptscriptstyle Y|X}(y|X;a)}{ \sqrt{f_{\scriptscriptstyle Y}(y;a)}}}_{(b_3)} \Bigg]  \Bigg)\nonumber\\
\hspace{-7mm} &	& \hspace{-3mm} -\frac{1}{a^2} \underbrace{\sqrt[4]{f_{\scriptscriptstyle Y}(y;a)} \log \sqrt[4]{f_{\scriptscriptstyle Y}(y;a)}}_{(b_1)} \nonumber\\
\hspace{-7mm} &	& \hspace{3mm}\times \Bigg(  \underbrace{y^2  \sqrt[4]{f_{\scriptscriptstyle Y}(y;a)}}_{(b_2)} \underbrace{\mathbb{E}_{\scriptscriptstyle X} \hspace{-1mm} \left[\frac{\frac{d}{dy} f_{\scriptscriptstyle Y|X}(y|X;a)}{ \sqrt{f_{\scriptscriptstyle Y}(y;a)}}\right]}_{(b_3)}  \nonumber\\
\hspace{-7mm} &	& \hspace{6mm} -2\underbrace{y  \sqrt[4]{f_{\scriptscriptstyle Y}(y;a)}}_{(b_2)} \mathbb{E}_{\scriptscriptstyle X} \hspace{-1mm} \Bigg[ \underbrace{X\frac{\frac{d}{dy} f_{\scriptscriptstyle Y|X}(y|X;a)}{ \sqrt{f_{\scriptscriptstyle Y}(y;a)}}}_{(b_3)} \Bigg] \Bigg) \nonumber\\
\hspace{-7mm} &&\hspace{-3mm} + \frac{1}{a^2} \hspace{-.7mm} \underbrace{\sqrt{\hspace{-.7mm} f_{\scriptscriptstyle Y}(y;a)} \log \hspace{-.7mm} \sqrt{\hspace{-.7mm} f_{\scriptscriptstyle Y}(y;a)}}_{(b_1)} \underbrace{\frac{\mathbb{E}_{\scriptscriptstyle X} \hspace{-1mm} \left[X f_{\scriptscriptstyle Y|X}(y|X;a)\right]}{ \sqrt{f_{\scriptscriptstyle Y}(y;a)}} }_{(b_4)} \hspace{-.7mm} \Bigg|_{\scriptscriptstyle y=-\infty}^{\scriptscriptstyle \infty}.\nonumber
\end{eqnarray}
Since $x\log x$ becomes zero as $x$ approaches zero and $f_Y(y;a)$ converges to zero as $y$ goes to $\pm \infty$, factor $(b_1)$ is zero as $y \rightarrow \pm \infty$. Due to assumptions (\ref{bruijn_eq12_1}c) and (\ref{bruijn_eq12_1}d), term $(b_2)$ becomes zero as $y \rightarrow \pm \infty$ and term $(b_3)$ is bounded. Also, factor $(b_4)$ must be bounded due to assumption (\ref{bruijn_eq12_1}e). Therefore, as $y \rightarrow \pm\infty$, the first part of equation (\ref{appd_eq29_2}) vanishes.

Then, equation (\ref{appd_eq29_2}) is further processed using integration by parts as follows:
\begin{eqnarray}
\hspace{-7mm}&   & \hspace{-3mm}\frac{1}{4a^2}  \int_{\scriptscriptstyle -\infty}^{\scriptscriptstyle \infty} \hspace{-1mm}\frac{d}{dy}\log \hspace{-.7mm}f_{\scriptscriptstyle Y}\hspace{-.7mm}(y;a) \frac{d}{dy} \mathbb{E}_{\scriptscriptstyle X} \hspace{-1mm}\left[(y \hspace{-.7mm} - \hspace{-.7mm} X)^{\scriptscriptstyle 2} f_{\scriptscriptstyle Y|X}\hspace{-.7mm}(y|X;a)\right]dy\nonumber\\
\label{appd_eq29_3}
\hspace{-7mm}& = & \hspace{-3mm}\frac{1}{4a^2} \frac{d}{dy}\log \hspace{-.7mm}f_{\scriptscriptstyle Y}\hspace{-.7mm}(y;a) \mathbb{E}_{\scriptscriptstyle X}\hspace{-1mm} \left[(y \hspace{-.7mm} - \hspace{-.7mm}X)^{\scriptscriptstyle 2} f_{\scriptscriptstyle Y|X}\hspace{-.7mm}(y|X;a)\right]\Big|_{\scriptscriptstyle y=-\infty}^{\scriptscriptstyle \infty}\nonumber\\
\hspace{-7mm}&	& \hspace{-3mm}- \frac{1}{4a^2} \int_{\scriptscriptstyle -\infty}^{\scriptscriptstyle \infty} \hspace{-1mm}\frac{d^2}{dy^2}\log \hspace{-.7mm}f_{\scriptscriptstyle Y}\hspace{-.7mm}(y;a) \mathbb{E}_{\scriptscriptstyle X}\hspace{-1mm} \left[(y\hspace{-.7mm}-\hspace{-.7mm}X)^{\scriptscriptstyle 2} f_{\scriptscriptstyle Y|X}\hspace{-.7mm}(y|X;a)\right]dy.\nonumber\\
\hspace{-7mm}&	& \hspace{-3mm}
\end{eqnarray}

Again, the first part of equation (\ref{appd_eq29_3}) is re-written as
\begin{eqnarray}\label{appd_eq29_4}
\hspace{-7mm} &   & \hspace{-3mm} \frac{1}{4a^2} \frac{d}{dy}\log f_{\scriptscriptstyle Y}(y;a) \mathbb{E}_{\scriptscriptstyle X} \hspace{-1.2mm} \left[(y \hspace{-.7mm} - \hspace{-.7mm} X)^{\scriptscriptstyle 2} f_{\scriptscriptstyle Y|X}(y|X;a)\right]\Big|_{\scriptscriptstyle y=-\infty}^{\scriptscriptstyle \infty}\nonumber\\
\hspace{-7mm} & = & \hspace{-3mm} \frac{1}{4a^2} \mathbb{E}_{\scriptscriptstyle X} \hspace{-1.2mm} \left[\hspace{-.8mm} \frac{\frac{d}{dy}f_{\scriptscriptstyle Y|X}(y|X;a)}{\sqrt{f_{\scriptscriptstyle Y}(y;a)}}\hspace{-.8mm} \right] \hspace{-1.2mm} \mathbb{E}_{\scriptscriptstyle X} \hspace{-1.2mm} \left[\hspace{-.8mm} (y \hspace{-.7mm} - \hspace{-.7mm} X)^{\scriptscriptstyle 2}  \hspace{-.8mm} \frac{f_{\scriptscriptstyle Y|X}(y|X;a)}{\sqrt{f_{\scriptscriptstyle Y}(y;a)}}\hspace{-.8mm} \right]\hspace{-1.2mm} \Bigg|_{\scriptscriptstyle y=-\infty}^{\scriptscriptstyle \infty}\nonumber\\
\hspace{-7mm} & = & \hspace{-3mm} \frac{1}{4a^2} \underbrace{\mathbb{E}_{\scriptscriptstyle X} \hspace{-1.2mm} \left[\frac{\frac{d}{dy}f_{\scriptscriptstyle Y|X}(y|X;a)}{\sqrt{f_{\scriptscriptstyle Y}(y;a)}}\right]}_{(c_1)} \underbrace{y^2 \sqrt{f_{\scriptscriptstyle Y}(y;a)}}_{(c_2)}\nonumber\\
\hspace{-7mm} & & -2 \frac{1}{4a^2} \underbrace{\mathbb{E}_{\scriptscriptstyle X} \hspace{-1.2mm} \left[\frac{\frac{d}{dy}f_{\scriptscriptstyle Y|X}(y|X;a)}{\sqrt{f_{\scriptscriptstyle Y}(y;a)}}\right]}_{(c_1)} \underbrace{y  \sqrt[4]{f_{\scriptscriptstyle Y}(y;a)}}_{(c_2)}\nonumber\\
\hspace{-7mm} &   & \hspace{9mm} \times \underbrace{\mathbb{E}_{\scriptscriptstyle X} \hspace{-1.2mm} \left[X \frac{f_{\scriptscriptstyle Y|X}(y|X;a)}{(f_{\scriptscriptstyle Y}(y;a))^{3/4}}\right]}_{(c_3)}\nonumber\\
\hspace{-7mm} &   &  + \frac{1}{4a^2} \underbrace{\mathbb{E}_{\scriptscriptstyle X} \hspace{-1.2mm} \left[\frac{\frac{d}{dy}f_{\scriptscriptstyle Y|X}(y|X;a)}{\sqrt{f_{\scriptscriptstyle Y}(y;a)}}\right]}_{(c_1)} \underbrace{\sqrt[4]{f_{\scriptscriptstyle Y}(y;a)}}_{(c_2)} \nonumber\\
\hspace{-7mm} &   & \hspace{9mm} \times \underbrace{\mathbb{E}_{\scriptscriptstyle X} \hspace{-1.2mm} \left[X^2  \frac{f_{\scriptscriptstyle Y|X}(y|X;a)}{(f_{\scriptscriptstyle Y}(y;a))^{3/4}}\right]}_{(c_3)}\Bigg|_{\scriptscriptstyle y=-\infty}^{\scriptscriptstyle \infty}.
\end{eqnarray}
Factors $(c_1)$ and $(c_3)$ are bounded due to assumptions (\ref{bruijn_eq12_1}c) and (\ref{bruijn_eq12_1}e), and, by assumption (\ref{bruijn_eq12_1}d),  factor $(c_2)$ approaches zero as $y \rightarrow \pm \infty$. Then,  equation (\ref{appd_eq29_3}) is expressed as
\begin{eqnarray}\label{appd_eq29_5}
\hspace{-7mm} &   & \hspace{-3mm} \frac{1}{4a^2} \hspace{-1.2mm} \int_{\scriptscriptstyle -\infty}^{\scriptscriptstyle \infty} \hspace{-1mm}\frac{d}{dy}\log f_{\scriptscriptstyle Y}(y;a) \frac{d}{dy} \mathbb{E}_{\scriptscriptstyle X} \hspace{-1.2mm}\left[(y \hspace{-.8mm}-\hspace{-.8mm}X)^2 f_{\scriptscriptstyle Y|X}(y|X;a)\right]dy\nonumber\\
\hspace{-7mm} & = & \hspace{-3mm} - \frac{1}{4a^2} \hspace{-1.2mm}\int_{\scriptscriptstyle -\infty}^{\scriptscriptstyle \infty} \hspace{-1mm}\frac{d^2}{dy^2}\log f_{\scriptscriptstyle Y}(y;a) \mathbb{E}_{\scriptscriptstyle X} \hspace{-1.2mm}  \left[(y\hspace{-.8mm}-\hspace{-.8mm}X)^2 f_{\scriptscriptstyle Y|X}(y|X;a)\right]dy.\nonumber\\
\hspace{-7mm} & & \hspace{-3mm}
\end{eqnarray}

Using equations (\ref{appd_eq28_1}) and (\ref{appd_eq29_5}), equation (\ref{appd_eq22_2}) is expressed as
\begin{eqnarray}
\hspace{-7mm} & & \hspace{-3mm}\frac{d^2}{da^2}h(Y)\nonumber\\
\hspace{-7mm}  & = &\hspace{-3mm} - J_a(Y) - \int_{-\infty}^{\infty} \log f_Y(y;a) \frac{d^2}{da^2} f_Y(y;a) dy\nonumber\\
\hspace{-7mm} & = &\hspace{-3mm} - J_a(Y) -\frac{1}{2a}\frac{d}{da}h(Y) \nonumber\\
\hspace{-7mm} & & -\frac{1}{4a^2}\mathbb{E}_Y \left[\frac{d}{dY}S_Y(Y)\mathbb{E}_{X|Y} \left[(Y-X)^2|Y \right]\right]\nonumber\\
\hspace{-7mm} & = &\hspace{-3mm} - J_a(Y)-\frac{1}{4a^2}\mathbb{E}_Y \left[ \frac{d}{dY}\mathbb{E}_{X|Y} \left[(Y-X)|Y\right] \right]\nonumber\\
\hspace{-7mm} & & -\frac{1}{4a^2}\mathbb{E}_Y \left[\frac{d}{dY}S_Y(Y) \mathbb{E}_{X|Y} \left[(Y-X)^2|Y \right]\right],\nonumber
\end{eqnarray}
and the proof is completed.
\end{proof}

% Proof of Lemma 1 (BCRLB)
\section{A proof of Lemma \ref{lem1}}\label{AppD}
\begin{proof}\label{pf_lem1} [Lemma \ref{lem1}]

Before we prove this lemma, we first introduce two lemmas which are necessary to prove Lemma \ref{lem1}.

% lemma 3 (De Bruijn extension)
\begin{lem}\label{lem3}
Given  the channel $Y=X+\sqrt{a}W$ in (\ref{int_eq1_1}), the following identity holds:
\begin{eqnarray}\label{appl_eq6_1}
\frac{d}{da} J(Y) & = & - \mathbb{E}_Y \left[\left(\frac{d}{dY}S_Y(Y)\right)^2\right],
\end{eqnarray}
where  $X$ is an arbitrary but fixed random variable with a finite second-order  moment, and $W$ is a Gaussian random variable with zero mean and unit variance.
\begin{proof}
In Theorems \ref{thm5}, \ref{thm4}, we showed the equivalence among De Bruijn, generalized Stein, and heat equation identities for specific conditions. Therefore, using one of the identities, this lemma can be proved. In this proof, Theorem \ref{thm3} (the heat equation identity) will be used with $g(y)=S_Y(y)^2$. Unlike the definition of $g(y)$ in Theorem \ref{thm3}, % here $g(y)$ is defined as $S_Y(y)^2$, and it is dependent on the parameter $a$. Therefore, we use the notation $g(y;a)$ instead of $g(y)$, and we should be careful to adapt the heat equation identity.
$g(y)$ is dependent on the parameter $a$. Therefore, we use the notation $g(y;a)$ instead of $g(y)$.
Since $J(Y)=\mathbb{E}[S_Y(Y)^2]$, the right-hand side of (\ref{appl_eq6_1}) is expressed as
\begin{eqnarray}
\hspace{-5mm} \frac{d}{da} J(Y) \hspace{-3mm} & = & \hspace{-3mm} \frac{d}{da} \mathbb{E}_{\scriptscriptstyle Y} \left[S_{\scriptscriptstyle Y}(Y)^2\right]\nonumber\\
\label{appl_eq7_2}
\hspace{-5mm} \hspace{-3mm} & = & \hspace{-3mm} \int_{\scriptscriptstyle -\infty}^{\scriptscriptstyle \infty} \frac{d}{da} f_{\scriptscriptstyle Y}(y;a) g(y;a) dy + \mathbb{E}_{\scriptscriptstyle Y}\hspace{-1mm}  \left[\frac{d}{da} g(Y;a)\right].
\end{eqnarray}
By the heat equation identity, the first term in equation (\ref{appl_eq7_2}) is expressed as
\begin{eqnarray}
\int_{-\infty}^{\infty} \frac{d}{da} f_Y(y;a) g(y;a) dy & = & \frac{1}{2} \mathbb{E}_Y \left[\frac{d^2}{dY^2}g(Y;a) \right].\nonumber
\end{eqnarray}
Using integration by parts, the second term in equation (\ref{appl_eq7_2}) is expressed as
\begin{eqnarray}
\mathbb{E}_{\scriptscriptstyle Y} \hspace{-1mm} \left[\frac{d}{da} g(Y;a)\right]\hspace{-3mm} & = & \hspace{-3mm} \frac{1}{2} \mathbb{E}_{\scriptscriptstyle Y}\hspace{-1mm}  \left[\frac{d^2}{dY^2}g(Y;a)\right] \hspace{-.5mm}  - \hspace{-.5mm}  \mathbb{E}_{\scriptscriptstyle Y} \hspace{-1mm} \left[\left(\frac{d}{dY}S_{\scriptscriptstyle Y}(Y)\right)^2\right]\nonumber\\
\hspace{-3mm} && \hspace{3mm} + 2\mathbb{E}_{\scriptscriptstyle Y} \hspace{-1mm} \left[S_{\scriptscriptstyle Y}(Y)^2 \frac{d}{dY}S_{\scriptscriptstyle Y}(Y)\right].\nonumber
\end{eqnarray}
Therefore, equation (\ref{appl_eq7_2}) takes the form:
\begin{eqnarray}
&   & \int_{-\infty}^{\infty} \frac{d}{da} f_Y(y;a) g(y;a) dy + \mathbb{E}_Y \left[\frac{d}{da} g(Y;a)\right]\nonumber\\
& = & -\mathbb{E}_Y\left[\left(\frac{d}{dY}S_Y(Y)\right)^2\right] \nonumber\\
&   & + \underbrace{\mathbb{E}_Y \left[\frac{d^2}{dY^2}g(Y;a)\right] + 2\mathbb{E}_Y \left[S_Y(Y)^2 \frac{d}{dY}S_Y(Y)\right]}_{(F)}.\nonumber
\end{eqnarray}
Performing an  integration by parts, the term  $(F)$ is shown to be equal to zero, and the proof is completed.
\begin{rem}
A vector version of this lemma was reported in \cite{Palomar:LinVecGauss}. The reasons why we introduce both this lemma and its proof are not only to present alternative proofs, but also to explain the usefulness of our novel results. For example, Lemma \ref{lem3} was proved based on the heat equation identity, which is a novel approach to prove this lemma. At the same time, this lemma can also be  alternatively proved using Theorem \ref{thm7} or Corollary \ref{cor5}.
\end{rem}
\end{proof}
\end{lem}

% Lemma 2 (Fisher Information Inequality)
\begin{lem} [Fisher Information Inequality] \label{lem2}
Consider the channel $Y=X+\sqrt{a}W$ in (\ref{int_eq1_1}), where the  random variable $X$ is assumed to have an arbitrary distribution but a fixed second-order moment and  $W$ is normally distributed with zero mean and  unit variance. Then, the following inequality is always satisfied:
\begin{eqnarray}
\frac{1}{J(Y)} & \geq & \frac{1}{J(X)} + \frac{1}{J(\sqrt{a}W)},\nonumber
\end{eqnarray}
where the equality holds if and only if $X$ is normally distributed.

\begin{proof}
Using Lemma \ref{lem3} (equivalently, Theorem \ref{thm7} or Corollary \ref{cor5} can be used),
\begin{eqnarray}\label{appl_eq11_1}
-\frac{d}{da}J(Y) & = & \mathbb{E}_Y \left[\left(\frac{d}{dY}S_Y(Y)\right)^2\right]\nonumber\\
& \geq & \mathbb{E}_Y \left[\left(\frac{d}{dY}S_Y(Y)\right)\right]^2\nonumber\\
& = & J(Y)^2.
\end{eqnarray}
Equation (\ref{appl_eq11_1}) is expressed as
\begin{eqnarray}
-\frac{d}{da}J(Y) & \geq & J(Y)^2,\nonumber
\end{eqnarray}
and it is equivalent to
\begin{eqnarray}\label{appl_eq13_1}
&&-\frac{\frac{d}{da}J(Y)}{J(Y)^2}  \geq  1\nonumber\\
&\Longleftrightarrow &\frac{d}{da} \left(\frac{1}{J(Y)}\right)  \geq   1.
\end{eqnarray}
Since inequality (\ref{appl_eq13_1}) is satisfied for any $a$,
\begin{eqnarray}\label{appl_eq14_1}
&&\int_0^a \frac{d}{dt} \left(\frac{1}{J(Y)}\right) dt  \geq \int_0^a 1 dt,\nonumber\\
&\Longleftrightarrow& \frac{1}{J(Y)}-\frac{1}{J(X)} \geq  a,\nonumber\\
&\Longleftrightarrow& \frac{1}{J(Y)} \geq  \frac{1}{J(X)} + \frac{1}{J(\sqrt{a}W)}.
\end{eqnarray}
Since $W$ is normally distributed with unit variance, $a=1/J(\sqrt{a}W)$, and the last equivalence holds. The last equation in (\ref{appl_eq14_1}) denotes the Fisher information inequality, and  the proof is completed.
\begin{rem}
This proof uses neither the convolutional inequality, the data processing inequality, nor the EPI, unlike previous proofs. The proof only relies on De Bruijn's identity, Stein's identity, or the heat equation identity. Namely, Theorem \ref{thm1}, \ref{thm2}, \ref{thm3}, or \ref{thm7} is the only adopted result, and Theorems \ref{thm5}, \ref{thm4} ensure Theorem  \ref{thm1}, \ref{thm2}, \ref{thm3}, or \ref{thm7}  can be equivalently adopted to the proof. Even though Lemma \ref{lem3} was used in this proof, Lemma \ref{lem3} itself was also proved using one of the above identities. Therefore, this proof only uses our results.
\end{rem}
\end{proof}
\end{lem}

Now, based on Lemma \ref{lem2}, the proof of Lemma \ref{lem1} is straightforward. From Lemma \ref{lem2},
\begin{eqnarray}\label{appl_eq15_1}
&& \frac{1}{J(Y)} \geq \frac{1}{J(X)}+\frac{1}{J(\sqrt{a}W)},\nonumber\\
&\Longleftrightarrow& J(Y) \leq \frac{J(X)J(\sqrt{a}W)}{J(X)+J(\sqrt{a}W)}.
\end{eqnarray}
Since $X$ and $W$ are independent, and $W$ is normally distributed,
\begin{eqnarray}
\hspace{-7mm} &   & \hspace{-3mm} \mathbb{E}_{\scriptscriptstyle X} \left[J(Y|X)\right] \nonumber\\
\hspace{-7mm} & = & \hspace{-3mm} \int_{\scriptscriptstyle -\infty}^{\scriptscriptstyle \infty} \hspace{-1mm}f_{\scriptscriptstyle X}(x) \hspace{-1mm} \int_{\scriptscriptstyle -\infty}^{\scriptscriptstyle \infty} \hspace{-1mm} \left(\frac{d}{dx} \log f_{\scriptscriptstyle Y|X}(y|x;a)\right)^2 \hspace{-1mm}f_{\scriptscriptstyle Y|X}(y|x;a) dy dx\nonumber\\
\hspace{-7mm} & = & \hspace{-3mm} \int_{\scriptscriptstyle -\infty}^{\scriptscriptstyle \infty} f_{\scriptscriptstyle X}(x)\int_{\scriptscriptstyle -\infty}^{\scriptscriptstyle \infty}\frac{1}{a^2}\left(y-x\right)^2 f_{\scriptscriptstyle Y|X}(y|x;a) dy dx\nonumber\\
\label{appl_eq15_2}
\hspace{-7mm} & = & \hspace{-3mm} \frac{1}{a}\\
\hspace{-7mm} & = & \hspace{-3mm} J(\sqrt{a}W).\nonumber
\end{eqnarray}
The equality in (\ref{appl_eq15_2}) is due to $\mathbb{E}_{Y|X} [(Y-X)^2 | X=x] = a$.

For a Gaussian random variable $W$,
\begin{eqnarray}\label{appl_eq4_1}
J(Y)  & = & \frac{1}{a}-\frac{1}{a^2} Var(X|Y),
\end{eqnarray}
where $Var(X|Y)$ stands for $\mathbb{E}_{X,Y} [(X-\mathbb{E}_{X|Y}[X|Y])^2]$ (\cite{inf_gauss:guo}, \cite{Rioul:EPI}).

Substituting $Var(X|Y)$ and $\mathbb{E}_X [J(Y|X)]$ for $J(Y)$ and $J(\sqrt{a}W)$, respectively, equation (\ref{appl_eq15_1}) is expressed as
\begin{eqnarray}
&& J(Y) \leq \frac{J(X)J(\sqrt{a}W)}{J(X)+J(\sqrt{a}W)},\nonumber\\
&\Longleftrightarrow& \frac{1}{a}-\frac{1}{a^2} Var(X|Y) \leq \frac{J(X)J(\sqrt{a}W)}{J(X)+J(\sqrt{a}W)},\nonumber\\
&\Longleftrightarrow& Var(X|Y) \geq \frac{1}{J(X)+J(\sqrt{a}W)},\nonumber\\
&\Longleftrightarrow& Var(X|Y) \geq \frac{1}{J(X)+\mathbb{E}_X \left[J(Y|X)\right]}.\nonumber
\end{eqnarray}
Since $Var(X|Y)$ is equal to the minimum mean square error,
\begin{eqnarray}
MSE(\hat{X}) & \geq & MMSE(\hat{X})\nonumber\\
& = & Var(X|Y)\nonumber\\
& \geq & \frac{1}{J(X)+\mathbb{E}_X \left[J(Y|X)\right]},\nonumber
\end{eqnarray}
where $\hat{X}$ denotes a Bayesian estimator, and the obtained inequality is the Bayesian Cram\'{e}r-Rao lower bound (BCRLB).
\end{proof}

% proof of lemma 4 (New lower bound)
\section{A Proof of Lemma \ref{lem4}}\label{AppE}
\begin{proof}\label{pf_lem4} [Lemma \ref{lem4}]

When $a$ is zero, the right-hand side of (\ref{appl_eq17_1}) is zero due to the following relations:
\begin{eqnarray}
N(X|Y) & = & \frac{1}{2\pi e} \exp(2h(X|Y))\nonumber\\
& = & \frac{1}{2\pi e} \exp(2(h(X) + h(Y|X) - h(Y)))\nonumber\\
& = & \frac{1}{2\pi e} \exp(2(h(X) + h(\sqrt{a}W) - h(Y)))\nonumber\\
& = & \frac{N(X)N(\sqrt{a}W)}{N(Y)}\nonumber\\
& = & \frac{a N(X)N(W)}{N(X+\sqrt{a}W)}.\nonumber
\end{eqnarray}
Therefore, when $a$ goes to zero,
\begin{eqnarray}
\lim\limits_{a\rightarrow 0} N(X|Y) & = & \lim\limits_{a\rightarrow 0} \frac{a N(X)N(W)}{N(X+\sqrt{a}W)}\nonumber\\
& = & 0.
\end{eqnarray}
The equality is due to the fact that $\lim\limits_{a \rightarrow 0}N(X+\sqrt{a}W) = N(X)$. Since the left-hand side of (\ref{appl_eq17_1}) is always greater than or equal to zero,  the inequality in (\ref{appl_eq17_1}) is satisfied when $a$ is zero.

Without loss of generality, from now on, we assume that $a>0$.

Since $h(X|Y)=h(X)+h(Y|X)-h(Y)$, by Theorem \ref{thm1} (De Bruijn's identity),
\begin{eqnarray}
&& \frac{d}{da}N(X|Y) \nonumber\\
& = & \frac{d}{da} \left(\frac{1}{2\pi e} \exp \left( 2h(X|Y) \right)\right)\nonumber\\
& = &2N(X|Y) \left\{\frac{d}{da}h(X)+\frac{d}{da}h(Y|X)-\frac{d}{da}h(Y)\right\}\nonumber\\
\label{appl_eq18_2}& = & 2N(X|Y) \left\{\frac{1}{2a}-\frac{1}{2} J(Y)\right\}\\
\label{appl_eq18_3}& = &  N(X|Y)\frac{1}{a^2} Var(X|Y).
\end{eqnarray}
Since $h(X)$ is independent of $a$ and $h(Y|X)=h(\sqrt{a}W)$, $(d/da) h(X)$ is zero, and $(d/da) h(Y|X) =1/2a$. Therefore, the equality in (\ref{appl_eq18_2}) is satisfied. The equality in (\ref{appl_eq18_3}) is due to equation (\ref{appl_eq4_1}).

Based on equation (\ref{appl_eq4_1}),
\begin{eqnarray}\label{appl_eq19_1}
%\frac{d}{da} \left\{a-a^2J(Y)\right\} & = &
\frac{d}{da} Var(X|Y) & = & \frac{d}{da} \left[ a-a^2 J(Y) \right]\nonumber\\
& = & \frac{d}{da} \left[ a-a^2 \left(2\frac{d}{da}h(Y)\right) \right].
\end{eqnarray}
The equality in (\ref{appl_eq19_1}) is due to Theorem \ref{thm1}.

Using Corollary \ref{cor5} and equation (\ref{appl_eq4_1}),  %and assuming equation (\ref{appl_eq19_1}) is satisfied,  the following inequality turns out:
equation (\ref{appl_eq19_1}) is further processed as
\begin{eqnarray}
&   & \frac{d}{da} \left[ a-a^2 \left(2\frac{d}{da}h(Y)\right) \right] \nonumber\\
& = & 1 - 2a\left(2\frac{d}{da}h(Y)\right)  + a^2 \left(-2\frac{d^2}{da^2}h(Y)\right) \nonumber\\
\label{appl_eq19_1_1} & = & 1-2aJ(Y)+a^2 \mathbb{E}_{Y}\left[\left(\frac{d}{dY}S_Y(Y)\right)^2\right]\\
\label{appl_eq19_1_2} & \geq & 1-2aJ(Y) + a^2 J(Y)^2\\
& = &  (1-aJ(Y))^2\nonumber\\
& = & \frac{1}{a^2} Var(X|Y)^2.\nonumber
\end{eqnarray}
The equality in (\ref{appl_eq19_1_1}) is due to Theorem \ref{thm1} and Corollary \ref{cor5}, and the inequality in (\ref{appl_eq19_1_2}) holds because
\begin{eqnarray}
\mathbb{E}_Y \left[ \left( \frac{d}{dY} S_Y(Y) \right)^2 \right] & \geq & \left(\mathbb{E}_Y \left[\frac{d}{dY} S_Y(Y) \right]\right)^2 \nonumber\\
& = & J(Y)^2.\nonumber
\end{eqnarray}
Therefore,
\begin{eqnarray}\label{appl_eq21_1}
\frac{d}{da} Var(X|Y) & \geq & \frac{1}{a^2} Var(X|Y)^2.
\end{eqnarray}
Using equations (\ref{appl_eq18_3}) and (\ref{appl_eq21_1}), we obtain the following inequality:
\begin{eqnarray}
\frac{d}{da} \log N(X|Y) & \leq & \frac{d}{da} \log Var(X|Y).\nonumber
\end{eqnarray}
Since $N(X_G|Y_G)=Var(X_G|Y_G)$, where $X_G$ and $Y_G$ denote Gaussian random variables whose variances are equal to $X$ and $Y$, respectively, the following inequality also holds:
\begin{eqnarray}\label{appl_eq22_2}
&& \frac{d}{da}\left( \log N(X_G|Y_G) - \log N(X|Y) \right)\nonumber\\
& \geq & \frac{d}{da}\left( \log Var(X_G|Y_G)-\log Var(X|Y)\right).
\end{eqnarray}

By performing  an integration, from $0$ to $a$, of both sides in (\ref{appl_eq22_2}), equation (\ref{appl_eq22_2}) is expressed as
\begin{eqnarray}\label{appl_eq23_1}
&& \int_{0}^{a} \frac{d}{dt} \left( \log N_t(X_G|Y_G) - \log N_t(X|Y) \right) d t \nonumber\\
&&\hspace{3mm} \geq \int_{0}^{a} \frac{d}{dt}\left( \log Var_t(X_G|Y_G)-\log Var_t(X|Y)\right) d t \nonumber\\
&\Leftrightarrow&  \log N_t(X_G|Y_G) - \log N_t(X|Y) \Bigg|_{t=0}^{a} \nonumber\\
&& \hspace{3mm} \geq  \log Var_t(X_G|Y_G)-\log Var_t(X|Y)\Bigg|_{t=0}^{a}\nonumber\\
&\Leftrightarrow&  \log N_a(X_G|Y_G) - \log N_a(X|Y) \nonumber\\
&&\hspace{6mm} - \lim\limits_{t\rightarrow 0} \left( \log N_t(X_G|Y_G) - \log N_t(X|Y)\right) \nonumber\\
&&  \hspace{3mm} \geq  \log Var_a(X_G|Y_G)-\log Var_a(X|Y) \nonumber\\
&&\hspace{6mm} - \lim\limits_{t\rightarrow0} \left( \log Var_t(X|Y) - \log Var_t(X_G|Y_G)\right)\\
\label{appl_eq24_1}&\Leftrightarrow& \log N_a(X|Y) \leq  \log Var_a(X|Y),
\end{eqnarray}
where $\Leftrightarrow$ stands for equivalence between before and after the notation, subscript $t$ or $a$ denotes that a function depends on a parameter $t$ or $a$, respectively (the subscript is only used when there may be a confusion between an actual parameter variable and a dummy variable).

The equivalence in (\ref{appl_eq24_1}) is due to the following: $N_a(X_G|Y_G) = Var_a(X_G|Y_G)$,
\begin{eqnarray}\label{appl_eq25_0}
&& \lim\limits_{t\rightarrow 0}\left( \log N_t(X_G|Y_G) - \log N_t(X|Y)\right)\nonumber\\
 & = & \lim\limits_{t\rightarrow 0}\log \frac{N_t(X_G|Y_G)}{N_t(X|Y)} \nonumber\\
 & = & \lim\limits_{t\rightarrow 0}\log \left( \frac{N(X_G)N_t(Y_G|X_G)}{N_t(Y_G)} \Bigg/ \frac{N(X)N_t(Y|X)}{N_t(Y)}\right)\nonumber\\
 & = & \lim\limits_{t\rightarrow 0}\log \left( \frac{N(X_G)N(\sqrt{t}W)}{N(X_G+\sqrt{t}W)} \Bigg/ \frac{N(X)N(\sqrt{t}W)}{N(X+\sqrt{t}W)}\right)\nonumber\\
 & = & \lim\limits_{t\rightarrow 0}\log \left( \frac{N(X_G)N(X+\sqrt{t}W)}{N(X)N(X_G+\sqrt{t}W)} \right)\nonumber\\
 & = & \log \left( \frac{N(X_G)N(X)}{N(X)N(X_G)} \right)\nonumber\\
 & = & 0,
\end{eqnarray}
and
\begin{eqnarray}
\hspace{-7mm} &   & \hspace{-3mm} \lim\limits_{\scriptscriptstyle t\rightarrow 0} \left( \log Var_t(X_{\scriptscriptstyle G}|Y_{\scriptscriptstyle G}) - \log Var_t(X|Y) \right)\nonumber\\
\label{appl_eq25_1}
\hspace{-7mm} & = & \hspace{-3mm} \lim\limits_{\scriptscriptstyle t\rightarrow 0}  \hspace{-.8mm} \left( \hspace{-.5mm} \log  \hspace{-.8mm} \left(t \hspace{-.8mm} - \hspace{-.8mm} t^2 J(X_{\scriptscriptstyle G} \hspace{-.8mm} +  \hspace{-.8mm} \sqrt{t}W)\hspace{-.5mm}\right)  \hspace{-.8mm} - \hspace{-.8mm}\log \hspace{-.8mm} \left(t \hspace{-.8mm} - \hspace{-.8mm} t^2 J(X  \hspace{-.8mm} +  \hspace{-.8mm}\sqrt{t}W)\right)  \hspace{-.8mm}\right) \nonumber\\
\hspace{-7mm} &  & \hspace{-3mm}\\
\hspace{-7mm} & = & \hspace{-3mm} \lim\limits_{\scriptscriptstyle t\rightarrow 0} \hspace{-.8mm} \left( \hspace{-.5mm} \log \hspace{-.8mm} \left(1 \hspace{-.8mm} - \hspace{-.8mm} t J(X_{\scriptscriptstyle G}\hspace{-.8mm} + \hspace{-.8mm} \sqrt{t}W)\hspace{-.5mm} \right) \hspace{-.8mm} - \hspace{-.8mm} \log \hspace{-.8mm} \left(1 \hspace{-.8mm} - \hspace{-.8mm} t J(X \hspace{-.8mm} + \hspace{-.8mm} \sqrt{t}W)\right) \hspace{-.8mm}\right) \nonumber\\
\hspace{-7mm} & = & \hspace{-3mm} \log (1) -\log(1)\nonumber\\
\hspace{-7mm} & = & \hspace{-3mm} 0,\nonumber
\end{eqnarray}
where $W$ is a Gaussian random variable. The equality in (\ref{appl_eq25_1}) is due to equation (\ref{appl_eq4_1}).

Since $\log x$ is an increasing function with respect to $x$, equation (\ref{appl_eq24_1}) is equivalent to
\begin{eqnarray}
N(X|Y) & \leq & Var(X|Y),\nonumber
\end{eqnarray}
and the proof is completed.
\end{proof}

% proof of lemma 4, tighter lower bound
\section{A Proof of Lemma \ref{lem5}}\label{AppF}
\begin{proof}\label{pf_lem5} [Lemma \ref{lem5}]

When $a=0$, both sides of the inequality in (\ref{appl_eq17_2}) are zero, and the inequality in (\ref{appl_eq17_2}) is satisfied. Therefore, without loss of generality, we assume that $a>0$.

\begin{eqnarray}
\frac{d}{da} \log N(X|Y) & = & \frac{1}{N(X|Y)} \frac{d}{da} N(X|Y)\nonumber\\
\label{appd_eq31_1}  & = &\frac{1}{a^2} Var (X|Y)\\
\label{appd_eq31_2} & \geq & \frac{1}{a^2} \frac{1}{J(X)+J(\sqrt{a}W)}\\
& = & \frac{d}{da} \log \left(\frac{1}{J(X)+J(\sqrt{a}W)}\right),\nonumber
\end{eqnarray}
where $W$ is a Gaussian random variable with zero mean and unit variance.
The equality in (\ref{appd_eq31_1}) is due to equation (\ref{appl_eq18_3}), the inequality in (\ref{appd_eq31_2}) is because of BCRLB.

Since $N(X_G|Y_G)$ is equal to $1/(J(X_G)+J(\sqrt{a}W))$, where $X_G$ and $Y_G$ are Gaussian random variables whose variances are equal to $X$ and $Y$, respectively, the following inequality is satisfied:
\begin{eqnarray}\label{appd_eq32_1}
\hspace{-8mm} && \hspace{-3mm} \frac{d}{da}\hspace{-1mm}\left(\log N(X_{\scriptscriptstyle G}|Y_{\scriptscriptstyle G}) \hspace{-.5mm}- \hspace{-.5mm}\log N(X|Y)) \right)\nonumber\\
\hspace{-8mm} & \leq & \hspace{-3mm} \frac{d}{da}\hspace{-1mm}\left(\log \hspace{-1mm}\frac{1}{J(X_{\scriptscriptstyle G})+J(\sqrt{a}W)} \hspace{-.5mm} - \hspace{-.5mm}\log \hspace{-1mm}\frac{1}{J(X)+J(\sqrt{a}W)}\right).
\end{eqnarray}

By integrating  both sides in  (\ref{appd_eq32_1}), equation (\ref{appd_eq32_1}) is equivalent to the following:
\begin{eqnarray}\label{appd_eq33_1}
\hspace{-6mm}&   & \hspace{-3mm} \int_{\scriptscriptstyle 0}^{\scriptscriptstyle a} \frac{d}{dt}\left(\log N_t(X_{\scriptscriptstyle G}|Y_{\scriptscriptstyle G}) - \log N_t(X|Y)) \right) dt \nonumber\\
\hspace{-8mm}&&\hspace{-3mm} \leq \hspace{-1mm}\int_{\scriptscriptstyle 0}^{\scriptscriptstyle a} \hspace{-1mm} \frac{d}{dt}\hspace{-1mm}\left(\hspace{-.5mm}\log \hspace{-.8mm} \frac{1}{J(X_{\scriptscriptstyle  \hspace{-.5mm} G}\hspace{-.5mm}) \hspace{-.8mm} + \hspace{-.8mm} J(\sqrt{t}W\hspace{-.5mm})} \hspace{-.8mm} - \hspace{-.8mm} \log \hspace{-.8mm} \frac{1} {J(X\hspace{-.5mm}) \hspace{-.8mm}  + \hspace{-.8mm} J(\sqrt{t}W\hspace{-.5mm}) } \right)\hspace{-.8mm}  dt\nonumber\\
\hspace{-8mm}& \Leftrightarrow & \hspace{-3mm}\log N_a(X_{\scriptscriptstyle G}|Y_{\scriptscriptstyle G}) - \log N_a(X|Y) \nonumber\\
\hspace{-8mm}&   & - \lim\limits_{t\rightarrow0} \left(\textstyle\log N_t(X_{\scriptscriptstyle G}|Y_{\scriptscriptstyle G}) - \log N_t(X|Y)\right)\nonumber\\
\hspace{-8mm}&& \hspace{-3mm} \leq \log \frac{1}{J(X_{\scriptscriptstyle G})+J(\sqrt{a}W)}- \log \frac{1}{J(X)+J(\sqrt{a}W)} \nonumber\\
\hspace{-8mm}&&  - \lim\limits_{t\rightarrow 0} \left(\log \hspace{-.8mm} \frac{1}{J(X_{\scriptscriptstyle G})\hspace{-.8mm} +\hspace{-.8mm} J(\sqrt{t}W)}- \log \hspace{-.8mm} \frac{1}{J(X)\hspace{-.8mm} +\hspace{-.8mm} J(\sqrt{t}W)}\right)\nonumber\\
\hspace{-8mm}& \Leftrightarrow & \hspace{-3mm} \log N(X|Y) \geq \log \frac{1}{J(X)+J(\sqrt{a}W)},
\end{eqnarray}
where $\Leftrightarrow$ denotes the equivalence between before and after the notation, and subscript $a$ or $t$ of a function means dependency of the function with respect to $a$ or $t$, respectively. The equivalence in (\ref{appd_eq33_1}) is due to the following: $N(X_G|Y_G)$ is equal to $1/(J(X_G)+J(\sqrt{a}W))$, and
\begin{eqnarray}
&   & \lim\limits_{t\rightarrow 0} \left(\log \frac{1}{J(X_G)+J(\sqrt{t}W)}- \log \frac{1}{J(X)+J(\sqrt{t}W)}\right)\nonumber\\
& = & \lim\limits_{t\rightarrow 0} \left(\log \frac{t}{t J(X_G)+J(W)}- \log \frac{t}{t J(X)+J(W)}\right)\nonumber\\
& = & \lim\limits_{t\rightarrow 0} \log \frac{t J(X)+J(W)}{t J(X_G)+J(W)}\nonumber\\
& = & \log \frac{J(W)}{J(W)}\nonumber\\
& = & 0,
\end{eqnarray}
and
\begin{eqnarray}
\lim\limits_{t\rightarrow0} \left(\textstyle\log N_t(X_G|Y_G) - \log N_t(X|Y)\right) = 0\nonumber
\end{eqnarray}
due to equation (\ref{appl_eq25_0}).

Since $\log x$ is a increasing function with respect to $x$, the inequality in (\ref{appd_eq33_1}) is equivalent to
\begin{eqnarray}
\label{appd_eq33_3} N(X|Y) \geq \frac{1}{J(X)+J(\sqrt{a}W)}.
\end{eqnarray}
Since we have already proved that $N(X|Y)$ is a lower bound for any Bayesian estimator in Lemma \ref{lem4}, the inequality in (\ref{appd_eq33_3}) means that the lower bound $N(X|Y)$, the left-hand side of (\ref{appd_eq33_3}), is tighter than BCRLB, the right-hand side of (\ref{appd_eq33_3}).
\end{proof}

% proof of Costa's EPI (Lemma7)
\section{A Proof of Lemma \ref{lem7} (Costa's EPI)}\label{AppG}
\begin{proof}\label{pf_lem7} [Lemma \ref{lem7}]

The proof will be conducted in two different ways.
\begin{enumerate}
\item
Instead of proving equation (\ref{appl_eq27_1}), we are going to prove the inequality in (\ref{appl_eq28_1}).

Using De Bruijn's identity,
\begin{eqnarray}
\frac{d^2}{da^2} N(Y) & = & 2\frac{d}{da}N(Y) \frac{d}{da} h(Y) + 2 N(Y) \frac{d^2}{da^2}h(Y),\nonumber\\
& = & N(Y) \left(J(Y)^2 + 2 \frac{d^2}{da^2} h(Y)\right),\nonumber
\end{eqnarray}
where $Y=X+\sqrt{a}W$.
Since $N(Y) \geq 0$, proving the inequality in (\ref{appl_eq28_1}) is equivalent to proving the following inequality:
\begin{eqnarray}\label{appl_eq30_1}
J(Y)^2 + 2 \frac{d^2}{da^2} h(Y) & \leq & 0.
\end{eqnarray}
Using Theorem \ref{thm7}, the inequality in (\ref{appl_eq30_1}) is expressed as
\begin{eqnarray}\label{appl_eq31_1}
\hspace{-8mm} && \hspace{-3mm} J(Y)^2 -2 J_a(Y)- \frac{1}{2a^2} \mathbb{E}_{\scriptscriptstyle Y} \left[\frac{d}{dY}\mathbb{E}_{\scriptscriptstyle X|Y}\left[Y-X|Y\right]\right]\nonumber\\
\hspace{-8mm} &&  -\frac{1}{2a^2}\mathbb{E}_{\scriptscriptstyle Y} \left[\frac{d}{dY}S_{\scriptscriptstyle Y}(Y) \mathbb{E}_{\scriptscriptstyle X|Y} \left[(Y-X)^2|Y\right]\right] \leq 0.
\end{eqnarray}
By Corollary \ref{cor5}, equation (\ref{appl_eq31_1}) is equivalent to
\begin{eqnarray}
\hspace{-8mm} && \hspace{-3mm} J(Y)^2 \hspace{-.5mm} - \hspace{-.5mm} 2 J_a(Y) \hspace{-.5mm} - \hspace{-.5mm} \frac{1}{2a^2} \mathbb{E}_{\scriptscriptstyle Y} \hspace{-.5mm} \left[\frac{d}{dY}\mathbb{E}_{\scriptscriptstyle X|Y}\hspace{-.5mm}\left[Y \hspace{-.5mm} - \hspace{-.5mm}X|Y\right]\right]\nonumber\\
\hspace{-8mm} &&\hspace{3mm} -\frac{1}{2a^2}\mathbb{E}_{\scriptscriptstyle Y} \left[\frac{d}{dY}S_{\scriptscriptstyle Y}(Y) \mathbb{E}_{\scriptscriptstyle X|Y} \left[(Y-X)^2|Y\right]\right]\nonumber\\
\hspace{-8mm} & = & \hspace{-3mm} J(Y)^2 -\mathbb{E}_{\scriptscriptstyle Y}\left[\left(\frac{d}{dY}S_{\scriptscriptstyle Y}(Y)\right)^2\right]\nonumber\\
\label{appl_eq32_3}
\hspace{-8mm} & = & \hspace{-3mm} -\mathbb{E}_{\scriptscriptstyle Y}\left[\left(J(Y)+\frac{d}{dY}S_{\scriptscriptstyle Y}(Y)\right)^2\right]\\
\hspace{-8mm} & \leq & \hspace{-3mm} 0.\nonumber
\end{eqnarray}

Since $J(Y)=-\mathbb{E}[(d/dY) S_Y(Y)]$ and $\mathbb{E}[S_Y(Y)]=0$, the equality holds in (\ref{appl_eq32_3}). Therefore,
\begin{eqnarray}
\frac{d^2}{da^2}N(Y) & = & -\mathbb{E}_Y\left[\left(J(Y)+\frac{d}{dY}S_Y(Y)\right)^2\right],\nonumber\\
& \leq & 0,\nonumber
\end{eqnarray}
and the proof is completed.
\begin{rem}
This proof mostly follows the proof in \cite{Villani:EPI}. However, by using Theorem \ref{thm7} to prove Costa's EPI, we show that Costa's EPI can be proved by De Bruijn-like identity without using the Fisher information inequality.
\end{rem}

\item In the second proof, the inequality (\ref{appl_eq28_1}) is proved by a slightly different method.

First, define a function $l(a)$ as follows:
\begin{eqnarray}\label{epi_eq0_0}
l(a) & = & - \frac{J(X)}{1+a J(X)} + J(Y) ,
\end{eqnarray}
where $Y=X+\sqrt{a}W$, $X$ is an arbitrary but fixed random variable, $W$ is a Gaussian random variable, and $X$ and $W$ are independent of each other.

For arbitrary non-negative real-valued $a$, $l(a) \leq 0$, and it is proved by the following procedure; using Lemma \ref{lem3} (Theorem \ref{thm7} or Corollary \ref{cor5} can be used instead of Lemma \ref{lem3}),
\begin{eqnarray}\label{epi_eq0_1}
-\frac{d}{da}J(Y) & = & \mathbb{E}_Y \left[\left(\frac{d}{dY}S_Y(Y)\right)^2\right]\nonumber\\
& \geq & \mathbb{E}_Y \left[\left(\frac{d}{dY}S_Y(Y)\right)\right]^2\nonumber\\
& = & J(Y)^2.
\end{eqnarray}
Equation (\ref{epi_eq0_1}) is equivalent to the following inequalities:
\begin{eqnarray}\label{epi_eq0_2}
&&-\frac{\frac{d}{da}J(Y)}{J(Y)^2}  \geq  1\nonumber\\
&\Longleftrightarrow &\frac{d}{da} \left(\frac{1}{J(Y)}\right)  \geq   1.
\end{eqnarray}
Since inequality (\ref{epi_eq0_2}) is satisfied for arbitrary non-negative real-valued $a$,
\begin{eqnarray}\label{epi_eq0_3}
&&\int_0^a \frac{d}{dt} \left(\frac{1}{J(Y)}\right) dt  \geq \int_0^a 1 dt\nonumber\\
&\Longleftrightarrow& \frac{1}{J(Y)}-\frac{1}{J(X)} \geq  a\nonumber\\
&\Longleftrightarrow& J(Y) \leq  \frac{J(X)}{1+ a J(X)},
\end{eqnarray}
and therefore, equation (\ref{epi_eq0_0}) is always non-positive.

Since $J(Y)$ converges to $J(X)$ as $a$ approaches zero, $l(0)=0$, and the following inequality holds for an arbitrary but fixed random variable $X$ and arbitrary small non-negative real-valued $\epsilon$:
\begin{eqnarray}
\label{epi_eq2_1}
\hspace{-6mm} l(\epsilon) - l(0) & = &  - \frac{J(X)}{1+\epsilon J(X)} + J(X+\sqrt{\epsilon}W)\\
\label{epi_eq2_2}
\hspace{-6mm} & \leq & 0.
\end{eqnarray}
Therefore,
\begin{eqnarray}\label{epi_eq3_1}
\frac{d}{d \epsilon} l(\epsilon) \Big|_{\epsilon=0} & \leq & 0,
\end{eqnarray}
for an arbitrary but fixed random variable $X$.

Since the inequality in (\ref{epi_eq3_1}) holds for an arbitrary random variable $X$, we define $X$ as $\tilde{X}+\sqrt{a}\tilde{W}$, where $\tilde{X}$ is an arbitrary but fixed random variable, $\tilde{W}$ is a Gaussian random variable whose variance is identical to the variance of $W$, and $\tilde{X}$, $\tilde{W}$, and $W$ are independent of one another. Then, the inequality in (\ref{epi_eq3_1}) is equivalent to the following inequalities:
\begin{eqnarray}
\hspace{-7mm}0 & \geq & \left( \frac{J(\tilde{X} + \sqrt{a} \tilde{W} )}{1+\epsilon J(\tilde{X} + \sqrt{a} \tilde{W} )} \right)^2\Bigg|_{\epsilon=0} \nonumber\\
&&\hspace{10mm} + \frac{d}{d \epsilon} J(\tilde{X} + \sqrt{a} \tilde{W} + \sqrt{\epsilon}W)  \Bigg|_{\epsilon=0} \nonumber\\
\label{epi_eq4_1}
\hspace{-7mm} \Leftrightarrow 0 & \geq &  \left( \frac{J(\tilde{X} + \sqrt{a} \tilde{W})}{1+\epsilon J(\tilde{X} + \sqrt{a} \tilde{W})} \right)^2 \Bigg|_{\epsilon=0}\nonumber\\
\hspace{-7mm} &&\hspace{10mm} + \frac{d}{d \epsilon} J(\tilde{X} + \sqrt{a + \epsilon} \tilde{W}) \Bigg|_{\epsilon=0}
\end{eqnarray}
\begin{eqnarray}
\label{epi_eq4_2}
\hspace{-7mm} \Leftrightarrow 0 & \geq &\left( \frac{J(\tilde{X} + \sqrt{a} \tilde{W})}{1+\epsilon J(\tilde{X} + \sqrt{a} \tilde{W})} \right)^2 \Bigg|_{\epsilon=0}\nonumber\\
\hspace{-7mm} && \hspace{10mm} + \frac{d}{d a} J(\tilde{X} + \sqrt{a + \epsilon} \tilde{W}) \Bigg|_{\epsilon=0}\\
\label{epi_eq4_3}
\hspace{-7mm} \Leftrightarrow 0  & \geq & J(\tilde{X} + \sqrt{a} \tilde{W})^2 + \frac{d}{d a} J(\tilde{X} + \sqrt{a } \tilde{W}),
\end{eqnarray}
where $\Leftrightarrow$ denotes the equivalence between before and after the notation. The equivalence in (\ref{epi_eq4_1}) is due to the fact that $J(\tilde{X} + \sqrt{a} \tilde{W} + \sqrt{\epsilon}W) = J(\tilde{X} + \sqrt{a + \epsilon} \tilde{W})$ for independent Gaussian random variables $W$ and $\tilde{W}$ whose variances are identical to each other. The inequality in (\ref{epi_eq4_2}) holds due to the following procedure: first, the Fisher information $J(\tilde{X} + \sqrt{a + \epsilon} \tilde{W})$ is expressed as
\begin{eqnarray}\label{epi_eq5_1}
\hspace{-7mm} && \hspace{-3mm} J(\tilde{X} + \sqrt{a + \epsilon} \tilde{W})\nonumber\\
\hspace{-7mm} & = & \hspace{-4mm} \int_{\scriptscriptstyle -\infty}^{\scriptscriptstyle \infty} \hspace{-1mm}\frac{d}{dy} f_{\scriptscriptstyle Y}(y;a,\epsilon) \frac{d}{dy} \log f_{\scriptscriptstyle Y}(y;a,\epsilon) dy\nonumber\\
\hspace{-7mm} & = & \hspace{-4mm} \int_{\scriptscriptstyle -\infty}^{\scriptscriptstyle \infty}\hspace{-1mm} \frac{d}{dy} \mathbb{E}_{\scriptscriptstyle \tilde{X}} \hspace{-1.5mm} \left[\hspace{-.5mm}f_{\scriptscriptstyle Y \hspace{-.5mm}|\hspace{-.5mm}\tilde{X}}\hspace{-.5mm}(y|\tilde{X};\hspace{-.5mm}a,\hspace{-.5mm}\epsilon)\hspace{-.5mm}\right] \hspace{-1mm}\frac{d}{dy} \hspace{-.8mm} \log \hspace{-.5mm}\mathbb{E}_{\scriptscriptstyle \tilde{X}} \hspace{-1.5mm} \left[\hspace{-.5mm}f_{\scriptscriptstyle Y\hspace{-.5mm}|\hspace{-.5mm}\tilde{X}}\hspace{-.5mm}(y|\tilde{X};\hspace{-.5mm}a,\hspace{-.5mm}\epsilon)\hspace{-.5mm}\right] \hspace{-1mm} dy\nonumber\\
\hspace{-7mm} & = & \hspace{-4mm} \int_{\scriptscriptstyle -\infty}^{\scriptscriptstyle \infty} \hspace{-1mm} \frac{d}{dy} \mathbb{E}_{\scriptscriptstyle \tilde{X}} \hspace{-1.5mm} \left[\frac{1}{\sqrt{2\pi (a \hspace{-.8mm} + \hspace{-.8mm} \epsilon)}} \exp \hspace{-.8mm} \left( - \frac{1}{2(a \hspace{-.8mm} + \hspace{-.8mm} \epsilon)} (y \hspace{-.8mm} - \hspace{-.8mm} \tilde{X})^2 \hspace{-1mm}\right) \hspace{-1mm}\right] \nonumber\\
\hspace{-7mm} && \hspace{-4mm}  \times \frac{d}{dy} \hspace{-.8mm}\log \hspace{-.5mm}\mathbb{E}_{\scriptscriptstyle \tilde{X}} \hspace{-1.5mm} \left[\hspace{-.8mm}\frac{1}{\sqrt{2\pi (a \hspace{-.8mm} + \hspace{-.8mm} \epsilon)}} \exp \hspace{-.8mm}\left( \hspace{-1mm}- \frac{1}{2(a \hspace{-.8mm} + \hspace{-.8mm} \epsilon)} (y\hspace{-.8mm}-\hspace{-.8mm}\tilde{X})^2 \hspace{-1mm}\right) \hspace{-1mm}\right] \hspace{-1mm}dy,\nonumber\\
\hspace{-7mm} && \hspace{-8mm}
\end{eqnarray}
where $Y=\tilde{X} + \sqrt{a + \epsilon} \tilde{W}$. Since $f_{Y|\tilde{X}}(y|\tilde{x};a,\epsilon)$ is a Gaussian density function with mean $\tilde{x}$ and variance $a+\epsilon$, the equality in (\ref{epi_eq5_1}) holds. In equation (\ref{epi_eq5_1}), $a$ and $\epsilon$ are symmetrically included in the equation, and therefore,
\begin{eqnarray}
\frac{d}{d \epsilon} J(\tilde{X} + \sqrt{a + \epsilon} \tilde{W}) = \frac{d}{d a} J(\tilde{X} + \sqrt{a + \epsilon} \tilde{W}).\nonumber
\end{eqnarray}

Since random variable $\tilde{X}$ is arbitrary and $a$ is an arbitrary non-negative real-valued number in equation (\ref{epi_eq4_3}), the proof is completed.
%\begin{rem}
%This proof is mainly based on Lemma \ref{lem2} (FII), and Lemma \ref{lem2} is proved mainly using  Theorem \ref{thm7} (or, equivalently, Theorem  \ref{thm1} or \ref{thm3}). Therefore, this proof is basically based on Theorem \ref{thm1}, \ref{thm3}, or \ref{thm7}.
%\end{rem}
\end{enumerate}
\end{proof}

\section{Derivation of Equation (\ref{appd_eq17_1})}\label{AppH}

Given the channel model (\ref{int_eq1_1}), random variables $X$ and $W$ are independent of each other, $a$ is a deterministic parameter, and random variable $Y$ is the summation of $X$ and $\sqrt{a}W$. Therefore, between the two probability density functions $f_{Y|X}(y|x;a)$ and $f_W(w)$, there exists a relationship that can be established as follows.
\begin{eqnarray}
f_{Y|X}(y|x;a) & = & \frac{1}{\sqrt{a}} f_W (w)\Bigg|_{w=\frac{y-x}{\sqrt{a}}}\nonumber\\
& = & \frac{1}{\sqrt{a}} f_W \left(\frac{y-x}{\sqrt{a}}\right).\nonumber
\end{eqnarray}
Therefore,
\begin{eqnarray}
\frac{d}{dy} f_{Y|X}(y|x;a) & = &  \frac{1}{\sqrt{a}} \left( \frac{d}{dy} f_W \left(\frac{y-x}{\sqrt{a}}\right) \right)\nonumber\\
& = & \frac{1}{\sqrt{a}} \left(\frac{1}{\sqrt{a}} \frac{d}{dw} f_W \left(w\right) \right)\Bigg|_{w=\frac{y-x}{\sqrt{a}}}, \nonumber
\end{eqnarray}
and
\begin{eqnarray}
\hspace{-6mm} && \frac{d}{da} f_{Y|X}(y|x;a) \nonumber\\
\hspace{-6mm} & = & \frac{d}{da} \left(\frac{1}{\sqrt{a}} f_W \left(\frac{y-x}{\sqrt{a}}\right) \right)\nonumber\\
\hspace{-6mm} & = & -\frac{1}{2a\sqrt{a}} f_W \left(\frac{y-x}{\sqrt{a}}\right) + \frac{1}{\sqrt{a}} \frac{d}{da}  f_W \left(\frac{y-x}{\sqrt{a}}\right)\nonumber\\
\label{appl_eq42_1}
\hspace{-6mm} & = & -\frac{1}{2a\sqrt{a}} f_W \left(\frac{y-x}{\sqrt{a}}\right)\nonumber\\
\hspace{-6mm}  && + \frac{1}{\sqrt{a}} \left(-\frac{1}{2a\sqrt{a}} (y-x) \frac{d}{dw}  f_W \left(w\right) \Bigg|_{w=\frac{y-x}{\sqrt{a}}}\right).
\end{eqnarray}
Equation (\ref{appl_eq42_1}) is further processed as
\begin{eqnarray}
\hspace{-3mm} &   & \hspace{-4mm} -\frac{1}{2a\sqrt{a}} f_{\scriptscriptstyle W}  \hspace{-1.5mm}\left(\frac{y \hspace{-.7mm}-\hspace{-.7mm} x}{\sqrt{a}}\right) \hspace{-1mm} + \hspace{-1mm}\frac{1}{\sqrt{a}} \hspace{-1mm} \left(\hspace{-.8mm}-\frac{1}{2a\sqrt{a}} (y \hspace{-.7mm} - \hspace{-.7mm}x) \frac{d}{dw}  f_{\scriptscriptstyle W}\hspace{-1.5mm}\left(w\right) \hspace{-1mm} \Bigg|_{\scriptscriptstyle w=\frac{y-x}{\sqrt{a}}}\hspace{-.8mm}\right)\nonumber\\
\hspace{-3mm} & = & \hspace{-3mm} -\frac{1}{2a}\hspace{-1mm}\left[ \frac{1}{\sqrt{a}} f_{\scriptscriptstyle W} \hspace{-1.5mm} \left(\frac{y \hspace{-.8mm} - \hspace{-.8mm} x}{\sqrt{a}}\right) \hspace{-1mm} + \hspace{-1mm}\frac{y \hspace{-.8mm} - \hspace{-.8mm} x }{\sqrt{a}} \left(\frac{1 }{\sqrt{a}} \frac{d}{dw}  f_{\scriptscriptstyle W} \hspace{-1.5mm}\left(w\right) \Bigg|_{\scriptscriptstyle w=\frac{y-x}{\sqrt{a}}}\right) \right]\nonumber\\
\hspace{-3mm} & = & \hspace{-3mm} -\frac{1}{2a}\hspace{-1mm}\left[  \left(\frac{d}{dy}(y\hspace{-.8mm}-\hspace{-.8mm}x)\right) f_{\scriptscriptstyle Y|X}(y|x;a)\hspace{-.8mm} + \hspace{-.8mm}(y\hspace{-.8mm}-\hspace{-.8mm}x)\frac{d}{dy} f_{\scriptscriptstyle Y|X}(y|x;a) \right]\nonumber\\
\hspace{-3mm} & = & \hspace{-3mm} -\frac{1}{2a}\frac{d}{dy} \left[  (y-x) f_{\scriptscriptstyle Y|X}(y|x;a)\right],\nonumber
\end{eqnarray}
and therefore,
\begin{eqnarray}
\frac{d}{da} f_{Y|X}(y|x;a) & = & -\frac{1}{2a}\frac{d}{dy} \left[  (y-x) f_{Y|X}(y|x;a)\right].\nonumber
\end{eqnarray}

\section{Explanation of Assumptions (\ref{bruijn_eq1_1}) in Corollaries \ref{cor2}, \ref{cor3}}\label{AppI}
\begin{enumerate}
\item Corollary \ref{cor2}\\%, \ref{cor6}
Given the channel $Y=X+\sqrt{a}W$ in (\ref{int_eq1_1}), $W$ is assumed to be exponentially distributed with unit parameter, i.e., its pdf $f_W(w)$ is defined as $\exp(-w)U(w)$, where $U(\cdot)$ denotes a unit step function. Since random variables $X$ and $W$ are independent of each other, conditional density function $f_{Y|X}(y|x;a)$ is expressed as
\begin{eqnarray}
f_{Y|X}(y|x;a) = \frac{1}{\sqrt{a}}\exp \left(\frac{y-x}{\sqrt{a}}\right) U(y-x),
\end{eqnarray}
and its derivatives with respect to $y$ and $a$ are respectively denoted as
\begin{eqnarray}
\label{appl_eq43_1}
\hspace{-9mm} &&\hspace{-3mm} \frac{d}{dy}f_{\scriptscriptstyle Y|X}(y|x;a)\nonumber\\
\hspace{-9mm} & = &\hspace{-3mm}  -\frac{1}{\sqrt{a}} f_{\scriptscriptstyle Y|X}(y|x;a) \hspace{-.8mm}+ \hspace{-.8mm}\frac{1}{\sqrt{a}} \exp \hspace{-.8mm} \left(\frac{y-x}{\sqrt{a}}\right)\hspace{-1mm} \delta(y-x),
\end{eqnarray}
\begin{eqnarray}
\label{appl_eq43_2}
\hspace{-9mm} && \hspace{-3mm} \frac{d}{da}f_{\scriptscriptstyle Y|X}(y|x;a)\nonumber\\
\hspace{-9mm} & = & \hspace{-3mm} -\frac{1}{2a} f_{\scriptscriptstyle Y|X}(y|x;a) + \frac{(y-x)}{2a\sqrt{a}}  f_{\scriptscriptstyle Y|X}(y|x;a),
\end{eqnarray}
where $\delta(\cdot)$ is a Dirac delta function.

The absolute values of equations (\ref{appl_eq43_1}), (\ref{appl_eq43_2}) are bounded as
\begin{eqnarray}
&& \left| \frac{d}{dy}f_{Y|X}(y|x;a) \right|\nonumber\\
& = & \left| -\frac{1}{\sqrt{a}} f_{Y|X}(y|x;a) + \frac{1}{\sqrt{a}} \exp \left(\frac{y-x}{\sqrt{a}}\right) \delta(y-x) \right|\nonumber\\
& \leq & \left| \frac{1}{\sqrt{a}} f_{Y|X}(y|x;a)\right|  + \left|\frac{1}{\sqrt{a}} \exp \left(\frac{y-x}{\sqrt{a}}\right) \delta(y-x)\right| \nonumber\\
\label{appl_eq43_3}& \leq & \frac{1}{a} + \frac{1}{\sqrt{a}} \exp \left(\frac{y-x}{\sqrt{a}}\right) \delta(y-x),
\end{eqnarray}
and
\begin{eqnarray}
\hspace{-8mm} && \left| \frac{d}{da}f_{Y|X}(y|x;a) \right|\nonumber\\
\hspace{-8mm} & = & \left| -\frac{1}{2a} f_{Y|X}(y|x;a) + \frac{(y-x)}{2a\sqrt{a}}  f_{Y|X}(y|x;a) \right|\nonumber\\
\label{appl_eq43_4_1}
\hspace{-8mm} & \leq & \left| \frac{1}{2a} f_{Y|X}(y|x;a) \right| + \left|\frac{(y-x)}{2a\sqrt{a}}  f_{Y|X}(y|x;a) \right|\\
\label{appl_eq43_4}
\hspace{-8mm} & \leq & \frac{1}{2a\sqrt{a}} + E,
\end{eqnarray}
where $E = \max_y [(y-x)f_{Y|X}(y|x;a)]$. Since $f_{Y|X}(y|x;a)$ is exponentially decreasing as $y$ approaches $\infty$, the real valued $E$ always exists. Also, $\max_y f(Y|X)(y|x;a) = 1/\sqrt{a}$, and therefore, the inequalities in (\ref{appl_eq43_3}) and (\ref{appl_eq43_4}) are satisfied.

The right-hand side of (\ref{appl_eq43_3}) and (\ref{appl_eq43_4}) are now integrable as follows:
\begin{eqnarray}
\hspace{-6mm} &  & \hspace{-3mm}\mathbb{E}_{\scriptscriptstyle X} \hspace{-1mm}\left[ \frac{1}{a} + \frac{1}{\sqrt{a}} \exp \left(\frac{y-X}{\sqrt{a}}\right) \hspace{-1mm} \delta(y-X) \right] = \frac{1}{a} + f_{\scriptscriptstyle X}(y),\nonumber\\
\hspace{-6mm} &  & \hspace{-3mm}\mathbb{E}_{\scriptscriptstyle X}\hspace{-1mm} \left[ \frac{1}{2a\sqrt{a}} + E \right] = \frac{1}{2a\sqrt{a}} + E.
\end{eqnarray}
If a function $f_X(x)$ is bounded, by dominated convergence theorem, assumption (\ref{bruijn_eq1_1}a) is verified.

Second, assumption (\ref{bruijn_eq1_1}b) is verified as follows.
\begin{eqnarray}
\label{appl_eq44_4}
\hspace{-5mm} && \hspace{-3mm} \left| \frac{d}{da} \left( f_{\scriptscriptstyle Y}(y;a) \log f_{\scriptscriptstyle Y}(y;a) \right)\right|\\
\hspace{-5mm} & \leq & \hspace{-3mm} \left| \log f_{\scriptscriptstyle Y}(y;a) \frac{d}{da} f_{\scriptscriptstyle Y}(y;a)  \right| + \left|\frac{d}{da} f_{\scriptscriptstyle Y}(y;a) \right|\nonumber\\
\hspace{-5mm} & = & \hspace{-3mm} \Bigg| \log f_{\scriptscriptstyle Y}(y;a) \mathbb{E}_{\scriptscriptstyle X} \Bigg[ -\frac{1}{2a} f_{\scriptscriptstyle Y|X}(y|X;a) \nonumber\\
\hspace{-5mm} && \hspace{10mm} + \frac{(y-X)}{2a\sqrt{a}}  f_{\scriptscriptstyle Y|X}(y|X;a) \Bigg]  \Bigg| + \left|\frac{d}{da} f_{\scriptscriptstyle Y}(y;a) \right|\nonumber
\end{eqnarray}
\begin{eqnarray}
\hspace{-5mm} & = & \hspace{-3mm} \Bigg| \sqrt{f_{\scriptscriptstyle Y}(y;a)} \log f_{\scriptscriptstyle Y}(y;a)\Bigg( -\frac{1}{2a}\sqrt{f_{\scriptscriptstyle Y} (y;a)}\nonumber\\
\hspace{-5mm} &  & \hspace{3mm}  + \frac{y}{2a\sqrt{a}}\sqrt{f_{\scriptscriptstyle Y}(y;a)} - \frac{\mathbb{E}_{\scriptscriptstyle X} \left[ X f_{\scriptscriptstyle Y|X}(y|X;a) \right]}{2a\sqrt{a}\sqrt{f_{\scriptscriptstyle Y}(y;a)}} \Bigg) \Bigg|\nonumber\\
\hspace{-5mm} && \hspace{-3mm} + \left|\frac{d}{da} f_{\scriptscriptstyle Y}(y;a)\right |\nonumber\\
\label{appl_eq44_3}
\hspace{-5mm} & = & \hspace{-3mm} \underbrace{\left| 2\sqrt{f_{\scriptscriptstyle Y}(y;a)} \log \sqrt{f_{\scriptscriptstyle Y}(y;a)} \right|}_{(d_1)} \nonumber\\
\hspace{-5mm} &  &  \times \Bigg| -\frac{1}{2a}\sqrt{f_{\scriptscriptstyle Y}(y;a)}\hspace{-.8mm} + \hspace{-.8mm}\frac{y}{2a\sqrt{a}}\sqrt{f_{\scriptscriptstyle Y}(y;a)}\hspace{-.8mm} \nonumber\\
\hspace{-5mm} &  & \hspace{3mm}\underbrace{\hspace{30mm}- \frac{\mathbb{E}_{\scriptscriptstyle X}\hspace{-1mm} \left[ X f_{\scriptscriptstyle Y|X}(y|X;a) \right]}{2a\sqrt{a}\sqrt{f_{\scriptscriptstyle Y}(y;a)}} \Bigg|}_{(d_2)}\nonumber\\
\hspace{-5mm} && \hspace{-3mm} + \underbrace{\left|\frac{d}{da} f_{\scriptscriptstyle Y}(y;a)\right |}_{(d_3)}\\
\label{appl_eq44_3_1}
\hspace{-5mm} & \leq & \hspace{-3mm} K \left| 2\sqrt{f_{\scriptscriptstyle Y}(y;a)} \log \sqrt{f_{\scriptscriptstyle Y}(y;a)} \right| + \left|\frac{d}{da} f_{\scriptscriptstyle Y}(y;a)\right |.\nonumber
\end{eqnarray}

The term $(d_3)$ is bounded by an integrable function due to equation (\ref{appl_eq43_4_1}), factor $(d_2)$ is bounded by a constant $K$ due to assumptions (\ref{bruijn_eq1_1}c) and (\ref{bruijn_eq1_1}d), which will be proved later, and factor $(d_1)$ is bounded, and it is integrable:
\begin{eqnarray}
\hspace{-5mm} && \hspace{-3mm} \int_{\scriptscriptstyle 0}^{\scriptscriptstyle \infty} \left| \sqrt{f_{\scriptscriptstyle Y}(y;a)} \log \sqrt{f_{\scriptscriptstyle Y}(y;a)} \right| dy\nonumber\\
\hspace{-5mm} & = & \hspace{-3mm} \frac{1}{2}\int_{\scriptscriptstyle 0}^{\scriptscriptstyle \infty} \left|\sqrt{f_{\scriptscriptstyle Y}(y;a)} \log f_{\scriptscriptstyle Y}(y;a)\right| dy\nonumber\\
\hspace{-5mm} & = & \hspace{-3mm} \frac{1}{2}\int_{\scriptscriptstyle 0}^{\scriptscriptstyle \infty}\hspace{-1.5mm} \left( \hspace{-.8mm} \mathbb{E}_{\scriptscriptstyle X}\hspace{-1.5mm}  \left[ \frac{1}{\sqrt{a}} \exp \hspace{-1mm} \left( -\frac{1}{\sqrt{a}} (y-X) \right) U(y-X) \right] \hspace{-.8mm} \right)^{\frac{1}{2}}\nonumber\\
\hspace{-5mm} && \hspace{-3mm} \times \log \mathbb{E}_{\scriptscriptstyle X} \hspace{-1.5mm} \left[ \frac{1}{\sqrt{a}} \exp \left( -\frac{1}{\sqrt{a}} (y-X) \right) U(y-X) \right] dy\nonumber\\
\hspace{-5mm} & = & \hspace{-3mm} \frac{1}{2}\int_{\scriptscriptstyle 0}^{\scriptscriptstyle \infty} \frac{1}{\sqrt[4]{a}} \exp \left( -\frac{1}{2\sqrt{a}} y \right)\nonumber\\
\hspace{-5mm} && \hspace{-3mm} \times \left( \mathbb{E}_{\scriptscriptstyle X} \left[\exp \left( \frac{1}{\sqrt{a}} X \right) U(y-X) \right] \right)^{\frac{1}{2}}\nonumber\\
\hspace{-5mm} && \hspace{-3mm} \times \Bigg| \log \Bigg( \frac{1}{\sqrt{a}} \exp \left( -\frac{1}{\sqrt{a}} y\right)\nonumber\\
\hspace{-5mm} && \hspace{15mm} \mathbb{E}_{\scriptscriptstyle X} \left[ \exp \left( \frac{1}{\sqrt{a}} X \right) U(y-X) \right] \Bigg)\Bigg|dy\nonumber\\
\hspace{-5mm} & \leq & \hspace{-3mm} \frac{1}{2}\int_{\scriptscriptstyle 0}^{\scriptscriptstyle \infty} \hspace{-1.5mm} \frac{1}{\sqrt[4]{a}} \exp \hspace{-1mm} \left( -\frac{1}{2\sqrt{a}} y \right)\hspace{-1.5mm}  \left( \mathbb{E}_{\scriptscriptstyle X} \hspace{-1.5mm} \left[\exp \hspace{-1mm} \left( \frac{1}{\sqrt{a}} X \right)  \right] \right)^{\frac{1}{2}}\nonumber\\
\hspace{-5mm} && \hspace{-3mm} \times \left| \log \left( \frac{1}{\sqrt{a}} \exp \hspace{-1mm} \left( -\frac{1}{\sqrt{a}} y\right) \mathbb{E}_{\scriptscriptstyle X} \hspace{-1.5mm} \left[ \exp \hspace{-1mm} \left( \frac{1}{\sqrt{a}} X \right) \right] \right)\right|dy\nonumber\\
\label{appl_eq44_3_2}
\hspace{-5mm} & \leq & \hspace{-3mm} \frac{1}{2}\int_{\scriptscriptstyle 0}^{\scriptscriptstyle \infty} \frac{1}{\sqrt[4]{a}} \exp \left( -\frac{1}{2\sqrt{a}} y \right) \left( M_{\scriptscriptstyle X} \left( \frac{1}{\sqrt{a}} \right)   \right)^{\frac{1}{2}}\nonumber\\
\hspace{-5mm} && \hspace{-3mm} \times \left| \log \left( \frac{1}{\sqrt{a}} \exp \left( -\frac{1}{\sqrt{a}} y\right) M_{\scriptscriptstyle X} \left( \frac{1}{\sqrt{a}} \right) \right)\right|dy,
\end{eqnarray}
where $M_X(\cdot)$ denotes the moment generating function of $X$. If the moment generating function of $X$ exists, then equation (\ref{appl_eq44_3_2}) is bounded and integrable, and so does the term $(d_1)$. Therefore, term $(d_1)$ is integrable with respect to $y$, and assumption (\ref{bruijn_eq1_1}b) is verified by dominated convergence theorem.

Similarly, assumption (\ref{bruijn_eq1_1}c) is verified as follows.
\begin{eqnarray}
\label{appl_eq45_1_1}\left| f_{Y|X}(y|x;a) \right| & = & \left|\frac{1}{\sqrt{a}}\exp \left(\frac{y-x}{\sqrt{a}}\right) U(y-x) \right|\nonumber\\
 & \leq & \frac{1}{\sqrt{a}}, \\
\label{appl_eq45_2_1}\left| x f_{Y|X}(y|x;a) \right| & = &\left| x \frac{1}{\sqrt{a}}\exp \left(\frac{y-x}{\sqrt{a}}\right) U(y-x)\right|\nonumber\\
 & \leq & \frac{1}{\sqrt{a}} x,
\end{eqnarray}
and the right hand-side terms of (\ref{appl_eq45_1_1}) and (\ref{appl_eq45_2_1}) are integrable as
\begin{eqnarray}
\mathbb{E}_X \left[ \frac{1}{\sqrt{a}} \right] & = & \frac{1}{\sqrt{a}},\nonumber\\
\label{appl_eq45_2_2}\mathbb{E}_X \left[ \frac{1}{\sqrt{a}} X \right] & = & \frac{1}{\sqrt{a}} \mathbb{E}_X [X],
\end{eqnarray}
and if $E_X[X]$ exists, assumption (\ref{bruijn_eq1_1}c) is satisfied.

Since $f_{Y|X}(y|x;a)$ is exponentially decreasing, $\lim\limits_{y \rightarrow \infty} y^2 f_Y(y;a)$ is zero. In addition,
\begin{eqnarray}\label{appl_eq45_2_3}
&& \lim\limits_{y \rightarrow 0} y^2 f_Y(y;a)\nonumber\\
& = & \lim\limits_{y \rightarrow 0} \mathbb{E}_X \left[ y^2 f_{Y|X}(y|X;a) \right]\nonumber\\
& = & \lim\limits_{y \rightarrow 0} \mathbb{E}_X \left[ y^2 \frac{1}{\sqrt{a}}\exp \left(\frac{y-x}{\sqrt{a}}\right) U(y-x) \right]\nonumber\\
& = & \mathbb{E}_X \left[ 0 \times \frac{1}{\sqrt{a}}\exp \left(\frac{-x}{\sqrt{a}}\right) U(-x) \right]\nonumber\\
& = & 0.
\end{eqnarray}

Assumption (\ref{bruijn_eq1_1}d) is expressed as
\begin{eqnarray}
\hspace{-9mm} && \hspace{-3mm} \frac{\mathbb{E}_{\scriptscriptstyle X} \left[ X f_{\scriptscriptstyle Y|X} (y|X;a) \right]}{\sqrt{f_{\scriptscriptstyle Y}(y;a)}}\nonumber\\
\hspace{-9mm} & = &  \hspace{-3mm}\frac{\mathbb{E}_{\scriptscriptstyle X} \left[ X f_{\scriptscriptstyle Y|X} (y|X;a) \right]}{f_{\scriptscriptstyle Y}(y;a)} \sqrt{f_{\scriptscriptstyle Y}(y;a)}\nonumber\\
\label{appl_eq44_1_1}
\hspace{-9mm} & = & \hspace{-3mm} \frac{\int_{\scriptscriptstyle 0}^{\scriptscriptstyle \infty} \hspace{-1mm} x f_{\scriptscriptstyle X} (x) \frac{1}{\sqrt{a}}\exp \hspace{-1mm} \left(\hspace{-.5mm} \frac{y-x}{\sqrt{a}}\right)\hspace{-1mm}  U\hspace{-.5mm} (y\hspace{-.5mm} -\hspace{-.5mm} x)dx }{\int_{\scriptscriptstyle 0}^{\scriptscriptstyle \infty} \hspace{-1mm} f_{\scriptscriptstyle X} (x) \frac{1}{\sqrt{a}}\exp \hspace{-1mm} \left(\hspace{-.5mm} \frac{y-x}{\sqrt{a}}\right) \hspace{-1mm} U\hspace{-.5mm} (y\hspace{-.5mm} -\hspace{-.5mm} x)dx}\hspace{-.5mm}  \sqrt{f_{\scriptscriptstyle Y}(y;a)}\\
\label{appl_eq44_2_1}
\hspace{-9mm} & \leq & \hspace{-3mm} \frac{y \int_{0}^{y}  f_{\scriptscriptstyle X} (x) \frac{1}{\sqrt{a}}\exp \left(\frac{y-x}{\sqrt{a}}\right) dx }{\int_{\scriptscriptstyle 0}^{\scriptscriptstyle y} f_{\scriptscriptstyle X} (x) \frac{1}{\sqrt{a}}\exp \left(\frac{y-x}{\sqrt{a}}\right) dx} \sqrt{f_{\scriptscriptstyle Y}(y;a)}\\
\hspace{-9mm} & = & \hspace{-3mm} y \sqrt{f_{\scriptscriptstyle Y}(y;a)}.\nonumber
\end{eqnarray}
The inequality in (\ref{appl_eq44_2_1}) is due to the fact that, in (\ref{appl_eq44_1_1}), the term inside integral is non-negative, $x$ is increasing, and integration is performed from $0$ to $y$.

Therefore, the assumptions in (\ref{bruijn_eq1_1}) require the following conditions: 1) existence of $\mathbb{E}_X [X]$, 2) existence of $M_X(\cdot)$, 3) bounded pdf $f_X(x)$, and these are further simplified into the existence of the moment generating function  of $X$ and bounded pdf $f_X(x)$. %due to equation (\ref{appl_eq45_2_2}).

\item Corollary \ref{cor3}\\%, \ref{cor7}
Given the channel  $Y=X+\sqrt{a}W$ in (\ref{int_eq1_1}), $W$ is assumed to be a gamma random variable, and its pdf is expressed as
\begin{eqnarray}
f_W(w) = \frac{1}{\Gamma(\alpha)} w^{\alpha-1} \exp(-w) U(w),\nonumber
\end{eqnarray}
where $\Gamma(\cdot)$ is a gamma function, $U(\cdot)$ denotes a unit step function, and $\alpha \geq 2$. Since random variables $X$ and $W$ are independent of each other,  the conditional density function $f_{Y|X}(y|x;a)$ is expressed as
\begin{eqnarray}
\hspace{-10mm}&&\hspace{-3mm}f_{Y|X}(y|x;a)\nonumber\\
\hspace{-10mm}& = & \hspace{-3mm}\frac{1}{\sqrt{a}\Gamma(\alpha)}\hspace{-.5mm}  \left(\frac{y-x}{\sqrt{a}}\right)^{\alpha-1} \hspace{-2mm}\exp\hspace{-.7mm} \left(-\frac{y-x}{\sqrt{a}}\right)\hspace{-.5mm}  U(y-x),
\end{eqnarray}
and its derivatives are denoted as
\begin{eqnarray}
\label{appl_eq46_1}
\hspace{-6mm}&& \hspace{-3mm}\frac{d}{dy} f_{\scriptscriptstyle Y|X}(y|x;a)\nonumber\\
\hspace{-6mm}& = &\hspace{-3mm} -\frac{1}{\sqrt{a}} f_{\scriptscriptstyle Y|X}(y|x;a)\nonumber\\
\hspace{-6mm}&&\hspace{-3mm} + \frac{1}{a \Gamma(\alpha \hspace{-.7mm} - \hspace{-.7mm} 1)} \left(\frac{y \hspace{-.7mm} - \hspace{-.7mm}x}{\sqrt{a}}\right)^{\alpha-2} \hspace{-3mm}\exp \hspace{-.8mm}\left( -\frac{y \hspace{-.7mm} - \hspace{-.7mm}x}{\sqrt{a}}\right) U(y \hspace{-.7mm} - \hspace{-.7mm} x),\\
\label{appl_eq46_2}
\hspace{-6mm}&&\hspace{-3mm} \frac{d}{da} f_{\scriptscriptstyle Y|X}(y|x;a)\nonumber\\
\hspace{-6mm}& = &\hspace{-3mm} -\frac{\alpha}{2a}f_{\scriptscriptstyle Y|X}(y|x;a) \nonumber\\
\hspace{-6mm}&&\hspace{-3mm} + \frac{\alpha}{2a}\hspace{-1mm} \left(\hspace{-.5mm} \frac{1}{\sqrt{a}\Gamma(\alpha \hspace{-.7mm}+ \hspace{-.7mm} 1)} \hspace{-1mm}\left( \frac{y \hspace{-.7mm} - \hspace{-.7mm} x}{\sqrt{a}} \right)^{\alpha}\hspace{-2mm} \exp \hspace{-1mm}\left(\hspace{-1mm} -\frac{y \hspace{-.7mm} - \hspace{-.7mm} x}{\sqrt{a}} \right)\hspace{-1mm}U\hspace{-.7mm}(y \hspace{-.7mm} - \hspace{-.7mm} x)\hspace{-.7mm}\right).\nonumber\\
\hspace{-6mm}&&\hspace{-3mm}
\end{eqnarray}

The absolute values of equations (\ref{appl_eq46_1}), (\ref{appl_eq46_2}) are bounded as
\begin{eqnarray}
\hspace{-4mm}&& \hspace{-3mm}\left| \frac{d}{dy} f_{Y|X}(y|x;a) \right|\nonumber\\
\hspace{-6mm}& = & \hspace{-3mm}\Bigg|-\frac{1}{\sqrt{a}} f_{Y|X}(y|x;a)\nonumber\\
\hspace{-6mm}&&\hspace{3mm} + \frac{1}{a \Gamma(\alpha-1)}\hspace{-1mm} \left(\frac{y-x}{\sqrt{a}}\right)^{\alpha-2}\hspace{-3mm} \exp \hspace{-1mm}\left(\hspace{-.8mm} -\frac{y-x}{\sqrt{a}}\right) \hspace{-1mm}U(y-x)\Bigg|\nonumber\\
\hspace{-6mm}& \leq &\hspace{-3mm} \left|\frac{1}{\sqrt{a}} f_{Y|X}(y|x;a)\right| \nonumber\\
\hspace{-6mm}&&\hspace{-3mm} + \left|\frac{1}{a \Gamma(\alpha-1)} \hspace{-1mm}\left(\frac{y-x}{\sqrt{a}}\right)^{\alpha-2}\hspace{-3mm} \exp \hspace{-1mm}\left(\hspace{-.8mm} -\frac{y-x}{\sqrt{a}}\right) \hspace{-1mm}U(y-x)\right|\nonumber\\
\hspace{-6mm}& = &\hspace{-3mm} \left|\frac{1}{\sqrt{a}} f_{Y|X}(y|x;a)\right| + \left|\frac{1}{\sqrt{a}}  f_{Y_{\alpha-1}|X}(y|x;a)\right|\nonumber\\
\label{appl_eq46_3}
\hspace{-6mm}& = &\hspace{-3mm} \frac{1}{\sqrt{a}} f_{Y|X}(y|x;a)+ \frac{1}{\sqrt{a}}  f_{Y_{\alpha-1}|X}(y|x;a),
\end{eqnarray}
where
\begin{eqnarray}
\hspace{-10mm} && \hspace{-3mm} f_{Y_{\alpha-1}|X}(y|x;a)\nonumber\\
\hspace{-10mm} &=& \hspace{-3mm} \frac{1}{\sqrt{a} \Gamma(\alpha   \hspace{-.7mm} -  \hspace{-.7mm} 1)}  \hspace{-1mm} \left(\frac{y \hspace{-.7mm} - \hspace{-.7mm} x}{\sqrt{a}}\right)^{\alpha-2}  \hspace{-3mm} \exp  \hspace{-1mm} \left( -\frac{y \hspace{-.7mm} - \hspace{-.7mm} x}{\sqrt{a}}\right)  \hspace{-1mm} U(y \hspace{-.7mm} - \hspace{-.7mm} x),
\end{eqnarray}
i.e., this is a gamma density function with two parameters defined as $\alpha-1$ and $1$,
and
\begin{eqnarray}
\hspace{-3mm} && \hspace{-3mm} \left| \frac{d}{da} f_{\scriptscriptstyle Y|X}(y|x;a) \right|\nonumber\\
\hspace{-7mm} & = & \hspace{-3mm} \Bigg| -\frac{\alpha}{2a}f_{\scriptscriptstyle Y|X}(y|x;a)\nonumber\\
\hspace{-7mm} && \hspace{3mm}  + \frac{\alpha}{2a} \left( \frac{1}{\sqrt{a}\Gamma(\alpha+1)}\hspace{-1mm} \left(\frac{y-x}{\sqrt{a}}\right)^{\alpha} \hspace{-1mm}\exp \hspace{-1mm}\left( -\frac{y-x}{\sqrt{a}} \right) \right) \Bigg|\nonumber\\
\hspace{-7mm} & \leq & \hspace{-3mm} \left| \frac{\alpha}{2a}f_{\scriptscriptstyle Y|X}(y|x;a) \right|\nonumber\\
\hspace{-7mm} && \hspace{-3mm} + \left|\frac{\alpha}{2a} \left( \frac{1}{\sqrt{a}\Gamma(\alpha+1)}\hspace{-1mm} \left(\frac{y-x}{\sqrt{a}}\right)^{\alpha}\hspace{-1mm} \exp \hspace{-1mm}\left( -\frac{y-x}{\sqrt{a}} \right) \right) \right|\nonumber\\
\hspace{-7mm} & = & \hspace{-3mm} \left| \frac{\alpha}{2a}f_{\scriptscriptstyle Y|X}(y|x;a) \right| + \left|\frac{\alpha}{2a}f_{\scriptscriptstyle Y_{\alpha+1}|X}(y|x;a)  \right|\nonumber\\
\label{appl_eq46_4}
\hspace{-7mm} & = & \hspace{-3mm} \frac{\alpha}{2a}f_{\scriptscriptstyle Y|X}(y|x;a)  + \frac{\alpha}{2a}f_{\scriptscriptstyle Y_{\alpha+1}|X}(y|x;a),
\end{eqnarray}
where
\begin{eqnarray}
\hspace{-7mm} &  & \hspace{-3mm} f_{Y_{\alpha+1}|X}(y|x;a) \nonumber\\
\hspace{-7mm} & = & \hspace{-3mm}\frac{1}{\sqrt{a} \Gamma(\alpha \hspace{-.7mm} + \hspace{-.7mm} 1)} \hspace{-1mm}\left(\frac{y \hspace{-.7mm} - \hspace{-.7mm} x}{\sqrt{a}}\right)^{\alpha} \hspace{-1mm} \exp \hspace{-1mm} \left( -\frac{y \hspace{-.7mm} - \hspace{-.7mm} x}{\sqrt{a}}\right)\hspace{-1mm} U(y \hspace{-.7mm} - \hspace{-.7mm}x),
\end{eqnarray}
i.e., this is a gamma density function with two parameters defined as $\alpha+1$ and $1$.

Since $f_{\scriptscriptstyle Y_{\alpha-1}|X}(y|x;a)$, $f_{\scriptscriptstyle Y|X}(y|x;a)$, and $f_{\scriptscriptstyle Y_{\alpha+1}|X}(y|x;a)$ are all integrable, the right-hand side of (\ref{appl_eq46_3}) and (\ref{appl_eq46_4}) are integrable as
\begin{eqnarray}
\hspace{-7mm}&& \hspace{-2mm}\mathbb{E}_X \left[ \frac{1}{\sqrt{a}} f_{Y|X}(y|X;a)+ \frac{1}{\sqrt{a}}  f_{Y_{\alpha-1}|X}(y|X;a) \right]\nonumber\\
\hspace{-7mm}& = & \hspace{-2mm}\frac{1}{\sqrt{a}} f_{Y}(y;a) + \frac{1}{\sqrt{a}}  f_{Y_{\alpha-1}}(y;a),\\
\hspace{-7mm}&& \hspace{-2mm}\mathbb{E}_X \left[ \frac{\alpha}{2a}f_{Y|X}(y|X;a)  + \frac{\alpha}{2a}f_{Y_{\alpha+1}|X}(y|X;a) \right]\nonumber\\
\hspace{-7mm}& = & \hspace{-2mm}\frac{\alpha}{2a} f_{Y}(y;a)  + \frac{\alpha}{2a}  f_{Y_{\alpha+1}}(y;a),
\end{eqnarray}
where $f_{\scriptscriptstyle Y_{\alpha-1}}(y;a) = \mathbb{E}_{\scriptscriptstyle X}[ f_{\scriptscriptstyle Y_{\alpha-1}|X}(y|X;a)]$, and $f_{\scriptscriptstyle Y_{\alpha+1}}(y;a) = \mathbb{E}_{\scriptscriptstyle X}[ f_{\scriptscriptstyle Y_{\alpha+1}|X}(y|X;a)]$. Therefore, assumption (\ref{bruijn_eq1_1}a) is verified by dominated convergence theorem.

Second, assumption (\ref{bruijn_eq1_1}b) is verified as  follows.
\begin{eqnarray}
\label{appl_eq46_5}
\hspace{-7mm} && \hspace{-3mm} \left| \frac{d}{da} \left( f_Y(y;a) \log f_Y(y;x) \right)\right|\\
\hspace{-7mm} & \leq & \hspace{-3mm} \left| \log f_Y(y;x) \frac{d}{da} f_Y(y;a)  \right| + \left|\frac{d}{da} f_Y(y;a) \right|\nonumber\\
\hspace{-7mm} & = & \hspace{-3mm} \Bigg| \log f_Y(y;x) \mathbb{E}_X \Bigg[ -\frac{1}{2a} f_{Y|X}(y|X;a)\nonumber\\
\hspace{-7mm} && \hspace{6mm}  + \frac{(y-X)}{2a\sqrt{a}}  f_{Y|X}(y|X;a) \Bigg]  \Bigg| + \left|\frac{d}{da} f_Y(y;a) \right|\nonumber\\
\hspace{-7mm} & = & \hspace{-3mm} \Bigg| 2\sqrt{f_Y(y;x)} \log \sqrt{f_Y(y;x)} \Bigg( -\frac{1}{2a}\sqrt{f_Y(y;x)}\nonumber\\
\hspace{-7mm} && \hspace{6mm} + \frac{y}{2a\sqrt{a}}\sqrt{f_Y(y;x)} - \frac{\mathbb{E}_X \left[ X f_{Y|X}(y|X;a) \right]}{2a\sqrt{a}\sqrt{f_Y(y;x)}} \Bigg) \Bigg|\nonumber\\
\hspace{-7mm} && \hspace{-3mm} + \left|\frac{d}{da} f_Y(y;a)\right |\nonumber
\end{eqnarray}
\begin{eqnarray}
\label{appl_eq46_5_1}
\hspace{-10mm} & = & \hspace{-3mm} \underbrace{\left| 2\sqrt{f_Y(y;x)} \log \sqrt{f_Y(y;x)} \right|}_{(e_1)}\nonumber\\
\hspace{-10mm} && \hspace{3mm}\times \Bigg| -\frac{1}{2a}\sqrt{f_Y(y;x)} + \frac{y}{2a\sqrt{a}}\sqrt{f_Y(y;x)}\nonumber\\
\hspace{-10mm} &&  \hspace{6mm} \underbrace{\hspace{35mm}- \frac{\mathbb{E}_X \left[ X f_{Y|X}(y|X;a) \right]}{2a\sqrt{a}\sqrt{f_Y(y;x)}} \Bigg|}_{(e_2)}\nonumber\\
\hspace{-10mm} && \hspace{-3mm} + \underbrace{\left|\frac{d}{da} f_Y(y;a)\right |}_{(e_3)}.
\end{eqnarray}
\end{enumerate}
The factors $(e_1)$, $(e_2)$, and $(e_3)$ can be verified using exactly the same reasons as the factors $(d_1)$, $(d_2)$, and $(d_3)$, in (\ref{appl_eq44_3}), respectively. Therefore, like equation (\ref{appl_eq44_3_2}), the existence of moment generating function of $X$ is required.

Assumption (\ref{bruijn_eq1_1}c) is confirmed by the following procedures.

Since $f_{Y|X}(y|x;a)$ is exponentially decreasing, $\lim\limits_{y \rightarrow \infty} y^2 f_Y(y;a)$ is zero. By the same procedure as equation (\ref{appl_eq45_2_3}), $y^2 f_Y(y;a)$ becomes zero as $y$ approaches zero. In addition,
\begin{eqnarray}
\label{appl_eq46_6}\left|f_{Y|X}(y|x;a) \right| \hspace{-2mm} & \leq &   \hspace{-2mm} f_{Y|X}(y|x;a) \Big|_{y=x+\sqrt{a}(\alpha-1)} , \\
\label{appl_eq46_7}\left| x f_{Y|X}(y|x;a) \right|  \hspace{-2mm} & \leq &   \hspace{-2mm} x f_{Y|X}(y|x;a) \Big|_{y=x+\sqrt{a}(\alpha-1)} .
\end{eqnarray}
The inequalities above are due to the fact that the function $ f_{Y|X}(y|x;a)$ is always nonnegative, and it is maximized at $y=x+\sqrt{a}(\alpha-1)$. Therefore, the right-hand sides of (\ref{appl_eq46_6}) and (\ref{appl_eq46_7}) are integrable as
\begin{eqnarray}
\hspace{-8mm}&& \mathbb{E}_X \left[ \frac{1}{\sqrt{a}\Gamma(\alpha)} (\alpha-1)^{\alpha-1} \exp(-(\alpha-1)) \right]\nonumber\\
\hspace{-8mm}& = & \frac{1}{\sqrt{a}\Gamma(\alpha)} (\alpha-1)^{\alpha-1} \exp(-(\alpha-1)),\nonumber\\
\label{appl_eq46_7_1}
\hspace{-8mm}&& \mathbb{E}_X \left[X \frac{1}{\sqrt{a}\Gamma(\alpha)} (\alpha-1)^{\alpha-1} \exp(-(\alpha-1)) \right]\nonumber\\
\hspace{-8mm}& = & \frac{1}{\sqrt{a}\Gamma(\alpha)} (\alpha-1)^{\alpha-1} \exp(-(\alpha-1))\mathbb{E}_X [X],
\end{eqnarray}
and, if $\mathbb{E}_X[X]$ exits, by dominated convergence theorem, assumption (\ref{bruijn_eq1_1}c) is verified.

Finally, assumption (\ref{bruijn_eq1_1}d) is expressed as
\begin{eqnarray}
\hspace{-8mm} &&\hspace{-3mm} \frac{\mathbb{E}_{\scriptscriptstyle X} \left[ X f_{\scriptscriptstyle Y|X} (y|X;a) \right]}{\sqrt{f_{\scriptscriptstyle Y}(y;a)}}\nonumber\\
\hspace{-8mm} & = & \hspace{-3mm} \frac{\mathbb{E}_{\scriptscriptstyle X} \left[ X f_{\scriptscriptstyle Y|X} (y|X;a) \right]}{f_{\scriptscriptstyle Y}(y;a)} \sqrt{f_{\scriptscriptstyle Y}(y;a)}\nonumber\\
\label{appl_eq46_8}
\hspace{-8mm} & = & \hspace{-3mm} \frac{\hspace{-1mm}\int_{\scriptscriptstyle 0}^{\scriptscriptstyle \infty}\hspace{-1mm}  x\hspace{-.7mm} f_{\scriptscriptstyle \hspace{-.5mm} X}\hspace{-.7mm} (x) \hspace{-.5mm}\frac{1}{\sqrt{a}\Gamma(\alpha)}\hspace{-1mm} \left(\frac{y-x}{\sqrt{a}}\right)^{\scriptscriptstyle \alpha-1} \hspace{-3mm} \exp\hspace{-1mm} \left(\hspace{-.7mm}-\frac{y-x}{\sqrt{a}}\right)\hspace{-1mm} U\hspace{-.7mm}(y\hspace{-.7mm}-\hspace{-.7mm}x)dx }{\hspace{-1mm}\int_{\scriptscriptstyle 0}^{\scriptscriptstyle \infty} \hspace{-1mm} f_{\scriptscriptstyle \hspace{-.5mm} X}\hspace{-.7mm} (x) \hspace{-.7mm}\frac{1}{\sqrt{a}\Gamma(\alpha)}\hspace{-1mm} \left(\frac{y-x}{\sqrt{a}}\right)^{\scriptscriptstyle \alpha-1} \hspace{-3mm} \exp\hspace{-1mm} \left(\hspace{-.7mm}-\frac{y-x}{\sqrt{a}}\right)\hspace{-1mm} U\hspace{-.7mm}(y \hspace{-.7mm}-\hspace{-.7mm}x)dx} \hspace{-1mm}\sqrt{\hspace{-.7mm}f_{\scriptscriptstyle Y}\hspace{-1mm}(y;a)}\nonumber\\
\hspace{-8mm} && \\
\label{appl_eq46_9}
\hspace{-8mm} & \leq & \hspace{-3mm} \frac{y \hspace{-.9mm}\int_{\scriptscriptstyle 0}^{\scriptscriptstyle y} \hspace{-1mm}  f_{\scriptscriptstyle \hspace{-.5mm}X} \hspace{-.7mm}(x) \hspace{-.5mm}\frac{1}{\sqrt{a}\Gamma(\alpha)} \hspace{-.9mm}\left(\frac{y-x}{\sqrt{a}}\right)^{\scriptscriptstyle \alpha-1} \hspace{-3mm} \exp\hspace{-1mm}  \left(-\frac{y-x}{\sqrt{a}}\right) dx }{\int_{\scriptscriptstyle 0}^{\scriptscriptstyle y} \hspace{-1mm} f_{\scriptscriptstyle \hspace{-.5mm}X} \hspace{-.7mm}(x) \hspace{-.5mm}\frac{1}{\sqrt{a}\Gamma(\alpha)} \hspace{-.9mm}\left(\frac{y-x}{\sqrt{a}}\right)^{\scriptscriptstyle \alpha-1} \hspace{-3mm} \exp\hspace{-1mm}  \left(-\frac{y-x}{\sqrt{a}}\right)  dx}\hspace{-1mm}  \sqrt{\hspace{-.7mm}f_{\scriptscriptstyle Y}\hspace{-.7mm}(y;a)}\\
\hspace{-8mm} & = & \hspace{-3mm} y \sqrt{f_{\scriptscriptstyle Y}(y;a)}.\nonumber
\end{eqnarray}
The inequality in (\ref{appl_eq46_9}) is due to the fact that, in (\ref{appl_eq46_8}), the term inside integral is non-negative, $x$ is increasing, and the integration with respect to $x$ is performed from $0$ to $y$.

Therefore, in this case, the assumptions in (\ref{bruijn_eq1_1}) require the existence of the mean and moment generating function of $X$, and these are further simplified to the existence of the moment generating function of $X$.% due to equation (\ref{appl_eq46_7_1}).

%\begin{biography}[{\includegraphics[width=1in,height=1.25in,clip,keepaspectratio]{mshell}}]{Michael Shell}

\begin{IEEEbiographynophoto}{Sangwoo Park}
Sangwoo Park received the B.S. degree in electrical engineering from Chung-Ang University (CAU), Seoul, Korea in 2004, and the M.S. degree in electrical engineering from Texas A\&M University, College Station in 2008. From 2004 to 2005, he has worked as a full-time assistant engineer for UMTS/WCDMA projects in Samsung Electronics. Currently, he is pursuing his PhD degree under supervision of Dr. Erchin Serpedin and Dr. Khalid Qaraqe. His research interests lie in wireless communications, information theory, and statistical signal processing.
\end{IEEEbiographynophoto}

\begin{IEEEbiographynophoto}{Erchin Serpedin}
Erchin Serpedin (SM¡¯04) received the specialization degree in signal processing and transmission of information from Ecole Superieure D'Electricite (SUPELEC), Paris, France, in 1992, the MSc degree from the Georgia Institute of Technology, Atlanta, in 1992, and the PhD degree in electrical engineering from the University of Virginia, Charlottesville, in January 1999. He is currently a professor in the Department of Electrical and Computer Engineering at Texas A\&M University, College Station. He is the author of two research monographs, one edited textbook, 85 journal papers, and 130 conference papers, and serves currently as associate editor for IEEE Transactions
on Information Theory, IEEE Transactions on Communications, Signal Processing (Elsevier), EURASIP Journal on Advances in Signal Processing, and EURASIP Journal on Bioinformatics and Systems Biology. His research interests include statistical signal processing, wireless communications, information theory, bioinformatics, and genomics.
\end{IEEEbiographynophoto}

%\newpage

\begin{IEEEbiographynophoto}{Khalid Qaraqe}
Dr Khalid A. Qaraqe (M'97-S'00 ) was born in Bethlehem. Dr Qaraqe received the B.S. degree in EE from the University of Technology, in 1986, with honors. He received the M.S. degree in EE from the University of Jordan, Jordan, in 1989, and he earned his Ph.D. degree in EE from Texas A\&M University, College Station, TX, in 1997. From 1989 to 2004 Dr Qaraqe has held a variety positions in many companies and he has over 12 years of experience in the telecommunication industry.  Dr Qaraqe has worked for Qualcomm, Enad Design Systems, Cadence Design Systems/Tality Corporation, STC, SBC and Ericsson.   He has worked on numerous GSM, CDMA, WCDMA projects and   has experience in product development, design, deployments, testing and integration.  Dr Qaraqe joined the department of Electrical Engineering of Texas A\&M University at Qatar, in July 2004, where he is now a professor.

Dr Qaraqe research interests include communication theory and its application to design and performance, analysis of cellular systems and indoor communication systems. Particular interests are in the development of 3G UMTS, cognitive radio systems,  broadband  wireless communications and diversity techniques.
\end{IEEEbiographynophoto}


%\vfill


\end{document}